\documentclass[traditabstract]{aa}
\usepackage{graphicx}
\usepackage{natbib}
\usepackage{amssymb}
\usepackage{amsmath}
\usepackage{color}
\bibpunct{(}{)}{;}{a}{}{,} % to follow the A&A style
\usepackage{txfonts}
\usepackage{hyperref}
\begin{document}
   \title{Gravothermal Catastrophe: the dynamical stability of a fluid model}

%   \subtitle{}

   \author{M. C. Sormani
          \inst{1}
          \and
          G. Bertin\inst{2}%\fnmsep%\thanks{Just to show the usage
          %of the elements in the author field}
          }

  \institute{Scuola Normale Superiore, Piazza dei Cavalieri 7, 56126 Pisa, Italy
   \thanks{Present address: Rudolf Peierls Centre for Theoretical Physics, 1 Keble Road, Oxford OX1 3NP}\\
              \email{mattia.sormani@physics.ox.ac.uk}
         \and
             Universit\`a degli Studi di Milano, Dipartimento di Fisica, via Celoria 16, I-20133 Milano, Italy\\
             \email{giuseppe.bertin@unimi.it}
            % \thanks{The university of heaven temporarily does not
              %       accept e-mails}
             }

\date{Received XXX; accepted YYY}

% \abstract{}{}{}{}{} 
% 5 {} token are mandatory
 
  \abstract{A re-investigation of the gravothermal catastrophe is presented. By means of a linear perturbation analysis, we study the dynamical stability of a spherical self-gravitating isothermal fluid of finite volume and find that the conditions for the onset of the gravothermal catastrophe, under different external conditions, coincide with those obtained from thermodynamical arguments. This suggests 
that the gravothermal catastrophe may reduce to Jeans instability, rediscovered in an inhomogeneous framework. We find normal modes and frequencies for the fluid system and show that instability develops on the dynamical time scale. We then discuss several related issues. In particular: (1) For perturbations at constant total energy and constant volume, we introduce a simple heuristic term in the energy budget to mimic the role of binaries. (2)  We outline the analysis of the two-component case and show how linear perturbation analysis can be carried out also in this more complex context in a relatively straightforward way. (3) We compare the behavior of the fluid model with that of the collisionless sphere. In the collisionless case the instability seems to disappear, which is at variance with the linear Jeans stability analysis in the homogeneous case; we argue that a key ingredient to understand the difference (a spherical stellar system is expected to undergo the gravothermal catastrophe only in the presence of some collisionality, which suggests that the instability is dissipative and not dynamical) lies in the role of the detailed angular momentum in a collisionless system. 

 Finally, we briefly comment on the meaning of the Boltzmann entropy and its applicability to the study of the dynamics of self-gravitating inhomogeneous gaseous systems.}
  % aims heading (mandatory)
%   {It is shown that stability
%   depends only upon the equations of state, the opacities and the local
%   thermodynamic state in the layer. Stability and instability can
%   therefore be expressed in the form of stability equations of state
%   which are universal for a given composition.}
%  % methods heading (mandatory)
%   {The stability equations of state are
%   calculated for solar composition and are displayed in the domain
%   $-14 \leq \lg \rho / \mathrm{[g\, cm^{-3}]} \leq 0 $,
%   $ 8.8 \leq \lg e / \mathrm{[erg\, g^{-1}]} \leq 17.7$. These displays
%   may be
%   used to determine the one-zone stability of layers in stellar
%   or planetary structure models by directly reading off the value of
%   the stability equations for the thermodynamic state of these layers,
%   specified
%   by state quantities as density $\rho$, temperature $T$ or
%   specific internal energy $e$.
%   Regions of instability in the $(\rho,e)$-plane are described
%   and related to the underlying microphysical processes.}
%  % results heading (mandatory)
%   {Vibrational instability is found to be a common phenomenon
%   at temperatures lower than the second He ionisation
%   zone. The $\kappa$-mechanism is widespread under `cool'
%   conditions.}
%  % conclusions heading (optional), leave it empty if necessary 
%   {}

   \maketitle
%
%________________________________________________________________
\section{Introduction} 

Core collapse in an $N$-body system is a problem relevant to the dynamics of globular clusters, because these are considered to be the only stellar systems that possess the necessary degree of collisionality to relax thermally.  The discovery of some clusters with cuspy cores was interpreted as a sign that they indeed experienced the gravothermal catastrophe. In the context of globular clusters, the three main phenomena caused by collisionality are: core collapse, evaporation, and mass segregation. The focus of this paper is on core collapse.

Consider the gravitational $N$-body problem, where $N \gg 1$. That is, consider a set of $N$ classical point masses, each of mass $m$, mutually interacting through Newtonian 
gravity. The particles may or may not be confined within a spherical volume of radius $R$.
We are interested in the following questions:
\begin{itemize}
\item Can the system reach some sort of \emph{equilibrium}? Can this equilibrium be called \emph{thermal}?
\item Can a \emph{model} (for example, collisionless or fluid) of the $N$-body problem reach equilibrium? How does the equilibrium of a model relate to the equilibrium of the pure $N$-body problem and of a real stellar system?
\item Are these equilibria stable?
\end{itemize}
This topic has been studied by many authors \citep[for recent reviews see for example][]{heggiehut,binneytremaine}. Different models of the $N$-body problem (in particular gaseous models, collisionless models, and fluid models) admit equilibrium configurations that are spatially truncated self-gravitating isothermal spheres, but it is not clear to what extent they can be considered as representative of the equilibrium states of the pure $N$-body problem. Several studies have addressed the stability problem of isothermal spheres using thermodynamical methods, starting with \cite{antonov} and \cite{lyndenbellwood} and continuing with a long list of papers \citep[for example,][see also \citealt{thirring}]{hachisu1, nakada,katz1,inagaki,padmana1, chavanis1,chavanis2}. 
In the thermodynamical approach the study of isothermal spheres is based on a form of entropy known as the Boltzmann entropy:
\begin{equation}
S_\mathrm{b}[f] \label{boltzmannentropy} \equiv-k \int\! f(\mathbf{r},\mathbf{v})\ln f(\mathbf{r},\mathbf{v})\, \mathrm{d}^3r\,\mathrm{d}^3v~,\ 
\end{equation}
where $f$ is the one-particle distribution function,\footnote{Normalized in such a way that $\int\!  f\, \mathrm{d}^3 r\, \mathrm{d}^3v=N$} $\mathbf{r}$ and $\mathbf{v}$ the position and velocity vector respectively, and $k$ the Boltzmann constant.

Unfortunately, the thermodynamics of self-gravitating systems still depends on a number of unresolved issues, partly because of the long-range nature of the force and partly because of the divergent behavior of the force at short distance \citep[e.g., see][]{padmana2,katz2002,chavareview,mukamel1,ruffo,mukamel2}. A critical 
analysis of the use of the Boltzmann entropy has been made by \cite{miller}. We will briefly comment on this point in Sect.~\ref{thermo}. 

Therefore, it would be interesting to study the gravothermal catastrophe by limiting as much as possible the use of thermodynamical arguments. 
In this paper we reconsider the gravothermal catastrophe and analyze the dynamical stability of self-gravitating fluids governed by the Euler equation by finding the normal modes and frequencies of spherical systems under various external conditions, in particular: constant total energy $E$ and volume $V$ \citep[this case corresponds to the
gravothermal catastrophe of][]{lyndenbellwood}, constant temperature $T$ and volume $V$ (isothermal collapse), constant temperature $T$ and boundary pressure $P$ \citep[isobaric collapse,][]{bonnor,ebert}. Note that, in contrast to thermodynamical arguments, the linear modal analysis 
automatically gives the time scale for the development of the instability. The constant $\{T, V\}$ case has been addressed by \cite{sanchez}, who studied the stability of the system numerically, and by \cite{chavanis1}, who found an analytical solution for the marginally stable perturbations using methods developed by \cite{padmana1}. The constant $\{T, P\}$ case has been addressed previously with synthetic arguments by \cite{bonnor} and \cite{ebert}, by \cite{yabushita}, who studied the stability of the system numerically, by \cite{lombardibertin}, who extended the analysis by \cite{bonnor} to the nonspherically symmetric case, and 
by \cite{chavanis2}, who gave an analytical solution for the case of marginally stable perturbations using the same method as he had used for the constant $\{T, V\}$ case. The constant $\{E,V\}$ case has been studied using a model based on the Smoluchowski-Poisson system (different from the Euler-Poisson system considered in this paper), by \cite{chavanis3}.

In this paper, we extend these analyses of the constant $\{T, V\}$ and $\{T, P\}$ cases to the general calculation of eigenfrequencies and eigenfunctions for conditions outside those of marginal stability, and study in detail the constant $\{E, V\}$ case. We use a Eulerian or Lagrangian representation of hydrodynamics as suggested by the boundary conditions to be imposed on the system and 
we provide a unified treatment for all cases. Surprisingly, we find that the system becomes dynamically unstable in all the cases considered exactly at the same points found by \cite{lyndenbellwood} by means of a thermodynamical analysis, that is for values of the density contrast (the ratio of the central density to the boundary density) $\approx$ 14, 32.1, and 709 respectively for the constant $\{T, P\}$, $\{T,V\}$, and $\{E,V\}$ case. These results suggest that the gravothermal catastrophe may reduce to the \cite{jeans} instability, rediscovered in the inhomogeneous context. 
Then we briefly outline the problem of the linear dynamical stability of a spherical fluid generalized to the two-component case, by perturbing the two-component spatially truncated isothermal sphere configurations previously considered by \cite{taff}, \cite{lightman}, \cite{yoshizawa}, \cite{vega2comp}, who studied the stability of these systems with a thermodynamical approach, and \cite{chava2comp}, who addressed also the dynamical stability with a model different from the one used in this paper. We show that the onset of thermodynamical and dynamical stability occurs at the same point in the simplest case (constant $\{T, V\}$) and confirm that the component made of heavier particles is the primary driver of the instability. This result is likely to be related to a similar finding by \cite{breen2,breen1} for two-component gravothermal oscillations.
Finally we focus on the following puzzling phenomenon: in the collisionless model the instability disappears. The phenomenon is at variance with what happens in homogeneous systems \citep[e.g., see][]{bertin}. In the last part of the paper we argue that the cause of the difference lies in the role of the angular momentum of the individual particles.

The paper is organized as follows. 
In Sect.~\ref{thermo} we comment on the meaning of the Boltzmann entropy. 
In Sect.~\ref{basicequations} we write the basic hydrodynamic equations in the Eulerian and Lagrangian representation. In Sect.~\ref{hydroanalysis} we show the results of the linear analysis of the hydrodynamic equations by studying the properties of spherically-symmetric perturbations. We find the relevant normal modes and frequencies under different boundary conditions. In the analysis of the gravothermal catastrophe, we propose a modified expression for the energy, which can be considered as a simple way to incorporate the energy generation from binaries in the linear regime; we show that the catastrophe can indeed be halted if we consider such a modified expression. In the last subsection and in Appendix \ref{appendix2comp}, we briefly consider the two-component case. 
In Sect.~\ref{timescales} we discuss the relevant time scales. 
In Sect.~\ref{collvscollless} we discuss the difference between the collisionless model and the fluid model of the $N$-body problem. 
In Sect.~\ref{conclusions} we draw our conclusions and identify some open questions and issues.

\section{The thermodynamical approach and the Boltzmann entropy} \label{thermo}

\cite{antonov} and \cite{lyndenbellwood} based their stability analysis of  
self-gravitating spatially-truncated isothermal spheres on the use of the Boltzmann entropy \eqref{boltzmannentropy}.
Starting from a kinetic description, they looked for stationary states of the Boltzmann entropy with respect to the distribution function $f$, at fixed\footnote{In terms of the distribution function $f$, we have $M=m \int f \, \mathrm{d}^3x\, \mathrm{d}^3v$ and $$E=\int\! \frac{m \mathbf{v}^2}{2} f(\mathbf{r},\mathbf{v}) \, \mathrm{d}^3r\,  \mathrm{d}^3v- \frac{1}{2}Gm^2\int \!\frac{f(\mathbf{r},\mathbf{v})f(\mathbf{r}',\mathbf{v}')}{|\mathbf{r}-\mathbf{r}'|} \, \mathrm{d}^3r\,  \mathrm{d}^3v\, \mathrm{d}^3r'\,  \mathrm{d}^3v',$$ where $m$ is the single-particle mass.} total mass $M$, total energy $E$, and volume $V=4\pi R^3/3$, where $R$ is the radius of a spherical box. These stationary states are spatially truncated isothermal spheres. It is found that, for a given single-particle mass $m$, each spatially truncated isothermal sphere is identified 
by two dimensional scales (e.g., the central density $\rho_0(0)$ and the temperature $T$) and one dimensionless parameter \{for example $\Xi=R/\lambda$, where $\lambda=[kT/m 4\pi G \rho_0(0)]^{1/2}$ is reminiscent of the Jeans length \citep{jeans}; another equivalent choice for the dimensionless parameter is the density contrast $\rho_0(0)/\rho_0(R)$, that is, the ratio of the central density to the density at the truncation radius $R$\}. Conversely, each choice of the three quantities $\{T,\rho(0),\Xi\}$ identifies one particular isothermal sphere. 
This one-to-one correspondence may be lost for other choices of the quantities characterizing the system.

In this thermodynamical approach the necessary condition for stability is that the stationary state corresponds to a (local at least) maximum of the Boltzmann entropy functional. \cite{antonov} found that if $\Xi<34.4$ [i.e., if $\rho_0(0)/\rho_0(R)<709$] the equilibrium configurations 
are local maxima, whereas for $\Xi>34.4$ they are saddle points. Therefore, he concluded that self-gravitating isothermal spheres are unstable if $\Xi>34.4$.

The work of \cite{antonov} was extended by \cite{lyndenbellwood}. These authors 
studied the thermodynamical stability of the system under various conditions by applying the relevant thermodynamical 
potentials (based on the Boltzmann entropy) and gave a physical interpretation of the instability in terms of \emph{negative specific heats}, 
as a characteristic feature of self-gravitating systems, thus
creating the paradigm of the gravothermal catastrophe. They found that the critical value of $\Xi$ that corresponds to the onset of instability depends on the adopted conditions. For example, 
$\Xi=34.4$ holds for a system at constant total energy $E$ and volume $V$, $\Xi=8.99$ for a system at constant temperature $T$ and volume $V$, and $\Xi=6.45$ 
for a system at constant temperature $T$ and external pressure $P$. In this paper, we will show that the three points can be identified by means of a dynamical stability analysis of the fluid model.

Now we briefly comment on the use of the Boltzmann expression for the entropy \eqref{boltzmannentropy}. If we refer to systems for which the traditional thermodynamic 
limit ($N\to\infty$ and $V\to\infty$ at $N/V$ fixed) is well defined and leads to homogeneous equilibria, it is well known \citep{jaynes1} that the 
Boltzmann entropy corresponds to the entropy defined 
in phenomenological thermodynamics only when the interparticle forces do not affect the thermodynamic properties; that is, the Boltzmann entropy neglects the interparticle potential energy and the effect of the interparticle forces on the pressure. If the equation of state is different from that of an ideal gas, the Boltzmann entropy is in error by a non-negligible amount. The correct expression for the entropy (corresponding to phenomenological thermodynamics) of systems with a well-defined thermodynamic limit is the one given by Gibbs [for the definition of Gibbs entropy, see \cite{jaynes1}].

The generalization of the above result to systems with inhomogeneous equilibria, such as self-gravitating systems, is that the Boltzmann entropy is valid (i.e., it corresponds to entropy as defined in phenomenological thermodynamics) if and only if (1) the equation of state of the ideal gas applies locally to the equilibrium states; (2) hydrostatic equilibrium holds. Indeed, the most elementary way to understand the Boltzmann entropy in the case of self-gravitating systems is to think of a fluid in hydrostatic equilibrium, with the equation of state of an ideal gas $p=\rho kT/m$, subject to reversible transformations between equilibrium states 
(i.e., between different truncated isothermal spheres in the spherically symmetric case). As shown by \cite{lyndenbellwood} in their Appendix I, the classical thermodynamic entropy defined from the relation 
$\mathrm{d}S_c=\mathrm{d}Q/T$ and calculated from such transformations coincides with $S_b$. 
For the general case of a system of particles interacting through an arbitrary two-body potential (not necessarily gravitational), it is possible to show that the stationary states of the Boltzmann entropy have the same density distribution as that of a fluid with the equation of state of an ideal gas
in hydrostatic equilibrium.

The above considerations suggest that the validity of the Boltzmann entropy is strictly related to the validity of the equation of state of an ideal gas.
In a kinetic description, the equation of state of an ideal gas is obtained by considering the local one-particle distribution function to be a Maxwellian (i. e., of the form $f(r,v)=A\exp[- mv^2/2kT(r)]$, where $A$ is a normalization constant) and by defining the pressure as $p\equiv \int\!  m f  (v^2/3)\,\mathrm{d}^3r\,\mathrm{d}^3v $. This definition of pressure, which is obtained by considering particles that would reverse their momentum when hitting an imaginary wall, ignores the effects of interparticle forces. If, for example, particles repelling one another were confined in a box, they would exert 
some pressure on the walls of the box even if at rest: this contribution is completely neglected by the equation of state of the ideal gas (the neglected pressure is similar to that of rigid spheres when packed too closely; thus the Boltzmann entropy cannot give a correct result for a gas of almost rigid spheres when their density is too high). 

In the case of particles interacting through gravity, 
the neglected contribution is attractive and should therefore decrease the pressure compared to that of an ideal gas. 
Therefore, it is unlikely that the true thermal equilibrium state of an $N$-body system when $t\to\infty$ ($t$ is time) has an effective equation of state\footnote{By effective equation of state we mean an equation of state that, by imposing hydrostatic equilibrium, would reproduce the density distribution of the equilibrium state.} not affected by the attractive nature of the gravitational force: thus the Boltzmann entropy may not be applicable to find the true thermal equilibrium of a self-gravitating $N$-body system, since it assumes the equation of state of an ideal gas. Moreover, because gravitational forces have long range, each particle feels the influence of all other distant particles. Thus, it may be that, strictly speaking, in the case of self-gravitating systems an equation of state \emph{cannot} be defined 
in terms of \emph{local} quantities. 

A different way to justify the use of the Boltzmann entropy is to show that it can be obtained from the Gibbs microcanonical entropy by means of the so-called mean field approximation  \citep{katz2002,padmana2}. However, the range of applicability of this approximation is still not clear, also because a small-scale cut-off is necessary to avoid divergences in the microcanonical entropy \citep{padmana2}. The Boltzmann entropy is also the $H$ quantity of the Boltzmann $H$-theorem; 
from this point of view, a critical analysis of the Boltzmann entropy in the context of the gravothermal 
catastrophe has been performed by \cite{miller}.

The discussion in this section supports the hypothesis that self-gravitating isothermal spheres are not true thermal equilibrium states of the pure $N$-body problem (i.e., the states that would be found in a spherical box containing $N$ gravitating particles eventually, after an infinite amount of time), but are only metastable states of which the significance is still not completely understood (for example, see \citealt{padmana2}, \citealt{chavareview} and references therein).

%__________________________________________________________________

\section{Basic equations of the dynamical approach} \label{basicequations}

%%                                     Two column figure (place early!)
%%______________________________________________ Gamma_1 (lg rho, lg e)
%   \begin{figure*}
%   \centering
%   \includegraphics[scale=0.4]{omega2minTV.pdf}
%   %%%\includegraphics{empty.eps}
%   %%%\includegraphics{empty.eps}
%   \caption{Adiabatic exponent $\Gamma_1$.
%               $\Gamma_1$ is plotted as a function of
%               $\lg$ internal energy $\mathrm{[erg\,g^{-1}]}$ and $\lg$
%               density $\mathrm{[g\,cm^{-3}]}$.}
%              \label{FigGam}%
%    \end{figure*}
%%
In this section we derive the linearized hydrodynamic equations that govern the evolution of a fluid system for small deviations from the truncated isothermal sphere equilibrium configurations. We assume spherical symmetry, that is, we consider only radial perturbations. The configurations are known to be stable against nonradial perturbations \citep{sanchez,chavanis1,binneytremaine}.

Consider a self-gravitating fluid governed by the Navier-Stokes and continuity equations, together with the equation of state of an ideal gas 
\citep[for example, see][]{landau}:
  \begin{equation} \label{idro1}
 \begin{alignedat}{1}
 &   \frac{\partial \rho}{\partial t} + \boldsymbol{\nabla}\cdot(\rho \mathbf{u}) =0~, \\
   & 			\frac{\partial \mathbf{u}}{\partial t}  + (\mathbf{u}\cdot \boldsymbol{\nabla})\mathbf{u} =
 - \frac{\boldsymbol{\nabla}p}{\rho} - \boldsymbol{\nabla}{\Phi} 
+\frac{\eta}{\rho} \nabla^2 \mathbf{u}+\frac{\zeta+\frac{\eta}{3}}{\rho}\boldsymbol{\nabla}
\left(\boldsymbol{\nabla}\cdot\mathbf{u}\right)~,
\\
& p=\rho \frac{kT}{m}~, 
  \end{alignedat}
\end{equation}
where $\rho$ is the density, $\mathbf{u}$ is the fluid velocity, $p$ is the local pressure, $\Phi$ is the gravitational potential, $\eta$ and $\zeta$ are viscosity coefficients, 
$m$ is the one-particle mass, $T$ is the temperature, $t$ is time, and $k$ is the Boltzmann constant. We will keep track of viscosity term until the end of Subsection \ref{basiceulerian}; then (from Subsection \ref{basiclagrangian} on), for simplicity, we will set $\eta=\zeta=0$. In this paper, we do not perform an analysis of the effects of viscosity. In particular we do not discuss whether it has a stabilizing or destabilizing effect: hopefully, the role of viscosity will be studied in detail in a subsequent paper.

Under the 
assumption of spherical symmetry, the gravitational potential obeys the following equations:
\begin{equation} \label{idro3}
\begin{alignedat}{1}
&  \boldsymbol{\nabla} \Phi = \frac{GM(r)}{r^2}\hat{r}, \\
& M(r)=\int_0^r\! \rho(s)4\pi s^2\, \mathrm{d}s,
\end{alignedat}
\end{equation}
which are equivalent to the Poisson equation.  

The hydrostatic equilibria of such fluid system are spatially truncated isothermal spheres \citep{chandra1} and are briefly described in Appendix \ref{appendixhydro}. As mentioned
 in Sect.~\ref{thermo}, each self-gravitating truncated isothermal sphere is identified by two dimensional scales (for example the central density $\rho_0(0)$ and the temperature $T$) and one dimensionless parameter $\Xi=R/\lambda$, which is 
 the value of the dimensionless radius $\xi\equiv r/\lambda$ at the truncation radius (the scale $\lambda$ was defined at the beginning of Sect.~2). Conversely, each choice of the three quantities $\{\rho_0(0), T, \Xi\}$ identifies a particular isothermal sphere. 

 We shall denote unperturbed quantities by subscript $0$ and the perturbations by subscript $1$. Thus the unperturbed density profile (the density profile of a truncated isothermal sphere) is, as a function of the dimensionless radius $\xi$ (see Appendix \ref{appendixhydro}):
 \begin{equation} \label{unpert}
 \begin{cases}
 \rho_0(\xi)=\rho_0(0)e^{-\psi(\xi)} & \mbox{if } \xi \leq \Xi\\
 0 & \mbox{if } \xi>\Xi~,
 \end{cases}
 \end{equation}
 where $\rho_0(0)$ is the unperturbed central density and $\psi(\xi)$ is the regular solution to the Emden equation (the symbol $'$ denotes derivative with respect to the argument $\xi$):
 \begin{equation} \label{emden}   \frac{\mathrm{d}}{\mathrm{d}\xi}(\xi^2 \psi')=\xi^2 e^{-\psi}, \hspace{6mm}  \psi(0)=\psi'(0)=0 .       \end{equation}
 %\begin{equation} \label{emden} \frac{\mathrm{d}}{\mathrm{d}\xi}\left( \xi^2 \psi' \right)=\xi^2 e^{-\psi}, \hspace{6mm}  \psi(0)=\psi'(0)=0 .\end{equation}
 When we perturb the hydrodynamic equations it is convenient to work in the Eulerian or Lagrangian representation of hydrodynamics, depending on which boundary conditions we 
impose. In the Eulerian representation the independent variables are the time $t$ and the position vector $\mathbf{r}$. In the Lagrangian representation the independent variables are the time $t$ and the original position $\mathbf{r}_0$ that the fluid element under consideration had at the initial time $t_0$. 

\subsection{Eulerian representation} \label{basiceulerian}

In this subsection we derive the linearized perturbation equations in the Eulerian representation. So we write each quantity as the sum of an unperturbed part (characterizing the truncated isothermal sphere) and a perturbation:
\begin{equation} \label{perturb} 
  \begin{array}{l l}
    \rho(r,t)=\rho_0(r)+\rho_1(r,t)  \\
 {u}(r,t)={u}_1(r,t)\\
T(t)=T_0+T_1(t)~,
  \end{array}
\end{equation}
where $u$ is the radial component of the fluid velocity (recall that we consider only radial perturbations). We substitute Eq.~ \eqref{perturb} in Eq.~\eqref{idro1} and expand to first order in the perturbed quantities. We allow the temperature to vary in time while remaining 
uniform in space: this constraint will be used in order to impose the condition of constant total energy (see Subsection \ref{constantenergy}). To find the normal modes of the system, 
we look for solutions of the following form:
\begin{equation} \label{perturb2}
  \begin{array}{l l}
    \rho_1(r,t)=\tilde{\rho}_1(r)e^{-i\omega t}  \\
 {u}_1(r,t)=\tilde{u}_1(r)e^{-i\omega t}\\
T_1(t)=\tilde{T}_1 e^{-i\omega t}~.
  \end{array}
\end{equation}
In the following, we shall drop the symbol \textasciitilde{} to keep the notation simpler.
After some manipulations (see Appendix \ref{euleriancalculations}) and using the unperturbed density profile \eqref{unpert}  we obtain the following linearized equation:

\begin{eqnarray} \label{linearized1}
Lf  &=& \mathcal{L}f
+\frac{i\omega}{4 \pi G \lambda} \frac{kT_1}{m} \psi' e^{-\psi} \\
&+& i \omega \frac{m}{\rho_0(0) k T_0 }\Bigg\{  \eta \frac{1}{\xi^2}[\xi^2 (f e^{\psi})']' 
+ (\zeta
+\frac{\eta}{3}) \left[ \frac{1}{\xi^2}\left(\xi^2 f e^{\psi}\right)'\right]' \Bigg\}~, \nonumber
\end{eqnarray}
where $f(\xi)=\rho_0(\xi)u_1(\xi)$ is the unknown function,  \begin{equation}L=\frac{\omega^2}{4\pi G \rho_0(0)}\end{equation} 
represents the dimensionless (squared) eigenfrequency and 
$\mathcal{L}$ is the following differential operator:
\begin{equation} \label{diffop1}
\mathcal{L}\equiv -\frac{\mathrm{d}^2}{\mathrm{d}\xi^2}-\left(\frac{2}{\xi}+\psi'\right)\frac{\mathrm{d}}{\mathrm{d}\xi}+\left(\frac{2}{\xi^2}-\frac{2\psi'}{\xi}-e^{-\psi}\right).
\end{equation}

Equation \eqref{linearized1} is the equation to be solved to find the normal modes and frequencies of the system, under the appropriate 
boundary conditions and constraints.
The boundary conditions for Eq. \eqref{linearized1} are defined in the following way. The absence of sinks and sources of mass, together with the assumption of spherical symmetry, implies $f(0)=0$. Since in the Eulerian representation we shall consider 
only systems in a spherical box of constant volume, the radial velocity at the edge must be zero, which implies $f(\Xi)=0$.
Thus the boundary conditions are:
\begin{equation} \label{boundary} f(0)=f(\Xi)=0. \end{equation}

To solve the equations, we still need to specify the function $T_1(t)$. This specification discriminates between the constant energy and 
the constant temperature case.
Once this specification is made, Eq.~\eqref{linearized1} is an eigenvalue problem: for fixed $\omega$, the equation admits a solution only for discrete 
values of $L$. These solutions are the radial normal modes of the system.

The operator $\mathcal{L}$ has the following properties, which can be proved directly from the Emden equation \eqref{emden}:
\begin{equation}\label{diffop2}\begin{array}{l}\mathcal{L} (\psi')=-e^{-\psi}\psi'\\
\mathcal{L} (\xi e^{-\psi})=-e^{-\psi}\psi'\end{array}~.\end{equation}
These properties allow us to obtain analytical solutions in some cases. They are equivalent to those found by \cite{padmana1} in his review of the thermodynamical analysis of \cite{antonov} and later also used by \cite{chavanis1,chavanis2} to obtain analytical solutions of the present hydrodynamic problem for the constant $\{T, V\}$ (Sect. \ref{isocollapse}) and constant $\{T, P\}$ (Sect. \ref{resultslagrangianTP}) cases. 

Note that viscosity disappears when $\omega=0$, which is the situation of marginal stability. Thus viscosity does not modify the points of the onset of instability \citep{sanchez,chavanis1}.

\subsection{Lagrangian representation} \label{basiclagrangian}
 
In this section we present the linearized perturbation equations in the Lagrangian representation in the inviscid case. 
In this representation, the independent variable is the position $\mathbf{r}_0$ of the fluid element under consideration at the initial time $t_0$. Thus in Eq. \eqref{idro1} we have to perform the following change of independent variables:
\begin{equation}
	\begin{cases} r \to r_0(r,t)\\
			  t \to t~.
	\end{cases} 
\end{equation}
In the Lagrangian representation, each quantity is a function of the new independent variables $r_0$ and $t$. 
 
The calculations are summarized in Appendix \ref{changevariables}. For linear perturbations, the resulting continuity and Euler equations in the Lagrangian representation are (neglecting the term quadratic in the velocity):
 
\begin{equation}\label{idrolagr}
  \begin{alignedat}{1}
&\left(\frac{\partial}{\partial t} - \frac{r^2}{r_0^2}\frac{\rho}{\rho_0}u\frac{\partial}{\partial r_0}\right) \rho+
\frac{1}{r_0^2}\frac{\rho}{\rho_0} \frac{\partial}{\partial r_0}   \left(r^2 \rho u\right)=0\\
&\left(\frac{\partial}{\partial t} - \frac{r^2}{r_0^2}\frac{ \rho}{ \rho_0}u   \frac{\partial}{\partial r_0} \right) u
=-\frac{r^2}{r_0^2} \frac{\partial \rho}{\partial r_0} \frac{1}{\rho_0}\frac{kT}{m}-\frac{GM(r_0)}{r^2} .
  \end{alignedat}
\end{equation}

As for the Eulerian case we separate each quantity in an unperturbed part and a perturbation:
\begin{equation} \label{perturblagr}
  \begin{array}{l l}
    \rho(r_0,t)=\rho_0(r_0)+\rho_1(r_0,t)  \\
 u(r_0,t)=u_1(r_0,t)\\
r(r_0,t)=r_0+r_1(r_0,t)\\
T(t)=T_0+T_1(t)~.
  \end{array}
  \end{equation}
 We substitute Eq.~\eqref{perturblagr} in Eqs.~\eqref{idrolagr} and expand to first order in quantities with subscript 1. Then 
 to find the normal modes we take:
\begin{equation}\label{lagr9}
  \begin{array}{l l}
    \rho_1(r_0,t)=\tilde{\rho}_1(r_0)e^{-i\omega t}  \\
 u_1(r_0,t)=\tilde{u}_1(r_0)e^{-i\omega t}\\
r_1(r_0,t)=\tilde{r}_1(r_0)e^{-i\omega t}~.
  \end{array}
\end{equation}
In the following we shall drop the symbol \textasciitilde{} for simplicity of notation. After introducing the dimensionless radius $\xi_0\equiv r_0/\lambda$, using the density profile \eqref{unpert} of the unperturbed state and after some manipulations we obtain the following equation 
for $\rho_1(\xi_0)$ (see Appendix \ref{linearizationlagrangian} for an outline of the calculations):
\begin{equation}\label{perturbedlagr}
-\frac{\rho_1}{\rho_0(0)}+\frac{e^{-\psi}}{\xi_0^2}\frac{\mathrm{d}}{\mathrm{d}\xi_0}\left[  
\frac{\left(\frac{\frac{\mathrm{d}}{\mathrm{d}\xi_0}(\frac{\rho_1}{\rho_0(0)})}{\frac{\mathrm{d}}{\mathrm{d}\xi_0}e^{-\psi(\xi_0)}}+\frac{T_1}{T_0}\right)}{\frac{4}{\xi_0^3}
+ \frac{L}{\xi_0^2 \psi'(\xi_0)} }\right]=0~,
\end{equation}
where $L={\omega^2}/{4\pi G \rho_0(0)}$ as for the Eulerian case.  
Boundary conditions are discussed in Subsection \ref{resultslagrangianTP}. Solving Eq.~\eqref{perturbedlagr} allows us to find the normal modes in the Lagrangian representation.
We have allowed the temperature to vary in time in Eq.~\eqref{perturbedlagr}, but in the following we shall analyze only the isothermal case with $T_1=0$.

\section{Modal stability of a self-gravitating inviscid fluid sphere under different boundary conditions} \label{hydroanalysis}

Here we analyze the equations obtained in the previous section by imposing different boundary conditions.

\subsection{Eulerian representation}

In this subsection we analyze Eq.~\eqref{linearized1} by imposing two kinds of boundary conditions: constant temperature $T$ and volume $V$ or constant total energy $E$ and volume $V$.

\subsubsection{Constant $\{T, V\}$ case (isothermal collapse)} \label{isocollapse}

Here we consider a fluid at constant temperature $T$ contained in a sphere of fixed radius $R$. The condition of constant temperature 
is satisfied by imposing $T_1=0$ in Eq.~\eqref{linearized1}. The condition of constant volume has been discussed in Sect.~\ref{basiceulerian} and leads to the boundary condition $f(\Xi)=0$. Thus Eq.~\eqref{linearized1} for 
the constant $\{T,V\}$ case becomes:
\begin{equation}\label{constTV} \mathcal{L}f=Lf\end{equation}
with the boundary conditions \eqref{boundary}.

Equation \eqref{constTV} is an eigenvalue equation that, at given $\Xi$, admits solutions only for discrete values of $L$. If the lowest value of $L$ at fixed $\Xi$ is positive, then all modes are stable (because $\omega$ is always real) and the system is stable. If the lowest value of $L$ at a given value of $\Xi$ is negative, then unstable modes are present and the system is unstable.

\begin{figure}
  \centering
  \includegraphics[width=8cm]{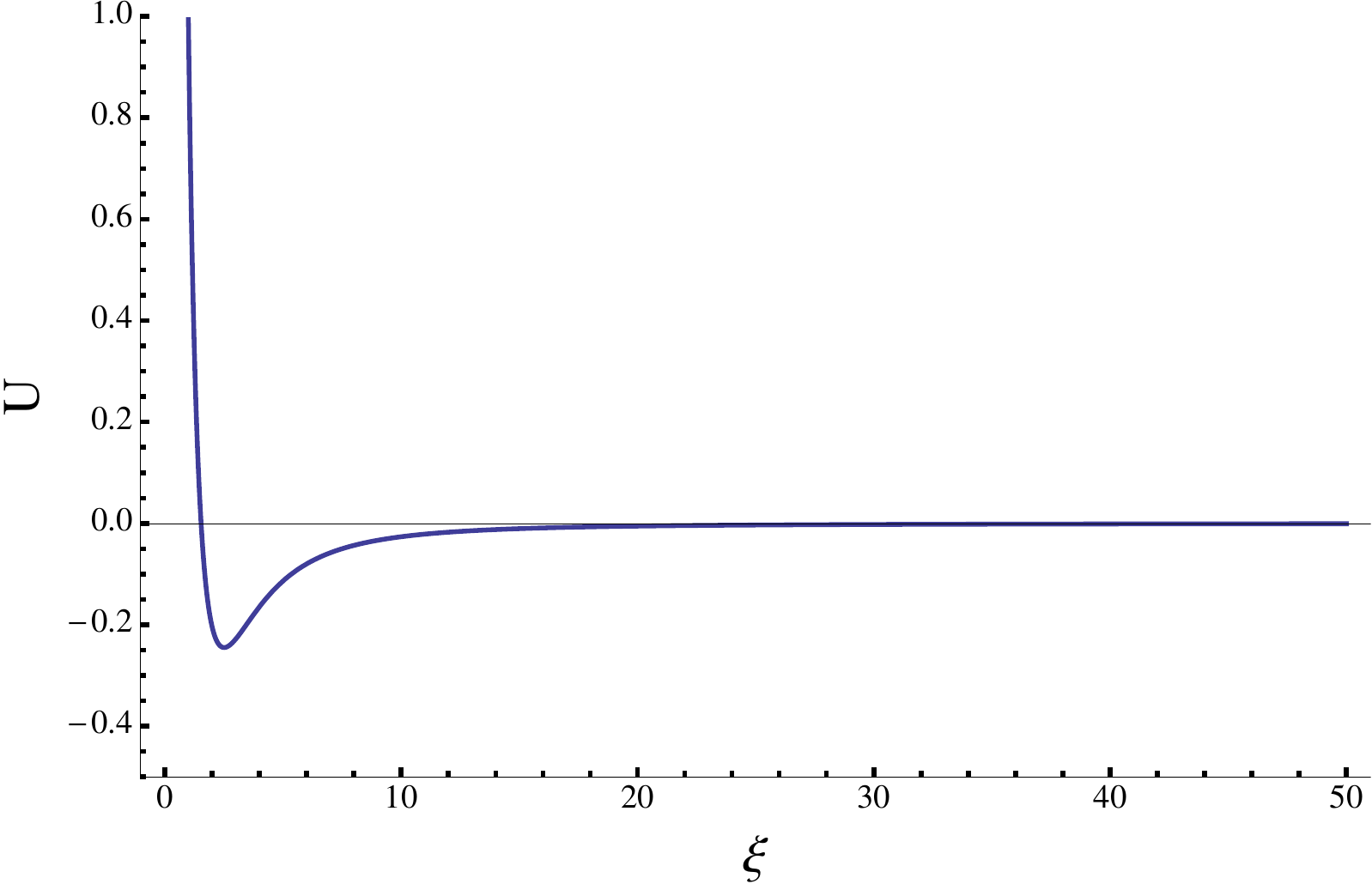}
     \caption{The effective potential $U$ [see Eqs.~\eqref{schrodingerequation} and \eqref{effectivepotential}] for the constant $\{T, V\}$ case. To find the frequencies of normal modes, the Schr\"odinger equation \eqref{schrodingerequation} has to be solved with boundary conditions \eqref{boundaryschrodinger}, which means that the effective potential is $U(\xi)$ for $\xi\leq\Xi$ and taken to be infinite for $\xi>\Xi$. States with negative energies, which exist for $\Xi\geq8.99$, correspond to unstable modes.               }
        \label{fig:schrodingerpotential}
  \end{figure}

By means of the standard transformation \begin{equation} f(\xi)=\tilde{f}(\xi) \exp{\left[- \int_{\bar{\xi}}^\xi \left(\frac{1}{\xi'} + \frac{\psi'(\xi')}{2}\right) \mathrm{d}\xi' \right]}~,\end{equation} 
where $\bar{\xi}$ is an arbitrary point in the domain of $f$, Eq.~\eqref{constTV} can be recast in the form of a Schr\"odinger equation:
\begin{equation} \label{schrodingerequation}-\tilde{f}''(\xi)+\tilde{f}(\xi)U(\xi)=L\tilde{f}(\xi)~, \end{equation}
where $U(\xi)$ is the effective potential given by: 
\begin{equation} \label{effectivepotential}  U(\xi) \equiv \frac{2}{\xi^2}+\frac{1}{4}\psi '(\xi)^2-\frac{2 }{\xi}\psi '(\xi)- \frac{1}{2}e^{-\psi (\xi)}~. \end{equation}
The boundary conditions become:
\begin{equation} \label{boundaryschrodinger} \tilde{f}(0)=\tilde{f}(\Xi)=0~.\end{equation} 
The effective potential is shown in Fig.~\ref{fig:schrodingerpotential}.
From the boundary conditions 
\eqref{boundaryschrodinger} we see that choosing a specific $\Xi$ requires us to consider a potential that is infinite for $\xi \geq \Xi$.
The existence of negative eigenvalues implies that the system is unstable. 
From the form of the potential it is clear that for small values of $\Xi$ negative eigenvalues (i.e., unstable modes) do not exist. Negative eigenvalues appear only for sufficiently large values of $\Xi$. For the present problem this occurs at $\Xi\geq 8.99$. When $\Xi\to\infty$, an infinite number of negative eigenvalues appear, precisely at the same points where new unstable modes appear in the thermodynamical approach \citep[see][]{katz1}.

     \begin{figure}
  \centering
  \includegraphics[width=9cm]{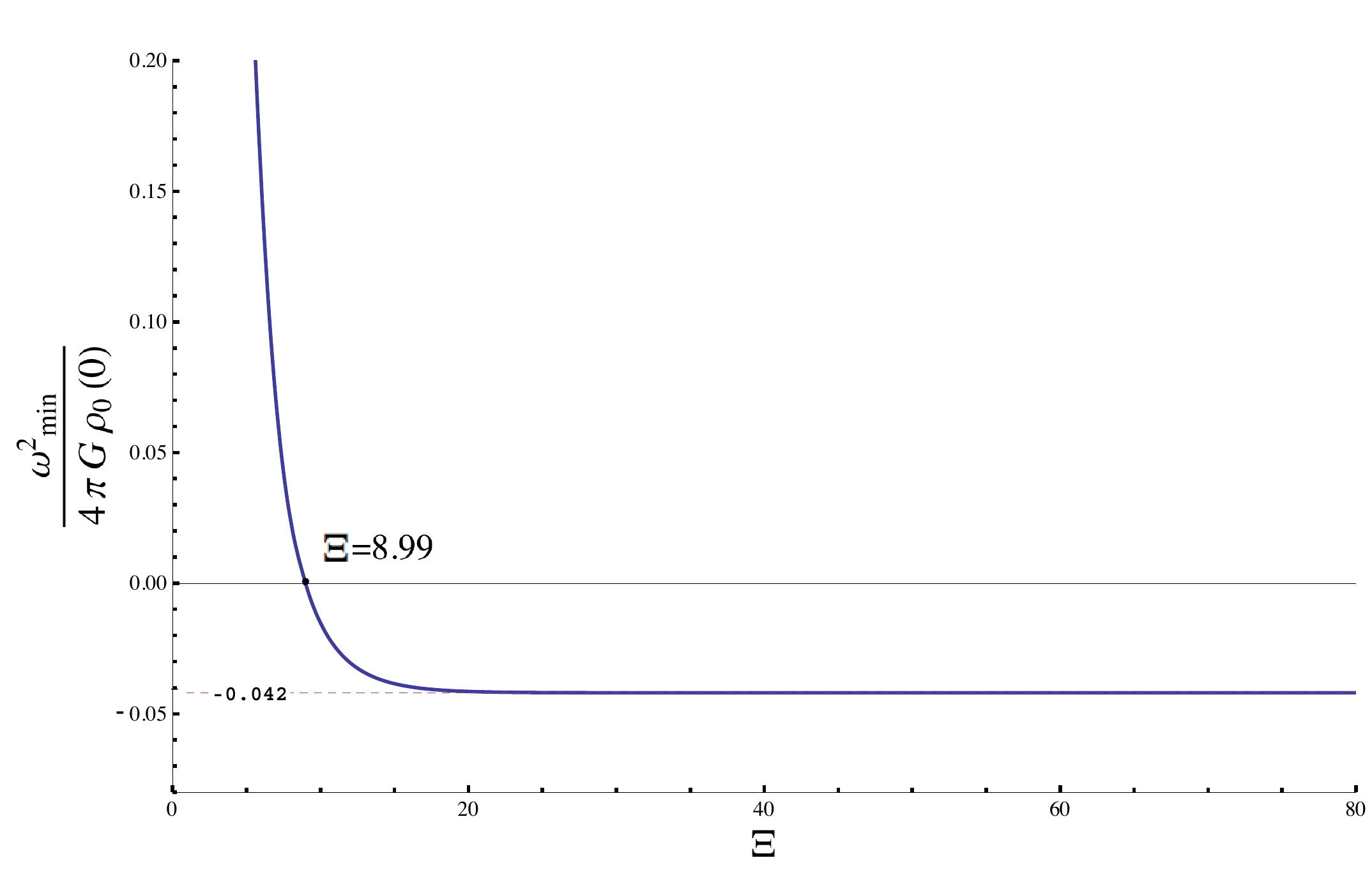}
     \caption{The minimum value of the eigenvalue $L$ at given $\Xi$ (the dimensionless radius characterizing the system) as a function of $\Xi$, for the constant $\{T, V\}$ case. If the minimum $L$ is negative then the system is unstable. The system becomes unstable at $\Xi=8.99$, the same value obtained from the thermodynamical approach. From the figure we can read the typical time scale of the instability. As $\Xi\to\infty$, the asymptotic value $L_{\infty}=-0.042$ is the same as for the other cases (see Figs.~\ref{fig:omegaEV} and \ref{fig:omegaTP}).
             }
        \label{fig:omegaTV}
  \end{figure}

Let us now consider the solutions to Eq.~\eqref{constTV} in greater detail. Figure \ref{fig:omegaTV} shows the minimum value of $L$ at given dimensionless radius $\Xi$ as a function of $\Xi$. We see that $L$ is negative, that is, the system is unstable, for $\Xi>8.99$, which is the same point found by \cite{lyndenbellwood} in the thermodynamical approach. Higher modes, that is, higher values of $L$ at given $\Xi$, would be represented by lines above the plotted curve. These lines would 
intersect the $\Xi$ axis at some points, which are the zeros of the analytical solution $G_{TV}$ described below [see Eq.~\eqref{GTV}].
In Appendix \ref{constantTVprofiles} a few density and velocity profiles of numerically calculated eigenfunctions are shown. Most of them are for modes of minimum $L$ at given $\Xi$. Density profiles of higher modes exhibit oscillations not present in the lowest mode.

For the case of marginal stability ($L=0$), with the help of properties \eqref{diffop2}, the relevant eigenfunction can be expressed analytically \citep{chavanis1}. The function
\begin{equation} \label{GTV} G_{TV}(\xi) \equiv \psi'(\xi) - \xi e^{-\psi(\xi)} \end{equation}
is indeed a solution to Eq.~\eqref{constTV} with $L=0$, which satisfies $G_{TV}(0)=0$. The values of $\Xi$ for which $G_{TV}(\Xi)=0$ are those for which the boundary conditions \eqref{boundary} are satisfied; thus they are the values of $\Xi$ at which each new unstable mode appears. The first zero of $G_{TV}$ occurs where the first unstable mode appears, at $\Xi=8.99$. From the asymptotic behavior of $\psi$ it is possibile to obtain an asymptotic approximation of the zeros: they approximately follow a geometric progression of ratio $e^{2 \pi/\sqrt{7}}$ \citep[see also][]{sanchez,chavanis1}.

\subsubsection{Constant $\{E, V\}$ case (gravothermal catastrophe)} \label{constantenergy}

Here we consider a self-gravitating fluid sphere at constant total energy $E$ and volume $V$. In this case the instability has been named \emph{gravothermal catastrophe} by \cite{lyndenbellwood}. The total energy $E$ of the fluid is defined as:
\begin{equation} \label{totalenergy1} E=\frac{3}{2}N k T +\frac{1}{2} \int\! \rho(\mathbf{r})\Phi(\mathbf{r})\, \mathrm{d}^3 r~.\end{equation}
It has two terms, which represent the thermal and gravitational contributions. The condition of constant energy is imposed in the following way. When the fluid is perturbed, its gravitational energy changes as a consequence of the redistribution of matter. We suppose that the temperature varies in time, while remaining uniform in space, so as to keep the total energy (thermal plus gravitational) constant (for a different model, based on the Smoluchowski-Poisson system of equations, the same nonstandard assumption has been made by \citealt{chavanis3}). The thermal energy expression at time $t$ is thus given by $3NkT(t)/2$. In doing so, we are assuming infinite thermal conductivity (see also Sect.~\ref{timescales} for the relation of this fact to the relevant time scales).

To reduce Eq.~\eqref{linearized1} to the constant $\{E, V\}$ case we need to find the expression for the temperature as a function of the density distribution at fixed total energy, in the linear regime of small perturbations.
Starting from Eq.~\eqref{totalenergy1} for the total energy and recalling that $\Phi(\mathbf{r})=-G\int\! \mathrm{d}^3r'\, \rho(\mathbf{r}')/|\mathbf{r}-\mathbf{r}'| $, we have:
\begin{equation} 
E=\label{totalenergy2} 
\frac{3}{2}N k T -\frac{G}{2} \int\! \frac{ \rho(\mathbf{r})\rho(\mathbf{r}')}{|\mathbf{r}-\mathbf{r}'|}\, \mathrm{d}^3 r\, \mathrm{d}^3 r'~.
\end{equation}
By substituting $T=T_0+T_1$, $\rho=\rho_0+\rho_1$ in Eq.~\eqref{totalenergy2}, keeping only first-order quantities and imposing that the energy remains constant, we obtain the following expression for $T_1$: 
\begin{equation} \label{T12}
T_1=-\frac{ \int\! \rho_1(\mathbf{r})\Phi_0(\mathbf{r})\, \mathrm{d}^3 r}{\frac{3}{2}Nk}~.
\end{equation}
Here $\Phi_0$ is the gravitational potential of the unperturbed density distribution $\rho_0$.
After some manipulations (see Appendix \ref{tempexpr}) we obtain:
\begin{equation} \label{T1}
T_1=- \frac{1}{i\omega}\frac{ \int_0^{\Xi}\! \{\mathrm{d}[\xi^2f(\xi)]/\mathrm{d}\xi\}\psi(\xi) \, \mathrm{d} \xi}{\frac{3}{2}\, \rho_0(0)\lambda\, \Xi^2 \psi'(\Xi)}T_0~.
\end{equation}
By substituting Eq.~\eqref{T1} in Eq.~\eqref{linearized1}, recalling the definition $\lambda=\left[kT_0/4\pi G \rho_0(0)  m\right]^{1/2}$, and neglecting viscosity, we obtain:
\begin{equation} \label{idroEV1}
L f =\mathcal{L}f
- \psi' e^{-\psi} \frac{1}{\frac{3}{2}\Xi^2\psi'(\Xi)} \int_0^{\Xi}\!  \psi(\xi)\frac{\mathrm{d}}{\mathrm{d}\xi}\left[\xi^2f(\xi)\right]\, \mathrm{d} \xi~.
\end{equation}
As discussed in Sect. ~\ref{basiceulerian} the boundary conditions are given by Eq.~\eqref{boundary}.
By integrating the last term of Eq.~\eqref{idroEV1} by parts under the boundary conditions \eqref{boundary}, we obtain:
\begin{equation} \label{idroEV2}
L f =\mathcal{L}f
+ \psi' e^{-\psi} \frac{1}{\frac{3}{2}\Xi^2\psi'(\Xi)} \int_0^{\Xi}\!  \xi^2\psi'(\xi)f(\xi)\, \mathrm{d} \xi~.
\end{equation}
Note that Eq.~\eqref{idroEV2} [or \eqref{idroEV1}] contains an integral \emph{global constraint}. Equation \eqref{idroEV2} is to be solved to find the normal modes and corresponds to Eq.~\eqref{constTV}.
The numerical procedure followed to solve Eq.~\eqref{idroEV2} is relatively straightforward and thus is not reported here.

\begin{figure}
  \centering
  \includegraphics[width=9cm]{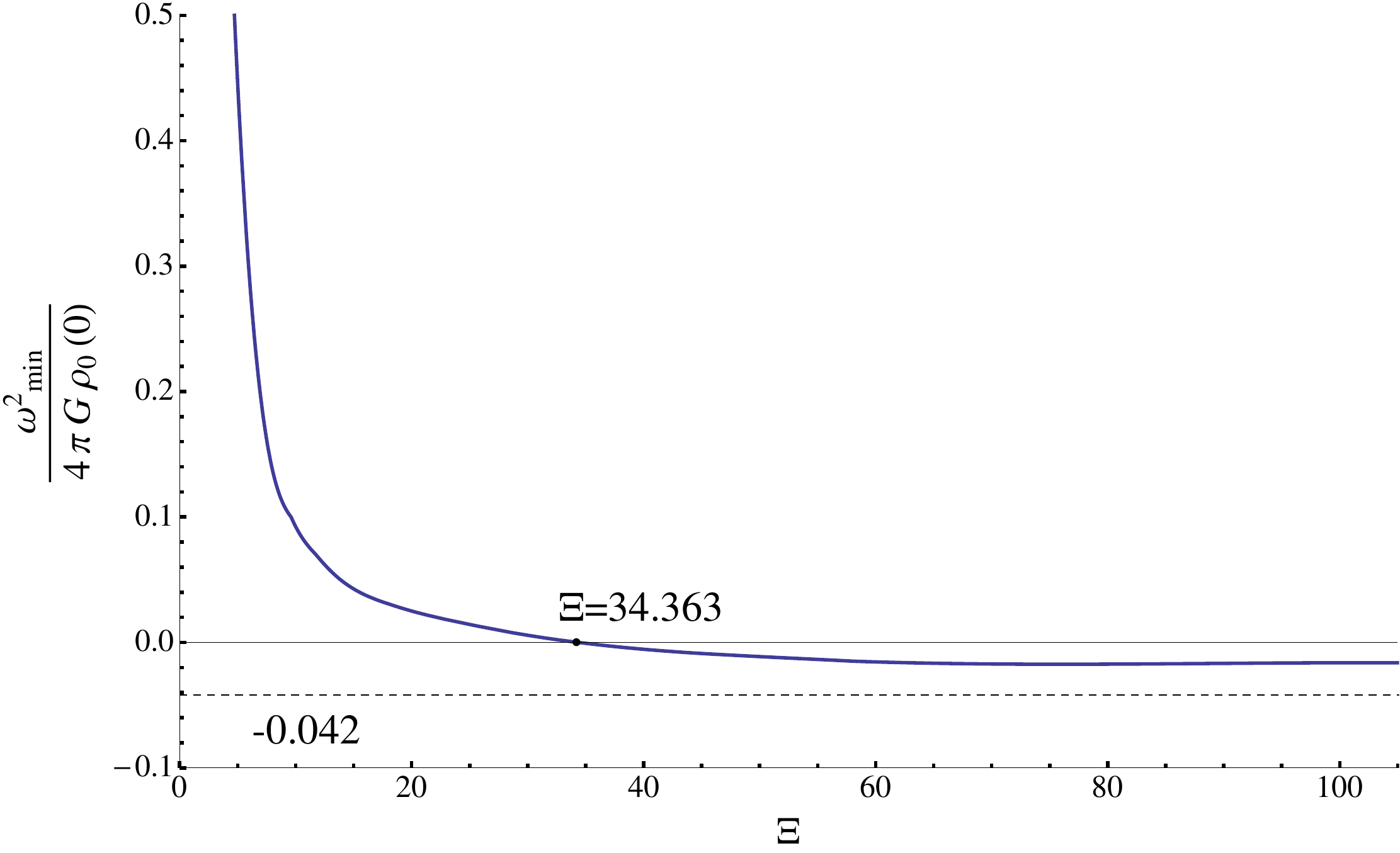}
     \caption{
     The minimum value of the eigenvalue $L$ at given $\Xi$, the dimensionless radius characterizing the system, as a function of $\Xi$, for the constant $\{E, V\}$ case. When the minimum $L$ is negative the system is unstable. The system becomes unstable at $\Xi=34.36$, which corresponds to a density contrast $\simeq709$, the same value obtained from the thermodynamical approach. From the figure we can read the typical time scale of the instability. As $\Xi\to\infty$, the asymptotic value of $L$, $L_{\infty}=-0.042$, is the same as for the other cases (see Figs.~\ref{fig:omegaTV} and \ref{fig:omegaTP}).
             }
        \label{fig:omegaEV}
  \end{figure}

Figure \ref{fig:omegaEV} shows the minimum value of $L$ at given $\Xi$.
We see that the system is unstable for $\Xi>34.36$, which corresponds to a density contrast $\rho_0(0)/\rho_0(\Xi)>709$ \citep{antonov}. 
As $\Xi$ increases, new unstable modes appear, similarly to the behavior observed in the constant $\{T, V\}$ case. 
As $\Xi\to\infty$, the asymptotic value of the minimum $L$ is found numerically to be the same as in the constant $\{T, V\}$ case, $L_{\infty}\simeq-0.042$.
Such asymptotic value is the same also in the constant $\{T, P\}$ case (see Sect.~\ref{resultslagrangianTP}).

In the case of marginal stability ($L=0$), Eq.~\eqref{idroEV2} 
is equivalent to that found and solved analytically by \cite{padmana1} in the thermodynamical approach. 
Here, for completeness, we record how the solution is derived, by adapting the method of \cite{padmana1} to our choice of variables and unknowns.

To solve analytically Eq.~\eqref{idroEV2} for $L=0$ we rewrite it in the following form:
\begin{equation}\label{idroEV3}
\mathcal{L}f=- C\psi'e^{-\psi}~,
\end{equation}
where \begin{equation} \label{C} C= \frac{1}{\frac{3}{2}\Xi^2\psi'(\Xi)} \int_0^{\Xi}\! \xi^2\psi'(\xi)f(\xi)\, \mathrm{d} \xi \end{equation} is a constant (with respect to $\xi$) that depends globally on $f$.
Using the properties 
\eqref{diffop2}, we look for a solution of the form
\begin{equation}\label{solEV} G_{EV}(\xi)=a\psi'(\xi)+b\xi e^{-\psi(\xi)}~,\end{equation}
where $a$ and $b$ are real numbers. Since Eq.~\eqref{idroEV3} is linear in $f$, only the ratio ${a}/{b}$ is relevant. 
Susbtituting \eqref{solEV} in \eqref{idroEV3} we obtain the first condition on $a$ and $b$:
\begin{equation}
a+b=C.
\end{equation}
Using the expression of $C$ \eqref{C} and dividing by $b$, we obtain:
\begin{equation} \label{ab1}
\frac{a}{b}+1= \frac{1}{\frac{3}{2}\Xi^2\psi'(\Xi)} \int_0^{\Xi}\!  \xi^2\psi'(\xi)\left(\frac{a}{b}\psi'(\xi)+\xi e^{-\psi(\xi)}\right)\, \mathrm{d} \xi~.
\end{equation}
A second condition on ${a}/{b}$ follows from the boundary condition $f(\Xi)=0$:
\begin{equation} \label{ab2} \frac{a}{b}\psi'(\Xi)+\Xi e^{-\psi(\Xi)}=0~.\end{equation}
Substituting ${a}/{b}$ from \eqref{ab2} in \eqref{ab1} we obtain: 
\begin{equation}\label{solEV2}\frac{3}{2}\Xi^2\left(\psi'(\Xi)-\Xi e^{-\psi(\Xi)} \right)=
\int_0^{\Xi}\!  \xi^2\psi'(\xi)\left(\xi e^{-\psi(\xi)}-\psi'(\xi)\frac{\Xi e^{-\psi(\Xi)}}{\psi'(\Xi)}\right)\, \mathrm{d} \xi  \end{equation}

The values of $\Xi$ satisfying Eq.~\eqref{solEV2} are those for which Eq.~\eqref{idroEV2} admits a solution for $L=0$. In particular,  the lowest value of $\Xi$ which 
satisfies Eq.~\eqref{solEV2} determines the threshold of instability: by numerically solving the algebraic equation \eqref{solEV2}, this minimum value is found to be $\Xi=34.36$ (as shown by Antonov 1962).
As for the constant $\{T, V\}$ case, the number of unstable modes for each $\Xi$ coincides with the results of the thermodynamical analysis [see \cite{katz1}]. Once a value of $\Xi$ is obtained, the value of $a/b$ and then an analytical solution $G_{EV}$ is determined.

Density profiles for modes of minimum $L$ at given $\Xi$ are shown in Appendix \ref{constantEVprofiles}. Since the equations involved are equivalent, the density profile for the marginally stable perturbation 
($L=0$) is the same as that found by \cite{padmana1}. He pointed out that it has a ``core-halo'' structure, that is, an oscillation in $\rho_1$: the density perturbation is positive 
in the inner part (nucleus), negative in the middle (emptying area), and then positive again (halo). The core-halo structure has been physically interpreted in the framework of the 
gravothermal catastrophe given by \cite{lyndenbellwood}, using the concept of negative specific heat. However, this interpretation is not applicable to the present context.

We found that for modes of minimum $L$ at given $\Xi$ the core-halo structure is present if $L<0.021$, disappears between $L=0.021$ and $L=0.022$, and is absent for $L>0.022$, as illustrated in the figures presented in Appendix \ref{constantEVprofiles}. Higher modes always exhibit one ore more oscillations, as in the constant $\{T, V\}$ case.

We have thus shown that for the present case the behavior of the eigenvalues $L$ as a function of $\Xi$ is similar to that of the constant $\{T, V\}$ case, with the difference that the instability threshold is higher in the constant 
$\{E, V\}$ case. The interpretation of this fact in the context of the fluid model is as follows. 
When the fluid is compressed, the gravitational potential energy decreases and the temperature 
increases in order to maintain the total energy constant. Therefore, the tendency toward collapse is weakened and instability can take place only at higher values of $\Xi$ (with respect to the case in which the temperature remains constant). 

In fact, if instead of expression
\eqref{totalenergy1} for the total energy we consider a modified expression in such a way that the temperature increase is greater, the collapse can be halted completely. Consider the following heuristic expression for the total energy:
\begin{equation} \label{defupsilon} E=\frac{3}{2}N k T +\frac{1}{2} \int\! \rho(\mathbf{r})\Phi(\mathbf{r})\, \mathrm{d}^3 r- \upsilon \sigma_0^2 R^3 \rho(0),\end{equation}
which contains an additional term proportional to the dimensionless parameter $\upsilon$ and to the central density $\rho(0)$. The quantity $\sigma_0^2=kT_0/m$ is the unperturbed thermal speed. We repeated the analysis of this subsection with such a modified expression for the energy and found the new value of $\Xi$ for the threshold of the instability. In this analysis the expression for $T_1$ \eqref{T1} obtained previously, to be substituted in Eq.~\eqref{linearized1}, is replaced with the expression 
for $T_1$ that is obtained from Eq.~\eqref{defupsilon} by substituting $T=T_0+T_1$, $\rho=\rho_0+\rho_1$ and by imposing that the variation of the total energy $E$ is zero.
We found that when $\upsilon>0$ (i.e., when the fluid is compressed the new term contributes to make the temperature increase) the (linear) instability is postponed to higher values of $\Xi$ for small $\upsilon$ and is completely halted for $\upsilon>5\times10^{-4}$. We also verified that if $\upsilon<0$ (i.e., when the fluid is compressed the new term contributes in the opposite direction), instability occurs at lower values of $\Xi$.

In the case of the pure $N$-body problem, it is believed that collapse can be halted by energy ``generation"  through binaries, giving rise to a phenomenon called \emph{gravothermal oscillations} \citep[for a review, see][]{heggiehut}. Gravothermal oscillations were discovered by \cite{sugimoto} using a gaseous model. These authors introduced in the model of \cite{lyndenbelleggleton} a phenomenological energy generation term (to represent the role of binaries), proportional to a power of 
the density and found that collapse can be halted and reversed. Similarly, our $\upsilon$ term shows in a simple manner that the instability can be halted in the linear regime with a suitable term in the total energy budget that mimics effects presumed to be associated with the presence of binaries.

Gravothermal oscillations have been confirmed by $N$-body simulations \citep{makino}. We will comment further on this topic in Sect.~\ref{collvscollless}.

\subsubsection{Constant $\{T, V\}$: two-component case}

We briefly studied the problem of dynamical stability by means of a linear modal analysis of a two-component ideal fluid. Each component  is assumed to interact with the other only through the common gravitational potential. The hydrostatic equilibrium configurations are spatially truncated two-component isothermal spheres, as considered by \cite{taff}, \cite{lightman}, and \cite{yoshizawa}; see also \cite{vega2comp} and \cite{chava2comp}. 

We call the single-particle masses $m_A$ and $m_B$, with $m_B>m_A$. Two-component isothermal spheres are characterized 
by two additional dimensionless parameters (with respect to the one-component case): the ratio of the single-particle masses $m_B/m_A$ and the ratio $M_B/M_A$ of the total masses associated with the two components. The reader is referred to Appendix \ref{appendix2comp} for a description of the equations used.

We considered the constant $\{T, V\}$ case, which is the simplest from the mathematical point of view, and found that the equivalence of the thermodynamical and dynamical approaches still holds. It is possible to show analytically that the onset of dynamical instability takes place at exactly the same points as those found with the thermodynamical approach. The analysis in the thermodynamical approach was performed by generalizing in a straightforward manner the analysis of \cite{chavanis1}.

An interesting result of this analysis is that the instability appears to be driven by the heavier component, in the following sense. Consider the ratios $\rho_{1A}/\rho_0$ and $\rho_{1B}/\rho_0$ of the density perturbation of each component to the total local unperturbed density. The ratio referring to the heavier component can be higher even if the total mass of the heavier component is very small. For example, for $m_B/m_A=3$, we found that the ratios $\rho_{1A}/\rho_0$, $\rho_{1B}/\rho_0$ were comparable for $M_B/M_A \simeq 0.066$ (see Fig. \ref{fig:2comp}). For higher values of $M_B/M_A$ the heavier component dominates. Independent indications that the heavier component dominates the collapse have been found by \cite{chava2comp} for a different model based on the Smoluchowski-Poisson system of equations. \cite{breen2,breen1} found that the heavier component dominates gravothermal oscillations in two-component clusters. This is likely to be related to the present analysis.

\subsection{Lagrangian representation}

\subsubsection{Constant $\{T, P\}$ case (isobaric collapse)} \label{resultslagrangianTP}

In this section we analyze perturbations at constant temperature $T$ and constant boundary pressure $P$. Therefore, $T_1=0$ and Eq.~\eqref{perturbedlagr} becomes:
\begin{equation}\label{perturbedTP}
-\frac{\rho_1}{\rho_0(0)}+\frac{e^{-\psi(\xi_0)}}{\xi_0^2}\frac{\mathrm{d}}{\mathrm{d}\xi_0}\left[ 
\frac{\frac{\frac{\mathrm{d}}{\mathrm{d}\xi_0}(\frac{\rho_1}{\rho_0(0)})}{\frac{\mathrm{d}}{\mathrm{d}\xi_0}e^{-\psi(\xi_0)}}}{\frac{4}{\xi_0^3}
+ \frac{L}{\xi_0^2 \psi'(\xi_0)} }\right]=0~.
\end{equation}
Boundary conditions are as follows. The condition $u_1(0)=0$ translates into the condition $\rho_1'(0)=0$ [using Euler equation \eqref{idrolagr}, under the assumption that ${\partial u}/{\partial r_0}$ does not diverge and ${\partial u}/{\partial t}=-i\omega u$]. The condition of constant pressure requires that a fixed Lagrangian fluid shell (which follows the fluid during the motion and thus does not have a fixed position in space) feels constant pressure. Constant 
pressure requires constant density because of the ideal gas equation of state. Therefore, the relevant boundary conditions are:
\begin{equation}
 \begin{array}{l} \label{condcontlagr}
\rho_1'(0)=0\\
\rho_1(\Xi)=0~.
 \end{array}
\end{equation}
From mass continuity and by imposing $u_1'(0)\neq 0$ (which is true in the Eulerian representation and thus we expect to be true in the present case), we also have the condition $\rho_1(0)\neq0$.

\begin{figure}
  \centering
  \includegraphics[width=9cm]{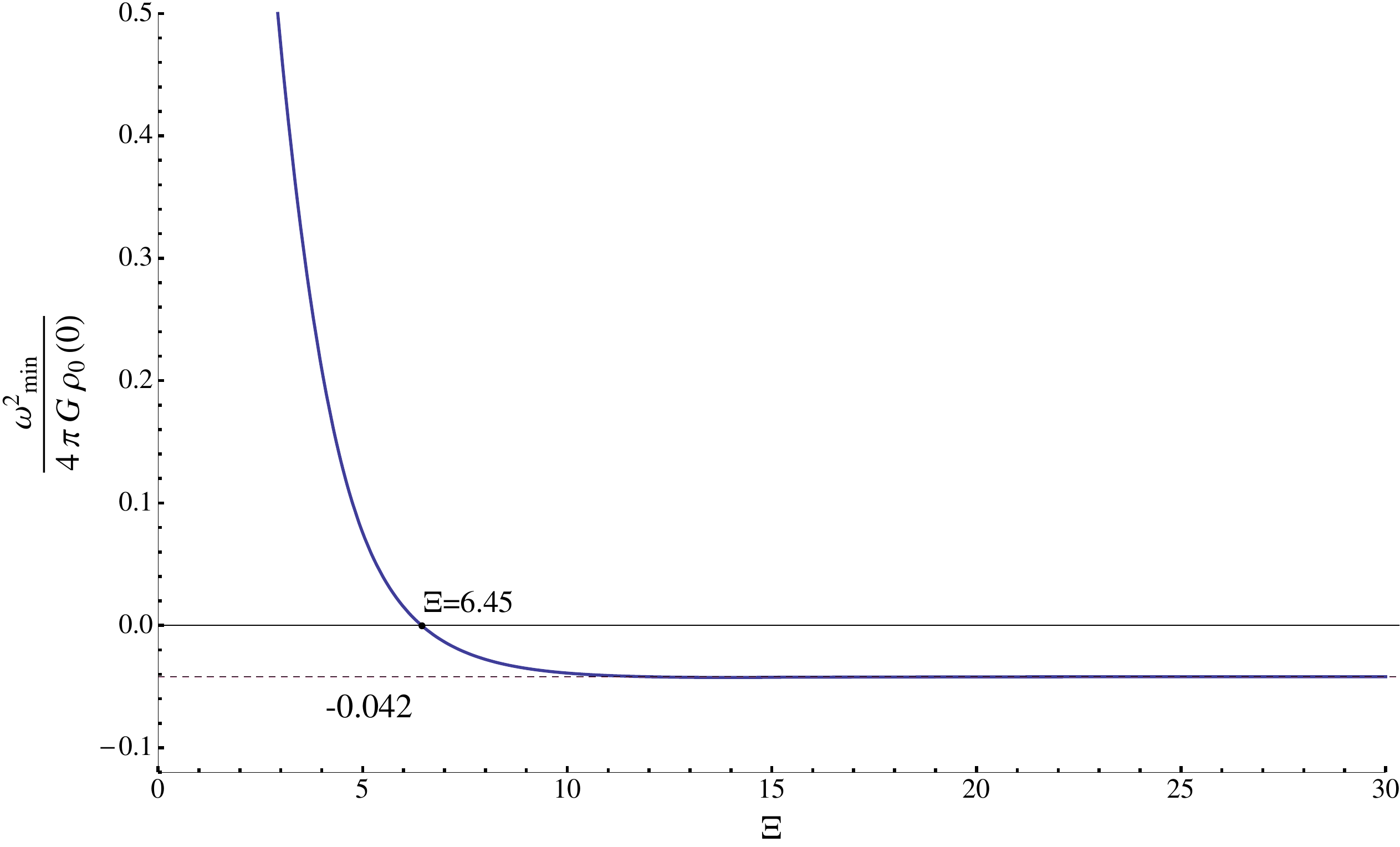}
     \caption{
       The minimum value of the eigenvalue $L$ at given $\Xi$, the dimensionless radius characterizing the system, as a function of $\Xi$, for the constant $\{E, V\}$ case. When the minimum $L$ is negative the system is unstable. The system becomes unstable at $\Xi=6.45$, the same value as found in the thermodynamical approach. From the figure we can read the typical time scale of the instability. As $\Xi\to\infty$, the asymptotic value $L_{\infty}=-0.042$ is the same as for the other cases (see Figs.~\ref{fig:omegaTV} and \ref{fig:omegaEV}).              }
        \label{fig:omegaTP}
  \end{figure}

The numerical procedure to solve Eq.~\eqref{perturbedTP} is relatively straightforward and is thus not reported here. Figure \ref{fig:omegaTP} represents the minimum value of $L$ at given $\Xi$. The system becomes unstable for $\Xi>6.45$, which is the same condition as found by \cite{bonnor}, \cite{ebert}, and \cite{lyndenbellwood}. As $\Xi$ increases, other unstable modes appear, similarly to the cases described previously. 
As $\Xi\to\infty$, the asymptotic value of the minimum $L$  is found numerically to be the same as for the constant $\{E, V\}$ and $\{T, V\}$ cases,  $L_{\infty}=-0.042$, but, in contrast to the constant $\{E,V\}$ and $\{T,V\}$ cases, it is reached from below. There is a minimum at $L\approx-0.043$ for $\Xi \approx 15$.

In Appendix \ref{constantTPprofiles} density profiles of normal modes are shown. Most of them are for modes of minimum $L$ at given $\Xi$. Higher modes present oscillations. It would be interesting to show whether the analysis of this subsection might be generalized to the nonspherically-symmetric case by following \cite{lombardibertin}.

As for previous cases, we can obtain analytical solutions for the marginally stable perturbations.
For $L=0$, Eq.~\eqref{perturbedTP} reads:
\begin{equation} \label{perturbedTP2}
\mathcal{L}\left(\frac{\rho_1}{\rho_0(0)}\right)=0~,
\end{equation}
where $\mathcal{L}$ is defined as:
\begin{equation}\label{opdifflagr1}
\mathcal{L}\equiv-1+\frac{e^{-\psi}}{\xi_0^2}\frac{\mathrm{d}}{\mathrm{d}\xi_0}\left[  
\frac{\frac{\frac{\mathrm{d}}{\mathrm{d}\xi_0}}{\frac{\mathrm{d}}{\mathrm{d}\xi_0}e^{-\psi}}}{\frac{4}{\xi_0^3}
}\right]=0~.
\end{equation}
From Emden Eq.~ \eqref{emden}, the operator $\mathcal{L}$ has the following properties:
\begin{equation}\label{opdifflagr2}\begin{array}{l}\mathcal{L} (\psi'^2)=-\frac{e^{-\psi}}{2}\\
\mathcal{L} ( e^{-\psi})=-\frac{e^{-\psi}}{4}~.\end{array}\end{equation}
Hence, an analytical solution of Eq.~\eqref{perturbedTP2} is the following (for a derivation in the Eulerian representation, see also \citealt{chavanis2}):
\begin{equation}G_{TP}(\xi_0)= 2 e^{-\psi}-\psi'^2~.\end{equation}
This solution satisfies the boundary conditions $\rho_1(0)=2$ and $\rho_1'(0)=0$, as required by Eq.~\eqref{condcontlagr}. The zeros of $G_{TP}(\xi_0)$ allow us to identify the marginally stable normal modes that satisfy the correct boundary conditions \eqref{condcontlagr}. The first zero of $G_{TP}(\xi_0)$ is at $\Xi=6.45$. Other zeros correspond to the values of $\Xi$ at which new unstable modes appear and are found to be the same as in the standard thermodynamical approach.

\section{The different behavior of a collisionless self-gravitating sphere}

\subsection{Time scales} \label{timescales}

In this subsection we briefly discuss the typical time scales that characterize gaseous and fluid models and compare them to those of globular clusters. To a large extent, our discussion follows that of \cite{inagaki}.

Let $\tau_{LTE}$ be the local relaxation time, $\tau_{GTE}$ the global relaxation time, and $\tau_{d}$ the dynamical time scale. For a gaseous system, we identify $\tau_{LTE}$ with the time (of the order of the inverse mean collision frequency) needed to reach a local Maxwellian distribution, $\tau_{GTE}$ with the time in which thermal conduction balances the temperatures of different parts of the system, and $\tau_{d}$ with the sound travel time. Gaseous models generally assume the following ordering:
\begin{equation} \label{hpgas} \tau_{LTE} \ll \tau_{d} \ll \tau_{GTE}~. \end{equation}For a globular cluster, because the mean free path is very large and stars can cross the system many times before actually colliding, the ordering of time scales is different. Stars do not collide significantly with neighboring stars, following the mechanisms that usually characterize thermal conduction; instead, they tend to release their energy through the entire cluster (also because of the long range nature of the force). Thus for globular clusters we have:
\begin{equation}\label{hpglobular} \tau_{d}\ll \tau_{LTE}\simeq\tau_{GTE}.\end{equation}
In the fluid model analyzed in this paper we assumed infinite thermal conductivity, because the temperature was always taken to be and to remain uniform. Moreover, we implicitly assumed that the distribution function is locally Maxwellian. Therefore, our fluid model follows the ordering:
\begin{equation}\begin{alignedat}{1} \label{hpfluid}
& \tau_{LTE} \ll \tau_{d}\\
& \tau_{GTE} \ll \tau_{d}~.
\end{alignedat}  \end{equation}
Note that assumptions \eqref{hpgas} and \eqref{hpfluid} are not applicable to the situation of a globular cluster \eqref{hpglobular}. This 
difference and the related limitations must be kept in mind if we wish to apply these models to understand the evolution of globular clusters.

\subsection{The dynamics of a collisionless self-gravitating isothermal sphere} \label{collvscollless}

Isothermal spheres are equilibrium configurations for several idealized models of the pure $N$-body problem. In particular, self-gravitating isothermal spheres can be studied as stationary states of the collisionless Boltzmann equation:
\begin{equation} \label{collisionless} \frac{\partial f(\mathbf{r},\mathbf{v},t)}{\partial t}+ \mathbf{v}\cdot \frac{\partial f(\mathbf{r},\mathbf{v},t) }{\partial \mathbf{r}}
-\frac{\partial{\Phi(\mathbf{r},t)}}{\partial\mathbf{r}} \cdot \frac{\partial f(\mathbf{r},\mathbf{v},t) }{\partial \mathbf{v}}=0~. \end{equation}
Therefore, it is natural to ask whether such collisionless isothermal spheres are or can be unstable. It can be shown \citep{binneytremaine} that the \emph{unbounded} collisionless isothermal sphere is linearly \emph{stable}, in contrast to the results of the fluid counterpart (in the fluid model we recover the unbounded case by taking $\Xi\to\infty$).
It is generally believed that the collisionless isothermal sphere is also stable in the nonlinear regime, although to our knowledge a rigorous proof of this statement is still lacking. 
Moreover, if only spherically symmetric perturbations are considered, it is possible to show, through an argument based on the conservation of the detailed angular momentum\footnote{By detailed angular momentum we mean the angular momentum of the individual particles.} (Appendix \ref{detailedangular}), that a collisionless sphere (bounded or unbounded) \emph{cannot} collapse, because each particle has a minimum radius it can attain. In contrast, the fluid system analyzed in Sects.~\ref{basicequations} and \ref{hydroanalysis} is unstable \emph{only} with respect to spherically symmetric perturbations, with the subsequent nonlinear evolution presumably leading to a collapse. The above considerations support the hypothesis that self-gravitating collisionless isothermal spheres (bounded or unbounded) are stable with respect to all kinds of perturbations and cannot collapse \citep{kandrup2,kandrup1,battmorrison}.

The different behavior, between the collisionless and the fluid case, emerged in this paper is actually quite surprising, because 
for the homogeneous case the collisionless and fluid models behave in the same way with respect to Jeans instability \citep[e.g., see][]{bertin}, in the sense that both the fluid and the collisionless systems are linearly unstable under the \emph{same} criterion for instability.

If the self-gravitating collisionless isothermal sphere were unstable, its instability would develop on the dynamical time scale; in turn, it is commonly believed that the gravothermal catastrophe may occur only if the system is at least weakly collisional and thus it is thought to develop on the collision time scale.

In the spherically symmetric case a difference between the two models is the following: while in the fluid model each fluid element is sustained against gravity by pressure of the inner parts, 
in the collisionless model stars are sustained by their individual angular momentum relative to the center, that is, 
by their velocity dispersion.\footnote{Even if the total angular momentum vanishes, the sum of the magnitudes of the angular momenta of the individual particles is different from zero.} In moving from a kinetic description to a fluid description, all the information about microscopic velocities and detailed angular momentum 
is lost: in particular, each fluid element has zero angular momentum (see also Appendix \ref{detailedangular}).
In this respect, the real situation of a globular cluster resembles more the collisionless case: strictly speaking, stars are not sustained by pressure, but rather by their velocity dispersion, because the mean free paths are long and stars cross the cluster many times before feeling the effects of collisions.

If we come back to the original pure $N$-body problem, it is therefore natural to ask: \emph{what is in this case the role of detailed angular momentum?}
We may argue that a quantity related to the detailed angular momentum should characterize the process of core collapse. If the collapse can happen in an $N$-body system, it may be accompanied by a slow decrease of the sum of the magnitudes of the angular momenta of the individual particles slowly sinking toward the center.

For a particle with angular momentum $J$ that moves in a gravitational potential generated by a mass $M$ a radius related to angular momentum (which is the radius of the circular orbit if $M$ is a point mass) can be defined as:
\begin{equation} d\equiv\frac{J^2}{Gm^2M}~. \end{equation}
Therefore, for a stellar system of $N$ stars and total mass $M$ we can define a radius related to detailed angular momentum in the following way:
\begin{equation} R_{\mathrm{am}}\equiv\frac{A}{NGm^2M}, \end{equation}
where $A$ is the sum of the squares of the angular momenta of the individual stars, that is, the following quantity: \begin{equation} A\equiv \sum_{i=1}^N \mathbf{J}^2_i~.\end{equation}

Thus we may argue that the typical time scale for core collapse should correlate with:
\begin{equation} t_\mathrm{cc}\equiv \frac{A}{{\mathrm{d}A}/{\mathrm{d}t}}~.\end{equation}
The main mechanism through which $A$ varies with time is expected to be that of two-body collisions.  
Thus $t_\mathrm{cc}$ should be of the order of the two-body relaxation time.

By monitoring the quantity $A(t)$ in $N$-body simulations, it would be interesting to test the role of detailed angular momentum in the mechanism of gravothermal oscillations \citep{makino} [$A(t)$ may reach an equilibrium value, around which the system gravothermally oscillates; this would be consistent with the fact that the typical time scale of gravothermal oscillations is the two-body relaxation time], its relevance to the studies of core-collapse in the gaseous model \citep{lyndenbelleggleton,sugimoto}, and its connection with the phenomenon of the gyro-gravothermal catastrophe \citep{hachisugyro}.

\section{Discussion and conclusions} \label{conclusions}

In this paper we have studied systematically the dynamical stability of a self-gravitating isothermal fluid sphere by means of a linear modal analysis with respect to spherically symmetric perturbations. In this sense, we have studied the Jeans instability in the inhomogeneous context of a sphere of finite size. 
Within a unified framework, by imposing the boundary conditions of constant $\{T,V\}$ (isothermal collapse), constant $\{E,V\}$ (gravothermal catastrophe), and constant $\{T, P\}$ (isobaric collapse), we have proved that the onset of dynamical instability occurs exactly at the points identified in the thermodynamical approach \citep[see][]{antonov,lyndenbellwood,bonnor,ebert} and, by adapting derivations from other authors \citep{padmana2,chavanis1,chavanis2}, we have provided an analytic expression for the eigenfunctions of the marginally stable modes.
Indeed, as noted in the Introduction, some results along these lines have been obtained previously by other authors. The main new results obtained in this paper are the following:
\begin{itemize}
\item Using the fluid model based on the Euler equation, we have extended previous studies of the constant $\{T,V\}$ and $\{T,P\}$ cases to the constant $\{E,V\}$ case (gravothermal catastrophe), proving that the onset of Jeans instability occurs exactly at the same point identified in the thermodynamical approach also in this case. For this constant $\{E, V\}$ case, we have introduced a heuristic term to incorporate effects akin to the stabilizing role of binaries.
\item For all the three cases described above, we have calculated numerically eigenfrequencies and eigenfunctions of the relevant modes also outside the conditions of marginal stability. The time scale for the instability that we have found is the dynamical time scale. The excitation of higher modes has been illustrated in a simple way, by referring to an effective potential that governs the structure of the linear modal analysis.
\item We have found that for all the cases treated in our investigation, as the dimensionless radius of the isothermal sphere $\Xi$ becomes larger and larger, the value of the dimensionless growth rate of the most unstable mode tends to a universal asymptotic constant value, independent of the adopted boundary conditions.
\item We have briefly shown that the correspondence between the stability in the dynamical and in the thermodynamical approach also holds for the two-component case and have found indications that the heavier component is the more important driver of the instability.
\item As a general discussion, we have commented on the meaning and applicability of the Boltzmann entropy for self-gravitating systems and argued that the main difference between the dynamical behavior of a fluid and a collisionless sphere, in relation to their application as models of the pure $N$-body problem or a real weakly collisional stellar system such as globular clusters, is to be ascribed to the role of the detailed angular momentum behavior in the collisionless and weakly collisional cases.
\end{itemize}
The role of the viscosity and its consequences outside the condition of marginal stability have not been examined; hopefully, this issue will be addressed in a future paper.

Another interesting question is how the instability depends on the particle-particle interaction, for non-Newtonian cases \citep{padmana1}. Potentials that exhibit a softening at small radii, such as $1/(r^2+r_0^2)^{1/2}$, where $r_0$ is a constant
\citep[][see also \citealt{casetti}]{ispolatov2}, or that decline with a different power law at large radii, such as $1/r^{\alpha}$, with $\alpha \neq 1$ \citep{ispolatov1}, have indeed been considered. Based on the present article, we may argue that the results obtained from the thermodynamical approach would be reinterpreted and clarified as the analogue of the Jeans instability, by studying a fluid model  for the potential considered, with the equation of state of a perfect gas.

In general, this paper strengthens the view that the applicability of different idealized models to describe the process of core collapse in systems made of a finite number of stars is more subtle than commonly reported and that, in general, the study of Jeans instability of inhomogeneous stellar systems still leaves a number of questions open.

%
%                                                One column figure
%----------------------------------------------------------- S_vib
%   \begin{figure}
%   \centering
%   \includegraphics[width=8cm]{omega2minTV.pdf}
%      \caption{Vibrational stability equation of state
%               $S_{\mathrm{vib}}(\lg e, \lg \rho)$.
%               $>0$ means vibrational stability.
%              }
%         \label{FigVibStab}
%   \end{figure}
%
%______________________________________________________________

\begin{acknowledgements}
   We would like to thank Marco Lombardi, Francesco Pegoraro and Steven N. Shore for many interesting comments and discussions.
\end{acknowledgements}

\bibliographystyle{aa} % style aa.bst
\bibliography{mybib}

\appendix

\section{Fluid truncated isothermal spheres} \label{appendixhydro}

In this appendix we summarize the properties of spatially truncated self-gravitating fluid isothermal spheres. 

The equations for the hydrostatic equilibrium of a spherically symmetric fluid with the equation of state of an ideal gas are:
  \begin{equation} \label{hydrostatic1}
       \frac{GM(r)\rho_0(r)}{r^2}=-\frac{\mathrm{d} p_0(r)}{\mathrm{d} r}~,      \end{equation}
   \begin{equation} \label{hydrostatic2}  M(r)=\int_0^r 4\pi s^2 \rho_0(s) \mathrm{d}s~,\end{equation}
\begin{equation}p_0(r)=\rho_0(r) k T/m~.\end{equation}

We express $M(r)$ by means of the condition of hydrostatic equilibrium 
\eqref{hydrostatic1}, differentiate to obtain $\mathrm{d}M$, and equate it to $\mathrm{d}M=4\pi \rho_0(r) r^2$. By making the change of variable $\rho_0(r)=\rho_0(0)e^{-\psi(r)}$, where $\rho_0(0)$ is the central density, and introducing the dimensionless radius $\xi=r/\lambda$, where $\lambda=\left[kT/4\pi G \rho_0(0) m\right]^{1/2}$, we obtain the differential equation for $\psi(\xi)$, recorded in the main text as Eq.~\eqref{emden}. [Because the constant $\rho_0(0)$ is interpreted as the central density, we are considering the boundary condition $\psi(0)=0$; the density is taken to be regular at the origin, so that the second boundary condition is $\psi'(0)=0$.] The solution of Eq.~\eqref{emden} (called Emden equation) is a monotonic increasing function characterized by  logarithmic behavior $\psi(\xi)\sim \ln \xi^2 $ and $\psi'(\xi) \sim 2/\xi$ as $\xi\to\infty$ .

From Eq.~$\eqref{emden}$ the mass enclosed within the radius $\xi$ is:
\begin{equation} \label{massaemden} M(\xi)= \frac{kT \lambda}{Gm}\ \xi^2 \psi'(\xi)~.\end{equation}
From Eq.~\eqref{massaemden} and the asymptotic behavior of $\psi$, it is clear that a solution with finite total mass is obtained only by truncating the system at a dimensionless radius $\xi=\Xi$. A truncated isothermal sphere is then identified by two scales $T$ and $\rho_0(0)$ and one dimensionless parameter $\Xi$. The density profile of a spatially truncated isothermal sphere is given by $\rho_0(r)=\rho_0(0)e^{-\psi(r)}$ where $\psi$ is the solution to Eq.~\eqref{emden}. 

\section{Linearization of the hydrodynamic equations}

\subsection{Eulerian representation} \label{euleriancalculations}

Here we record the calculations leading to the linearized Eq.~\eqref{linearized1}. The unperturbed density profile is given by Eq.~\eqref{unpert}. We substitute Eqs.~\eqref{perturb} in Eq.~\eqref{idro1} and expand to first order in quantities with subscript 1 to obtain:
\begin{equation}\label{idro3app} 
\frac{\partial \rho_1}{\partial t}+ \boldsymbol{\nabla}\cdot(\rho_0 \mathbf{u}_1)=0~,
\end{equation}
\begin{equation} \label{idro4app}\begin{alignedat}{1}
\frac{\partial \mathbf{u_1}}{\partial t}=\left( \frac{\boldsymbol{\nabla}\rho_0}{\rho_0}\frac{\rho_1}{\rho_0}-
\frac{\boldsymbol{\nabla}\rho_1}{\rho_0} \right)\frac{kT_0}{m} - \frac{\boldsymbol{\nabla}\rho_0}{\rho_0}\frac{k T_1}{m} 
-4\pi G \frac{\int_0^r \! \rho_1(s,t)s^2 \, \mathrm{d}s}{r^2} \\
+\frac{\eta}{\rho_0} \nabla^2 \mathbf{u}_1+\frac{\zeta
+\frac{\eta}{3}}{\rho_0}\boldsymbol{\nabla}\left(\boldsymbol{\nabla}\cdot\mathbf{u}_1\right)~.
\end{alignedat} \end{equation}
Then we look for solutions of the form \eqref{perturb2}. From Eq.~\eqref{idro3app} we obtain $\rho_1$:
\begin{equation} \label{idro11app}
\rho_1=\frac{ \boldsymbol{\nabla}\cdot(\rho_0 \mathbf{u}_1)}{i\omega},
\end{equation} and thus eliminate it from Eq.~\eqref{idro4app} to find the radial component of the Navier-Stokes equation:  
\begin{eqnarray}
\omega^2 \rho_0 u_1& = & \nonumber \left\{ \frac{  {\partial \rho_0}/{\partial r} }{\rho_0}\frac{1}{r^2}\frac{\partial}{\partial r}\left(r^2 \rho_0 u_1\right)
-\frac{\partial}{\partial r} \left[ \frac{1}{r^2}\frac{\partial}{\partial r}\left(r^2 \rho_0 u_1\right) \right]\right\}\frac{kT_0}{m} \\
& &-i\omega \frac{\partial \rho_0}{\partial r}  \frac{kT_1}{m} -4 \pi G \rho_0^2 u_1 \nonumber \\
& &\nonumber + i\omega \left\{ \eta \frac{1}{r^2}\frac{\partial}{\partial r}\left(r^2 \frac{\partial u_1}{\partial r}\right) +(\zeta
+\frac{\eta}{3})\frac{\partial}{\partial r} \left[ \frac{1}{r^2}\frac{\partial}{\partial r}(r^2 u_1)\right] \right\}~,\\ \label{idro5app} & &
\end{eqnarray}
where $u_1$ is the radial component of the velocity. By defining $f(r)\equiv \rho_0(r) u_1(r)$ and introducing the dimensionless radius $\xi=r/\lambda$ we obtain Eq.~\eqref{linearized1}.

\subsection{Lagrangian representation} \label{lagrangiancalculations}

\subsubsection{Change of variables} \label{changevariables}

Here we show the change of variables leading from Eqs.~\eqref{idro1}  (Eulerian representation) to Eqs.~\eqref{idrolagr} (Lagrangian representation). Let us assume spherical symmetry and neglect viscosity. By dropping the nonlinear term $(\mathbf{u}\cdot\boldsymbol{\nabla})\mathbf{u}$, the Euler  equation and the continuity equation \eqref{idro1} become, in the Eulerian representation:
\begin{equation}\label{idro2lagr}
\begin{alignedat}{1}
& \frac{\partial \rho}{\partial t} +\frac{1}{r^2}\frac{\partial}{\partial r}(r^2 \rho u)=0~,   \\
& \frac{\partial u}{\partial t} = - \frac{ \partial \rho/\partial r}{\rho}\frac{kT}{m}-\frac{GM(r)}{r^2}~.  \end{alignedat}
\end{equation}
Now we perform a change of variables. In the Lagrangian representation, each quantity is described as a function of the new independent variables $r_0$ and $t$, where $r_0$ is the position of the fluid element at $t=t_0$. The standard rules to transform derivatives of a generic function $f$ are:
\begin{equation}\label{rules}
	\begin{alignedat}{1} 
	& \frac{\partial f(r,t)}{\partial r}=\frac{\partial r_0(r,t)}{\partial r} \frac{\partial f(r_0,t)}{\partial r_0} \\
	&		  \frac{\partial f(r,t)}{\partial t}=\frac{\partial r_0(r,t)}{\partial t} \frac{\partial f(r_0,t)}{\partial r_0} +  \frac{\partial f(r_0,t)}{\partial t}~. 
	\end{alignedat} 
\end{equation}
The partial derivatives of $r_0$ are obtained from the following relation, which expresses the condition that two fluid shells do not cross each other:  
\begin{equation} \label{changeapp}
\int_0^{r_0(r,t)}\!\!\!\!\! \rho_0(s)4 \pi s^2 \, \mathrm{d}s=\int_0^r\! \rho(s,t)4 \pi s^2 \, \mathrm{d}s~.
\end{equation}
By taking the partial derivative with respect to $r$ of Eq.~\eqref{changeapp} (each side of the equation is considered as a function of $r$, $t$) we obtain:
\begin{equation}\label{lagr4app}
r_0^2 \rho_0(r_0) \frac{\partial r_0(r,t)}{\partial r}   = r^2 \rho(r,t)~.
\end{equation}
By taking the partial derivative of Eq.~\eqref{changeapp} with respect to $t$ we obtain:
\begin{equation}\label{lagr2app}
r_0^2 \rho_0(r_0)  \frac{\partial r_0(r,t)}{\partial t} =\int_0^r\!  s^2 \frac{\partial \rho(s,t)}{\partial t} \, \mathrm{d}s~.
\end{equation}
From the continuity equation, Eq.~\eqref{lagr2app} then becomes:
\begin{equation}\label{lagr3app}
r_0^2 \rho_0(r_0)  \frac{\partial r_0(r,t)}{\partial t} =- r^2 \rho(r,t) u(r,t)~.
\end{equation}
From Eqs.~\eqref{lagr3app}, \eqref{lagr4app}, and \eqref{rules} we obtain the equations 
of hydrodynamics in the Lagrangian representation \eqref{idrolagr}.

\subsubsection{Linearization} \label{linearizationlagrangian}

Here we approximate Eqs.~\eqref{idrolagr} to first order for small perturbations around the hydrostatic equilibrium states. We substitute Eqs.~\eqref{perturblagr} in Eqs.~\eqref{idrolagr} 
and expand to first order in quantities with subscript 1. 
By noting that $M(r_0,t)=M(r_0,t=t_0)$, we obtain:
\begin{equation}
\begin{alignedat}{1}\label{lagr8app}
&\frac{\partial \rho_1}{\partial t} +
\frac{\rho_0}{r_0^2}\frac{\partial}{\partial r_0}(r_0^2  u_1)=0~, \\
&\frac{\partial u_1}{\partial t}
=-\frac{kT_0}{m} \frac{\partial \rho_0/\partial {r_0} }{ \rho_0}\left(2\frac{r_1}{r_0} 
+\frac{\partial \rho_1/\partial {r_0}}{\partial \rho_0/\partial {r_0}}+
\frac{T_1}{T_0}\right)+2\frac{GM(r_0)}{r_0^3}r_1~.
\end{alignedat}
\end{equation}
From the usual rules of derivation, we have:
\begin{equation} \label{partials}
\frac{\partial r(r_0,t)}{\partial t}= - \frac{   {\partial r_0(r,t)}/{\partial t}      }{   {\partial r_0(r,t)}/{\partial r}   }~.
\end{equation}
By applying Eqs.~\eqref{lagr3app} and \eqref{lagr4app}, Eq.~\eqref{partials} can be written as:
\begin{equation}\label{lagr7app}
\frac{\partial r_1}{\partial t} = u_1~.
\end{equation}
Now we wish to obtain one equation involving only the unknown $\rho_1$, by combining the three Eqs.~\eqref{lagr8app} and \eqref{lagr7app}. We assume the modal dependence \eqref{lagr9}. We eliminate $r_1$ from \eqref{lagr7app} and the 
second of \eqref{lagr8app}. We then eliminate $u_1$ from the resulting equation and the first of \eqref{lagr8app}, to obtain the following equation
\{we also used hydrostatic equilibrium to replace $[({\mathrm{d} \rho_0/\mathrm{d}{r_0}})/{\rho_0}]
({kT_0}/{m})=-[{GM(r_0)}/{r_0^2}]$\}:
\begin{equation}\label{lagr11}
-\rho_1+\frac{\rho_0}{r_0^2}\frac{\mathrm{d}}{\mathrm{d}r_0}\left[  
\frac{GM(r_0)\left(\frac{{\mathrm{d} \rho_1 }/{\mathrm{d} r_0 }}{{\mathrm{d}  \rho_0 }/{\mathrm{d} r_0  }}+\frac{T_1}{T_0}\right)}{4\frac{GM(r_0)}{r_0^3}+\omega^2}\right]=0~.
\end{equation}
By referring to the dimensionless radius $\xi_0\equiv r_0/\lambda$,  we obtain Eq.~\eqref{perturbedlagr}.

\subsection{Temperature expression for the constant $\{E, V\}$ case} \label{tempexpr}

Here we show the steps leading from Eq.~\eqref{T12} to Eq.~\eqref{T1}.

From the Poisson and Emden equations, the dimensionless potential $\psi$ is related to the gravitational potential $\Phi_0$ of the unperturbed density distribution in the following way: \begin{equation} \label{psirelation} \psi(r)=\left[\Phi_0(r)-\Phi_0(0)\right]/(kT_0/m)~.\end{equation}  From Eq.~\eqref{psirelation}, by recalling that we consider only perturbations that do not change the total mass ($\int\! \rho_1(\mathbf{r}) \, \mathrm{d}^3 r=0$), we obtain:
\begin{equation}
T_1=-\frac{ \int\! \rho_1(r)\psi(r)\, \mathrm{d}^3 r}{\frac{3}{2}Nm}T_0~.
\end{equation}
From Eq.~\eqref{idro11app}, by setting $f=\rho_0u_1$ and $\xi=r/\lambda$, we obtain Eq.~\eqref{T1}.

\section{Detailed angular momentum conservation and collapse} \label{detailedangular}

Here, for completeness, we show in detail that the angular momentum barrier 
prevents a spherically symmetric collisionless system from collapsing. We consider only perturbations that do not break the assumed spherical symmetry of the system.

Consider a collisionless system of particles of individual mass $m$ and total mass $M$. From the assumption of spherical symmetry, each particle is confined to a plane and we can write the single-particle Lagrangian as:
\begin{equation} \label{lagrangian} \mathcal{L}=\frac{1}{2}m (\dot{r}^2 +r^2\dot{\theta}^2)-V(r,t)~,\end{equation}
where $V(r,t)$ is a time-dependent potential, $t$ is time, $r$ is the distance from the center and $\theta$ is the angular coordinate. 
The Lagrangian \eqref{lagrangian} conserves the angular momentum of the particle, that is, $r^2\dot{\theta}=C$, where $C$ is a constant. 
Then the equation of the motion is:
\begin{equation}
\ddot{r}=\frac{C^2}{r^3}-\frac{1}{m}\frac{\partial V(r,t)}{\partial r}\equiv  \frac{C^2}{r^3}  - \frac{GM(r,t)}{r^2}, \label{onedimparticle}
\end{equation}
where $M(r,t)$ is the total mass contained in the sphere of radius $r$. Equation \eqref{onedimparticle} is the equation of the motion of a particle moving in one dimension and subject to the force $F_r = m{C^2}/{r^3}- {GmM(r,t)}/{r^2}$. 
The following inequality holds:
\begin{equation}
0\leq M(r,t)\leq M~.\label{ineqmasses}
\end{equation}
By multiplying by $\dot{r}$ and integrating both sides of Eq.~\eqref{onedimparticle}, we obtain:
\begin{equation} \label{firstint}
 \frac{\dot{r}^2(t)}{2}=\frac{\dot{r}^2(t_0)}{2}+\frac{C^2}{2}\left[\frac{1}{r^2(t_0)}-\frac{1}{r^2(t)}\right] - \int_{r(t_0)}^{r(t)} \frac{GM(s,t(s))}{s^2}\,\mathrm{d}s.
\end{equation}
Since   ${\dot{r}^2(t)}/{2}$ is positive, the right-hand side of Eq.~\eqref{firstint} must be positive. By taking $r(t)\leq r(t_0)$ (we are not interested in the case in which $r(t)$ is greater than the initial radius) and using Eqs.~\eqref{firstint} and \eqref{ineqmasses}, we find:
\begin{eqnarray} \label{firstint}
 \frac{\dot{r}^2(t)}{2} &=&\frac{\dot{r}^2(t_0)}{2}+\frac{C^2}{2}\left[\frac{1}{r^2(t_0)}-\frac{1}{r^2(t)}\right] + \int_{r(t)}^{r(t_0)} \frac{GM(s,t(s))}{s^2}\,\mathrm{d}s \nonumber \\
 & \leq  & \frac{\dot{r}^2(t_0)}{2}+\frac{C^2}{2}\left[\frac{1}{r^2(t_0)}-\frac{1}{r^2(t)}\right] + \int_{r(t)}^{r(t_0)} \frac{GM}{s^2}\,\mathrm{d}s \label{diseguality} \\
 &=&\frac{\dot{r}^2(t_0)}{2}+\frac{C^2}{2}\left[\frac{1}{r^2(t_0)}-\frac{1}{r^2(t)}\right] + GM\left[\frac{1}{r(t)}-\frac{1}{r(t_0)} \right] .\nonumber
\end{eqnarray}
For given values of $r(t_0)$ and $\dot{r}(t_0)$, the quantity appearing in the last line of Eq.~\eqref{diseguality} tends to $-\infty$ as $r(t)\to 0$. Hence, for given initial conditions the particle cannot reach arbitrarily small values of $r(t)$.

\section{Density and velocity profiles of the linear modes}

In this appendix we show density and velocity profiles of the normal modes for the linear stability analysis presented in Sect.~\ref{hydroanalysis}; $\rho_1/\rho_0$ and $u_1$ are meant to be in arbitrary scales.

\clearpage

   \begin{figure*}[h!]
 \subsection{Constant $\{T, V\}$ profiles} \label{constantTVprofiles}
   \centering
   \includegraphics[width=9cm]{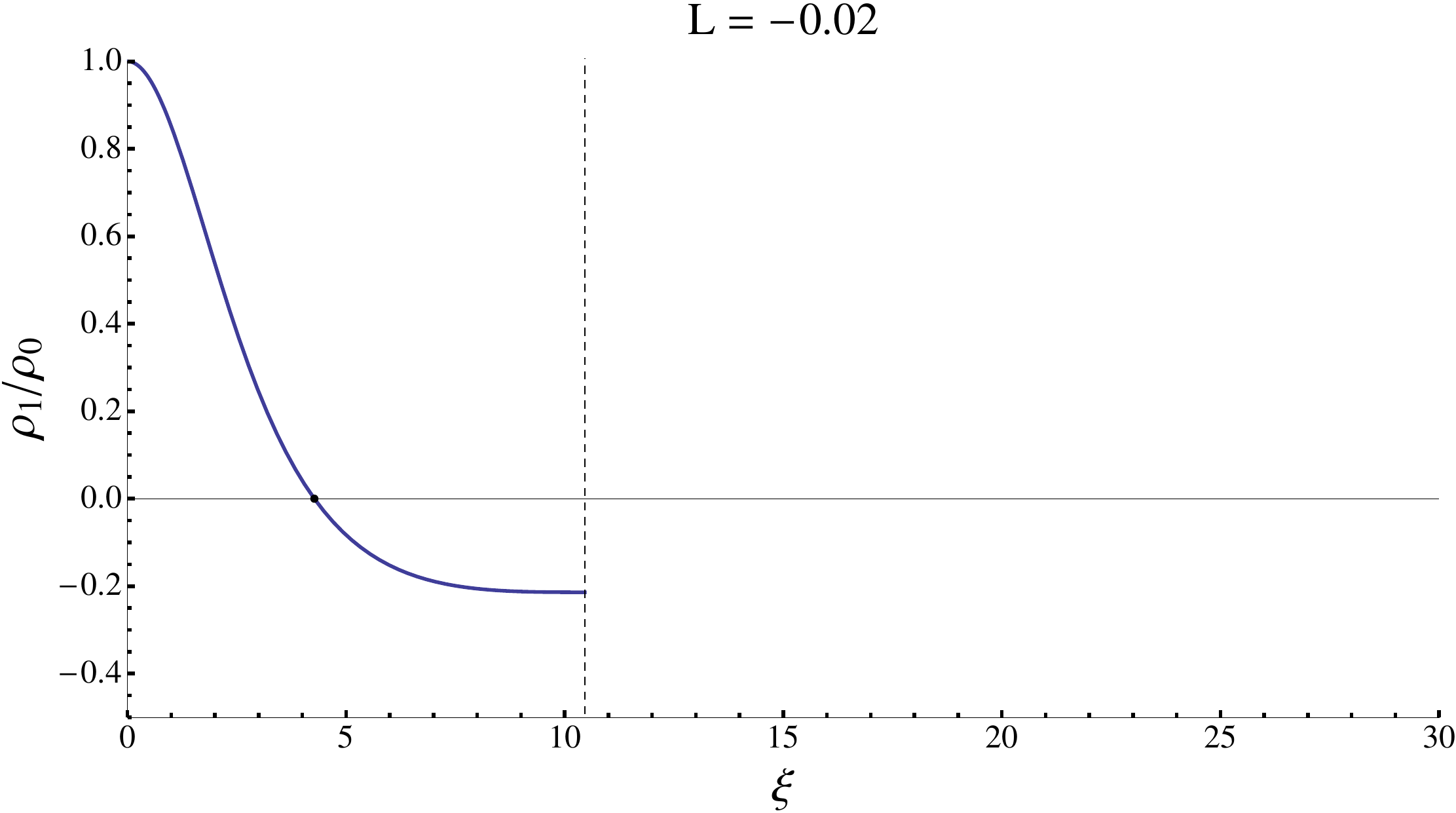}
   \includegraphics[width=9cm]{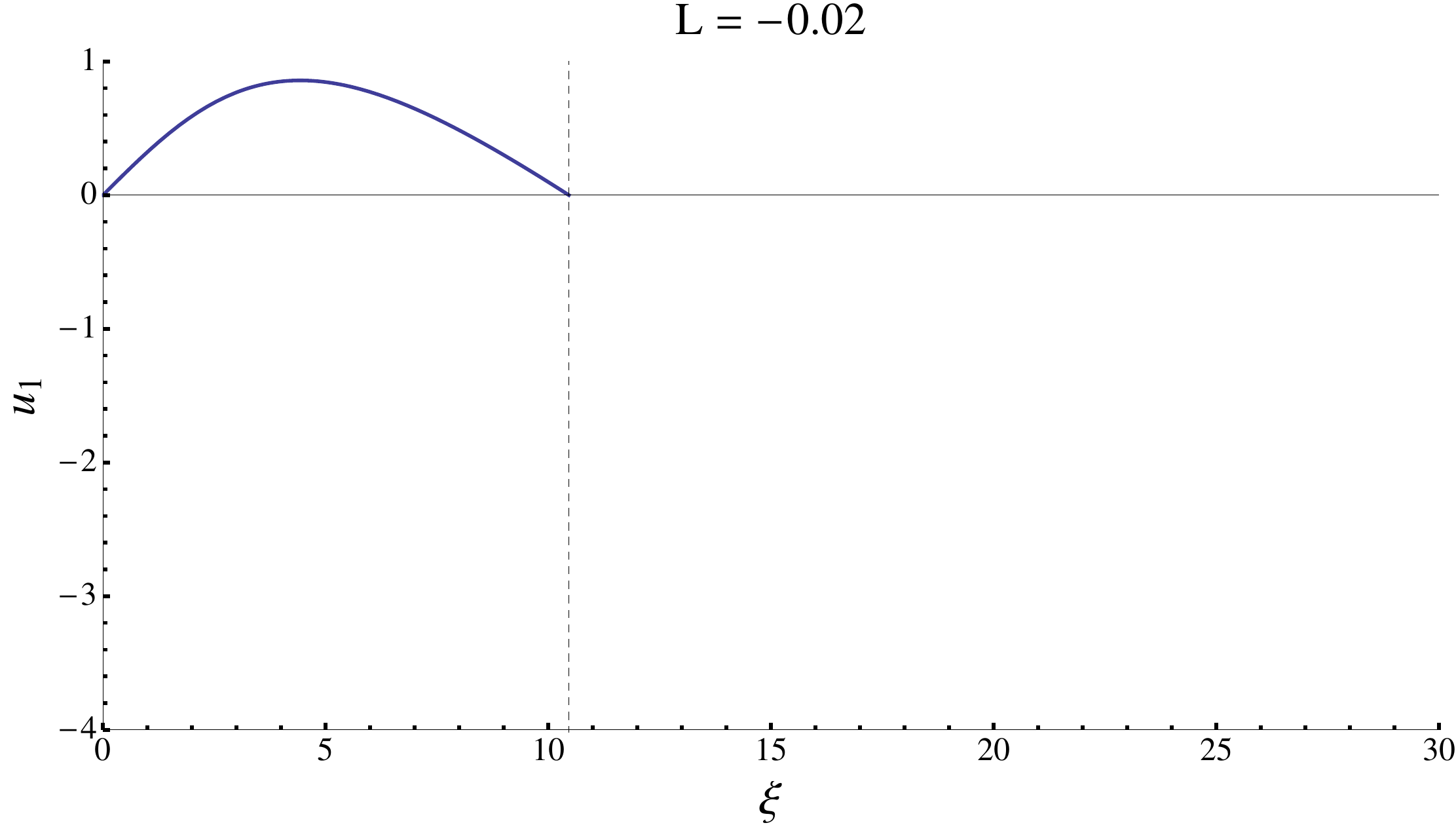}
   \includegraphics[width=9cm]{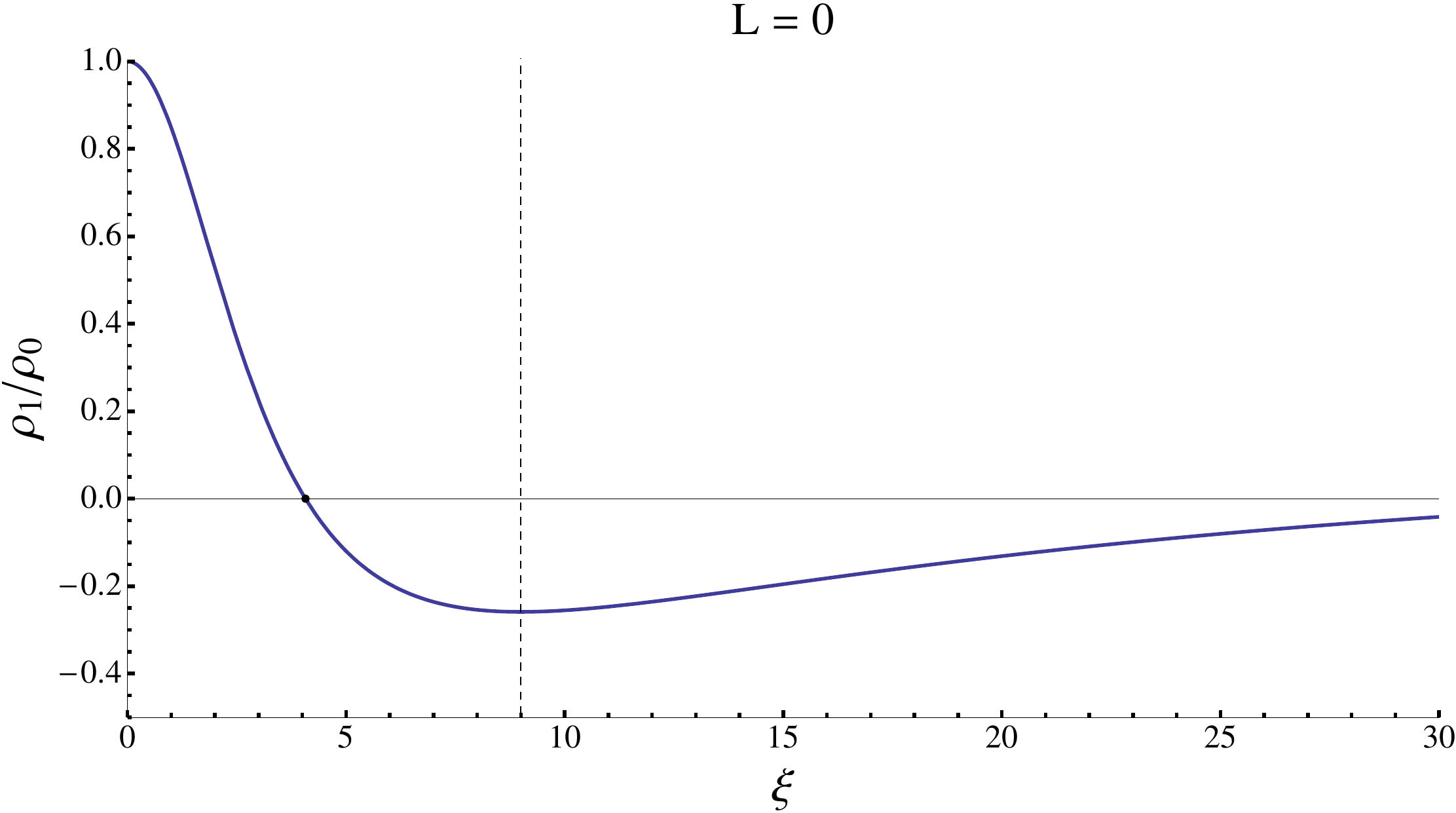}
   \includegraphics[width=9cm]{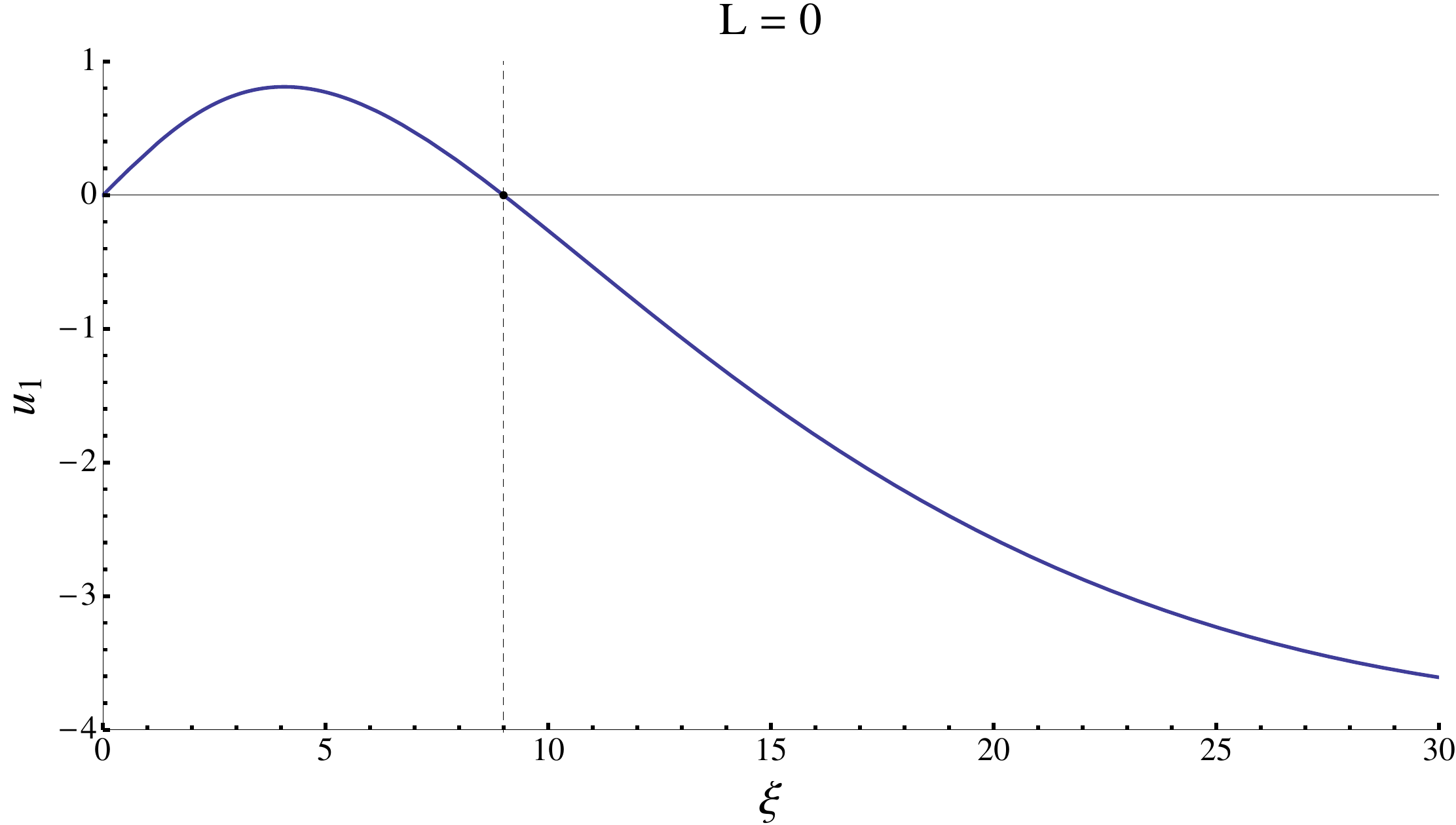}
   \includegraphics[width=9cm]{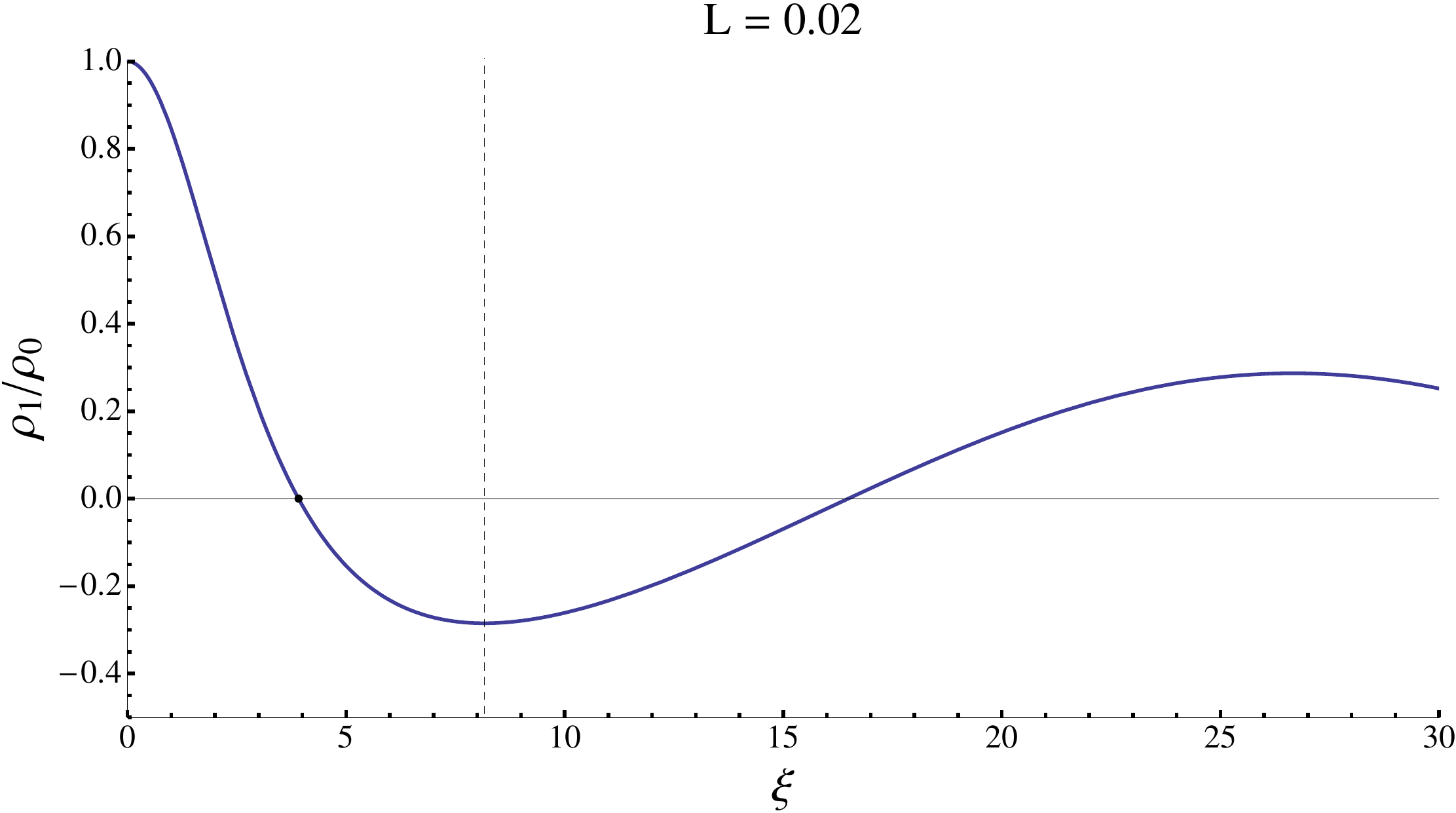}
   \includegraphics[width=9cm]{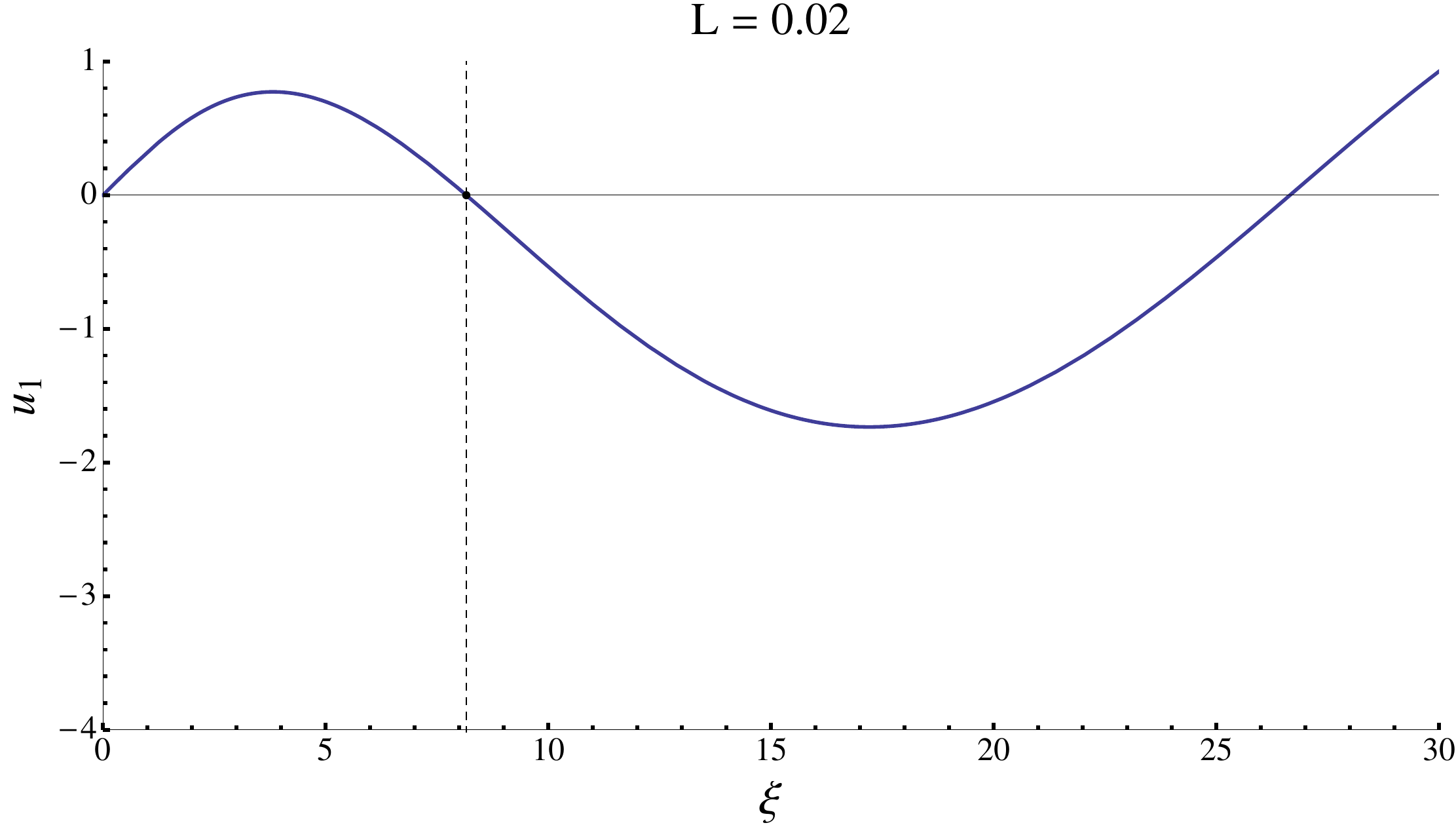}
      \caption{Relative density perturbation profiles $\rho_1(\xi)/\rho_0(\xi)$ and velocity profiles of the normal modes for the constant $\{T,V\}$ case, obtained by solving Eq.~\eqref{constTV}. The profiles should be truncated at a value $\xi=\Xi$ where the velocity profile vanishes, to satisfy boundary conditions \eqref{boundary}. The vertical dotted line indicates where the system should be truncated to obtain the mode of lowest $L$ for fixed $\Xi$: for $L=-0.02$ and $L=0$ only this mode is entirely displayed, while for $L=0.02$ two modes are displayed, depending on which zero of the velocity profile is chosen. In the case $L=0$, the total density $\rho(t)=\rho_0+\rho_1(t)$ at the point $\xi=4.07$ remains unchanged, that is, unperturbed; this is one of the relevant points listed by \cite{lyndenbellwood}.
              }
    \end{figure*}

       \begin{figure*}[h!]
\subsection{Constant $\{E, V\}$ profiles} \label{constantEVprofiles}
 \centering
 \includegraphics[width=9cm]{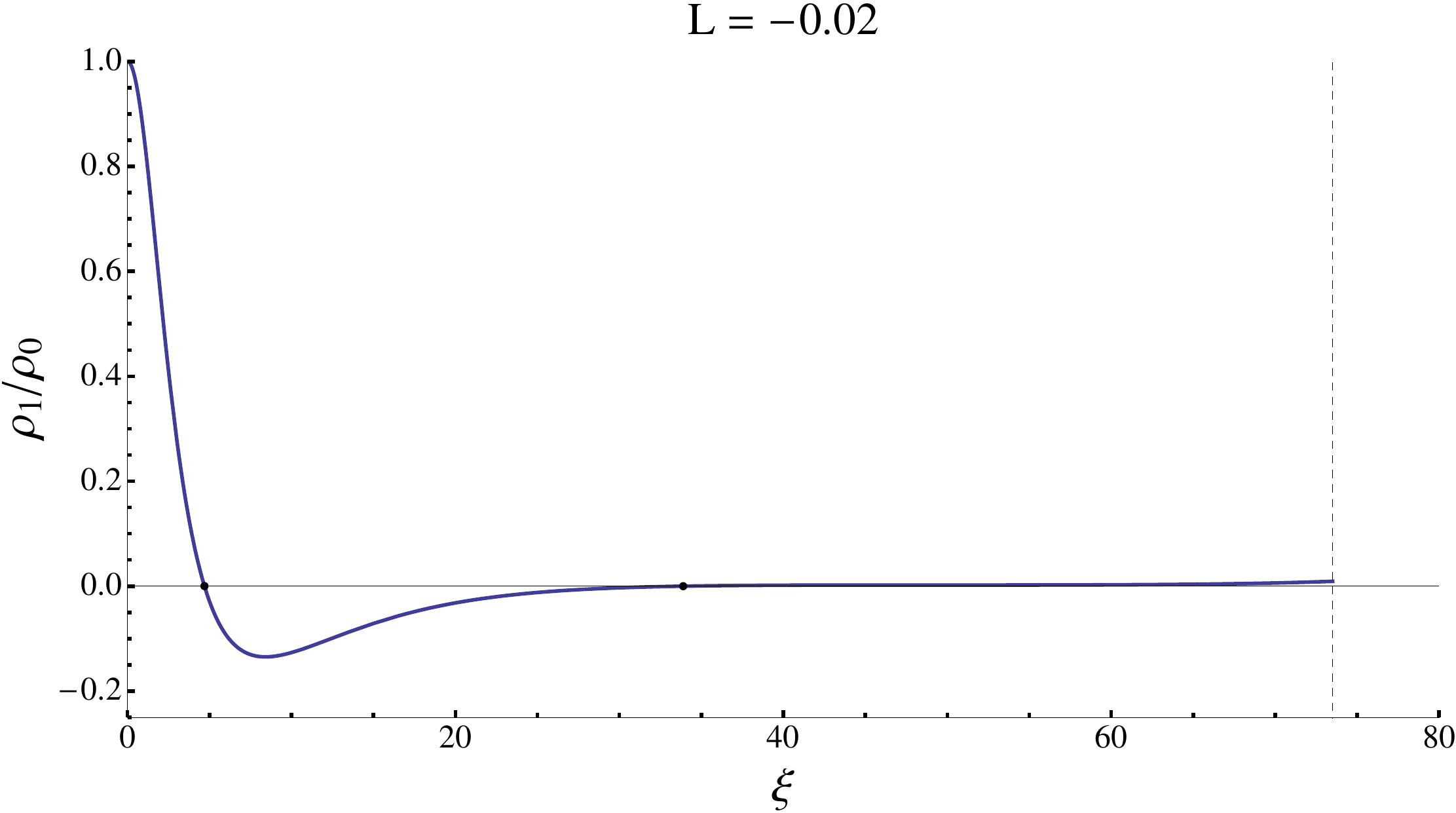}
 \includegraphics[width=9cm]{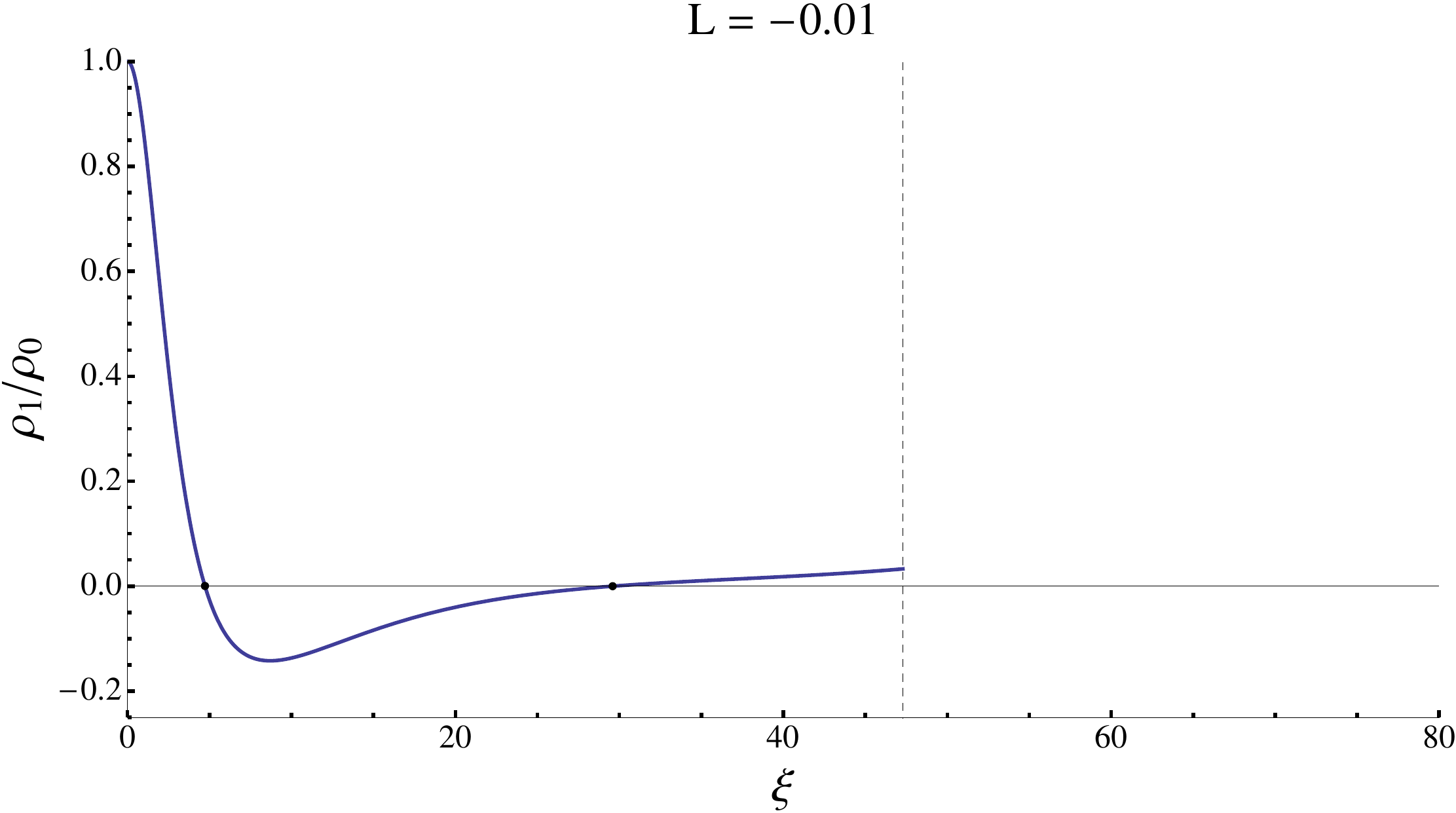}
 \includegraphics[width=9cm]{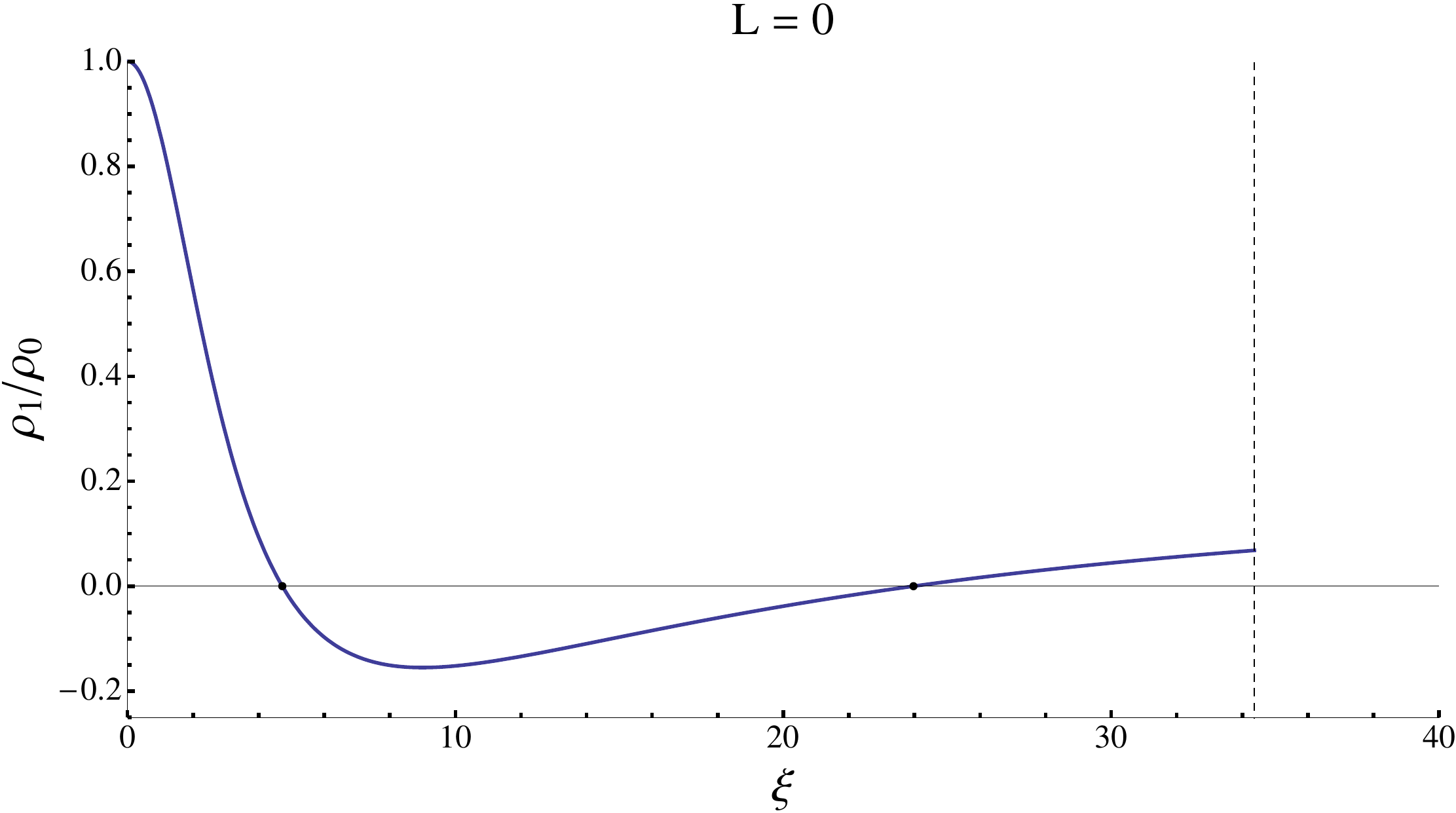}
 \includegraphics[width=9cm]{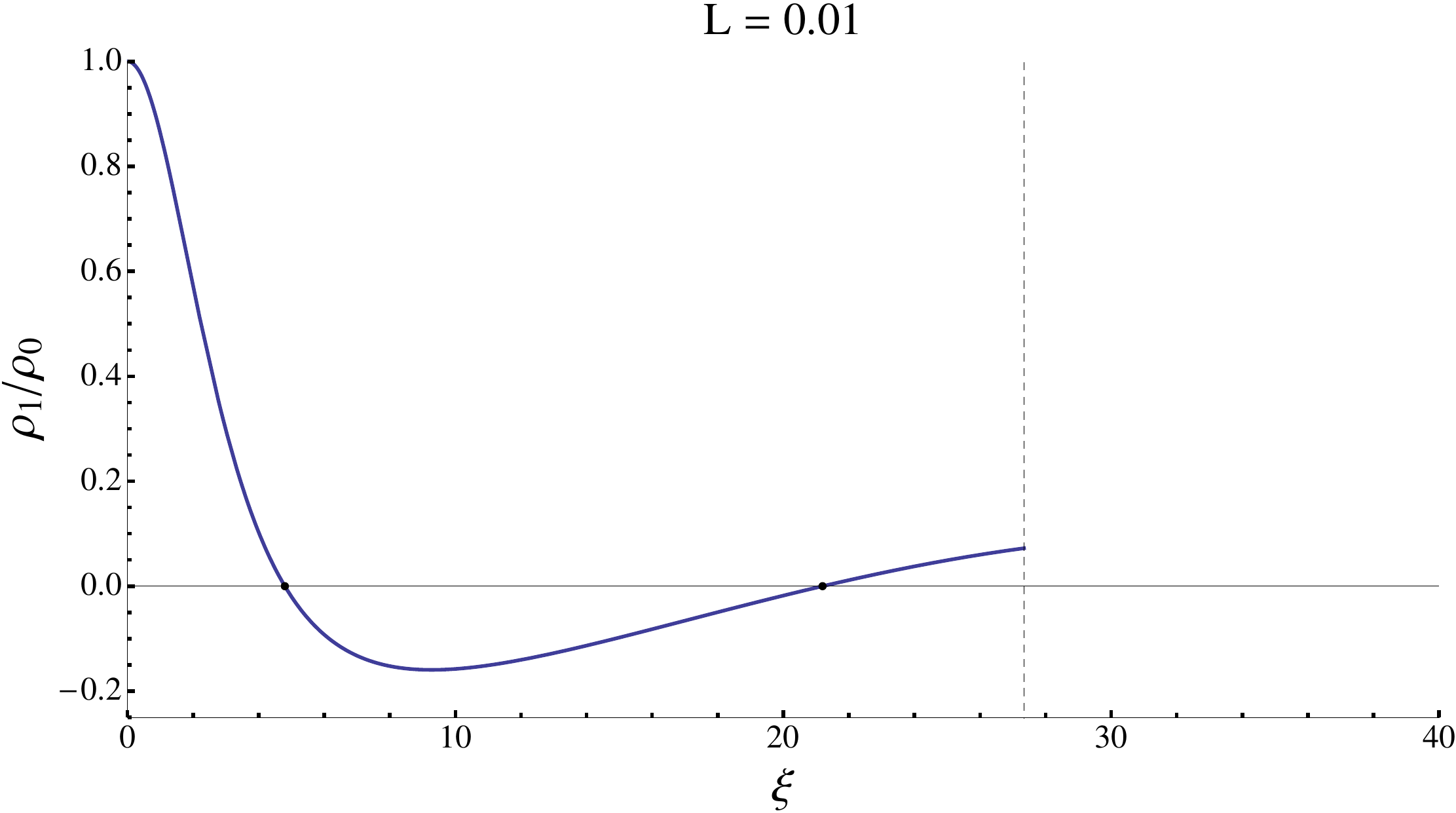}
\includegraphics[width=9cm]{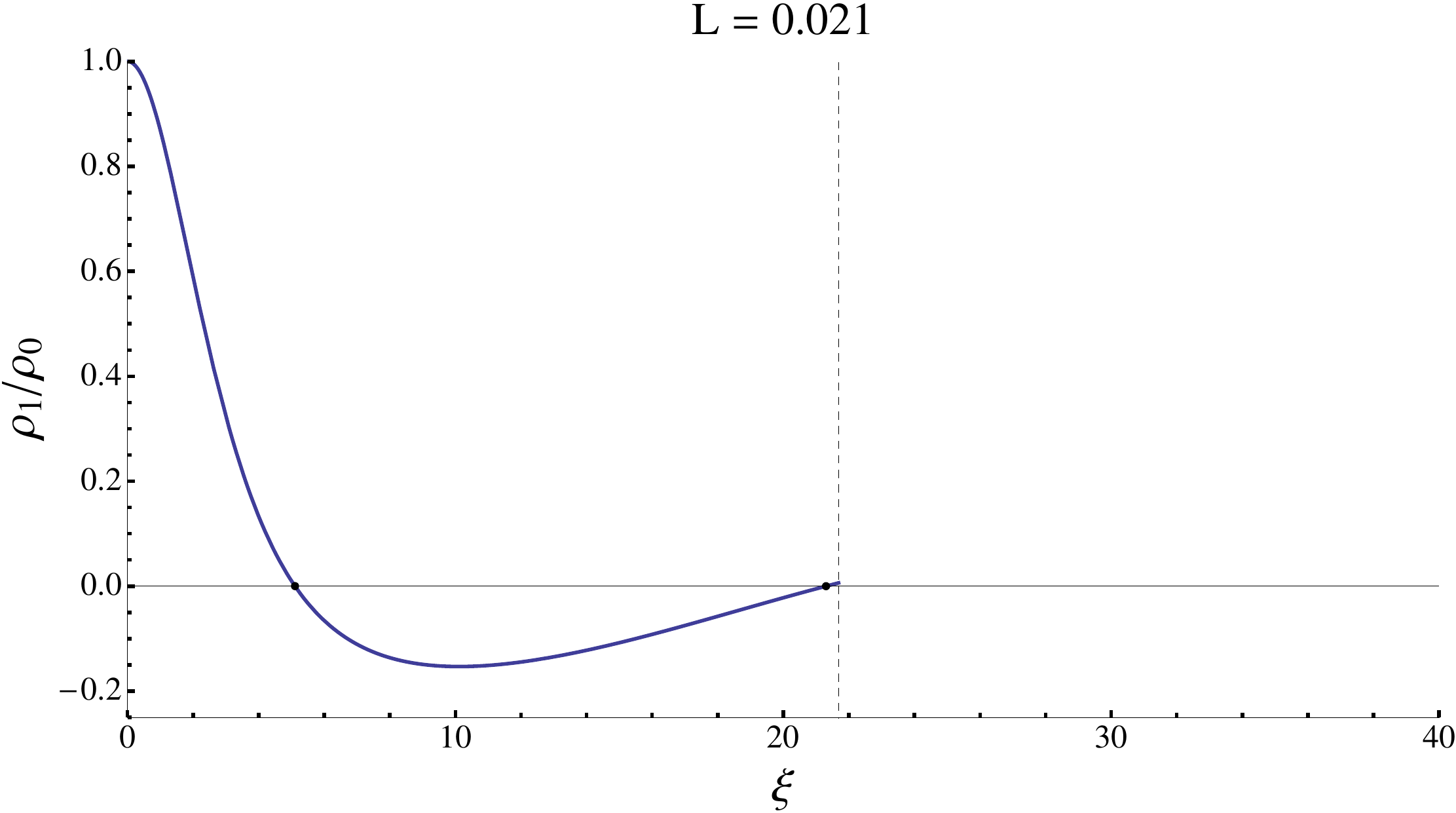}
 \includegraphics[width=9cm]{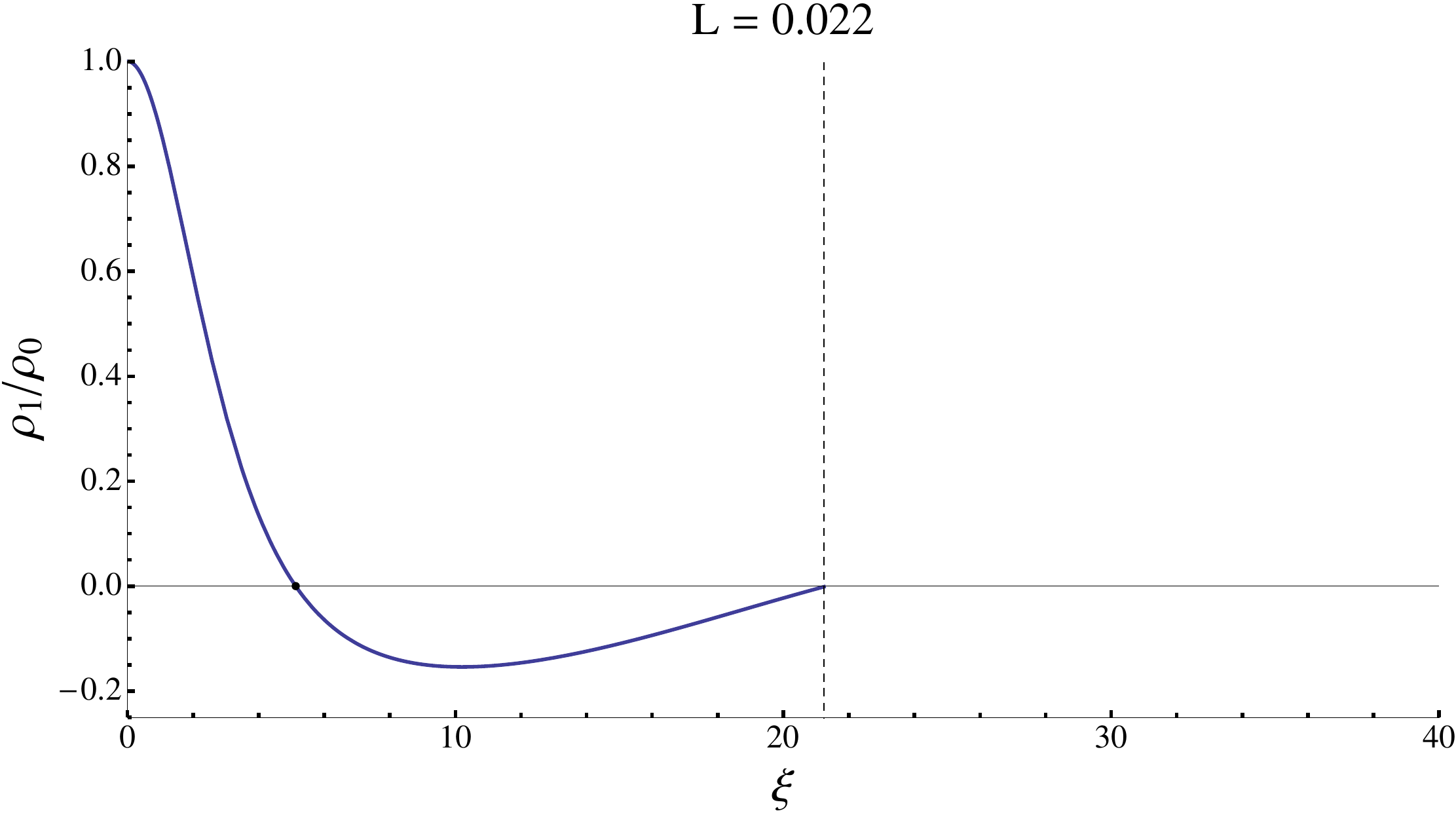}
 \includegraphics[width=9cm]{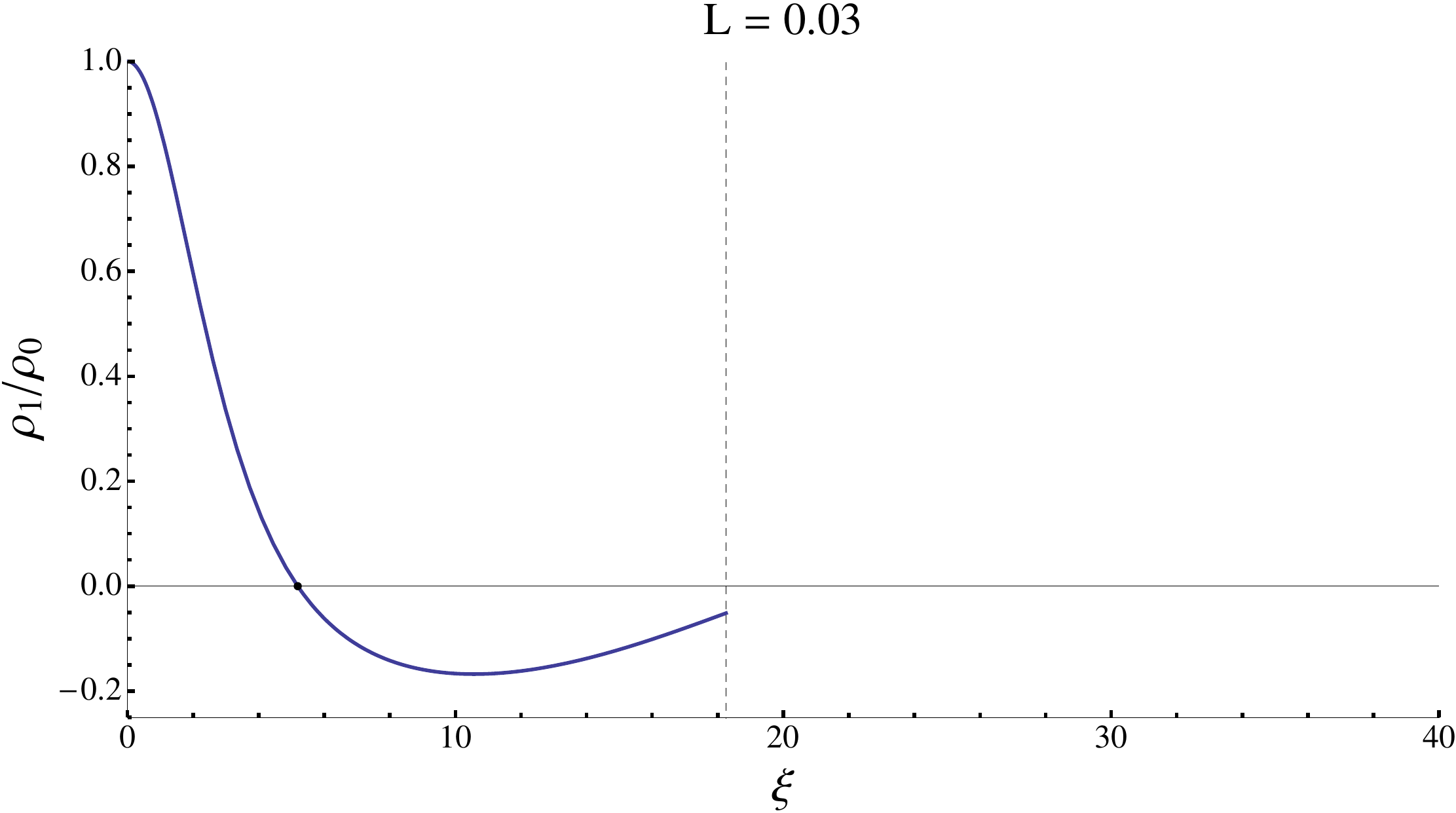}
   \includegraphics[width=9cm]{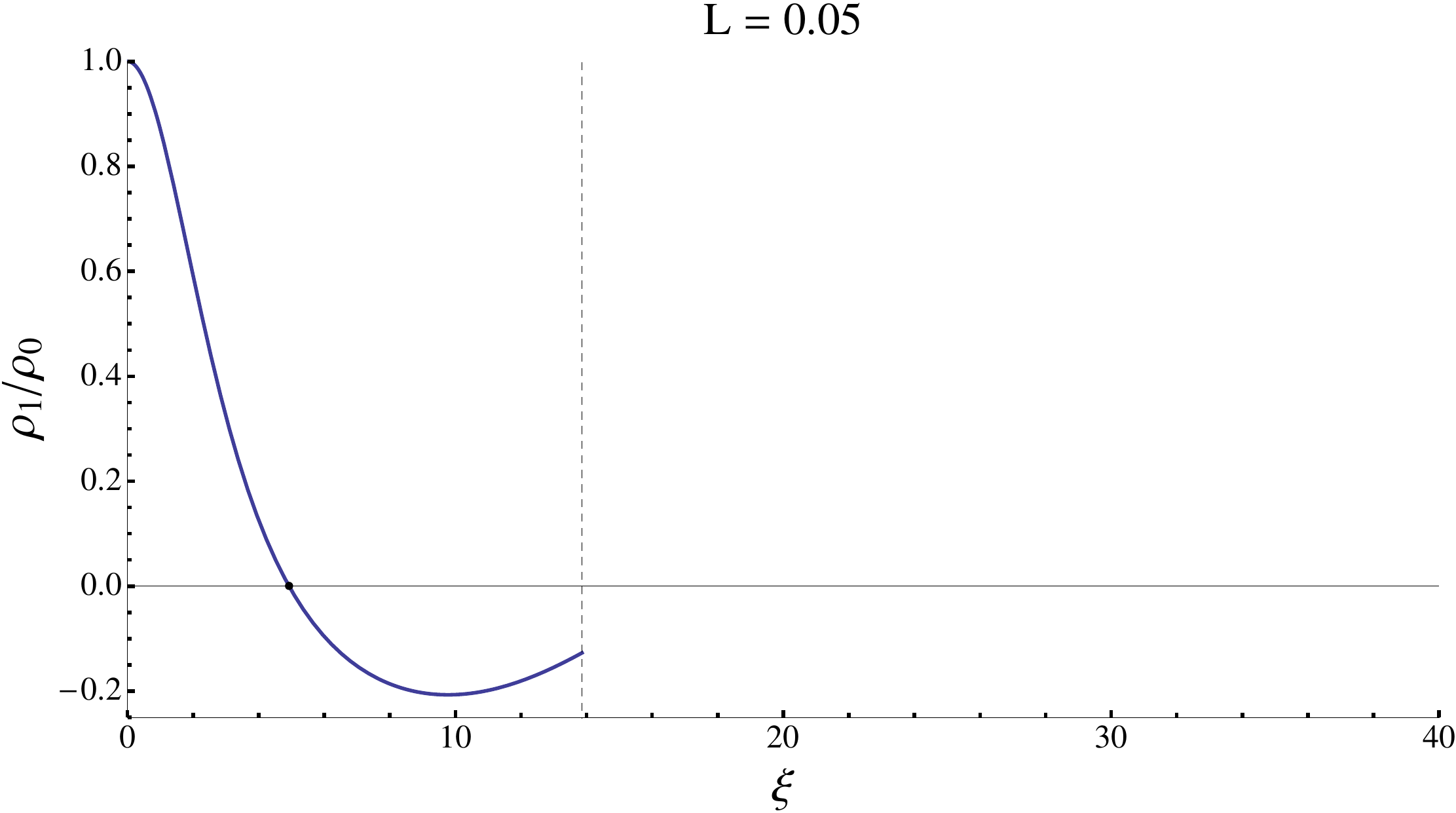}
   \caption{Relative density perturbation profiles $\rho_1(\xi)/\rho_0(\xi)$ of the normal modes for the constant $\{E,V\}$ case, obtained by solving Eq.~\eqref{idroEV1}. The modes should be truncated at a value $\xi=\Xi$ where the corresponding velocity profile shown in Fig.~\ref{fig:profvEV} vanishes, as marked by the vertical dotted lines, in order to satisfy the boundary conditions \eqref{boundary}. Zeros are displayed. Only modes of lowest $L$ at given $\Xi$ are shown. Note that the core-halo structure described in Subsection \ref{constantenergy} disappears between $L=0.021$ and $L=0.022$.
            }
 \end{figure*}

     \begin{figure*}[h!]
 \centering
 \includegraphics[width=9cm]{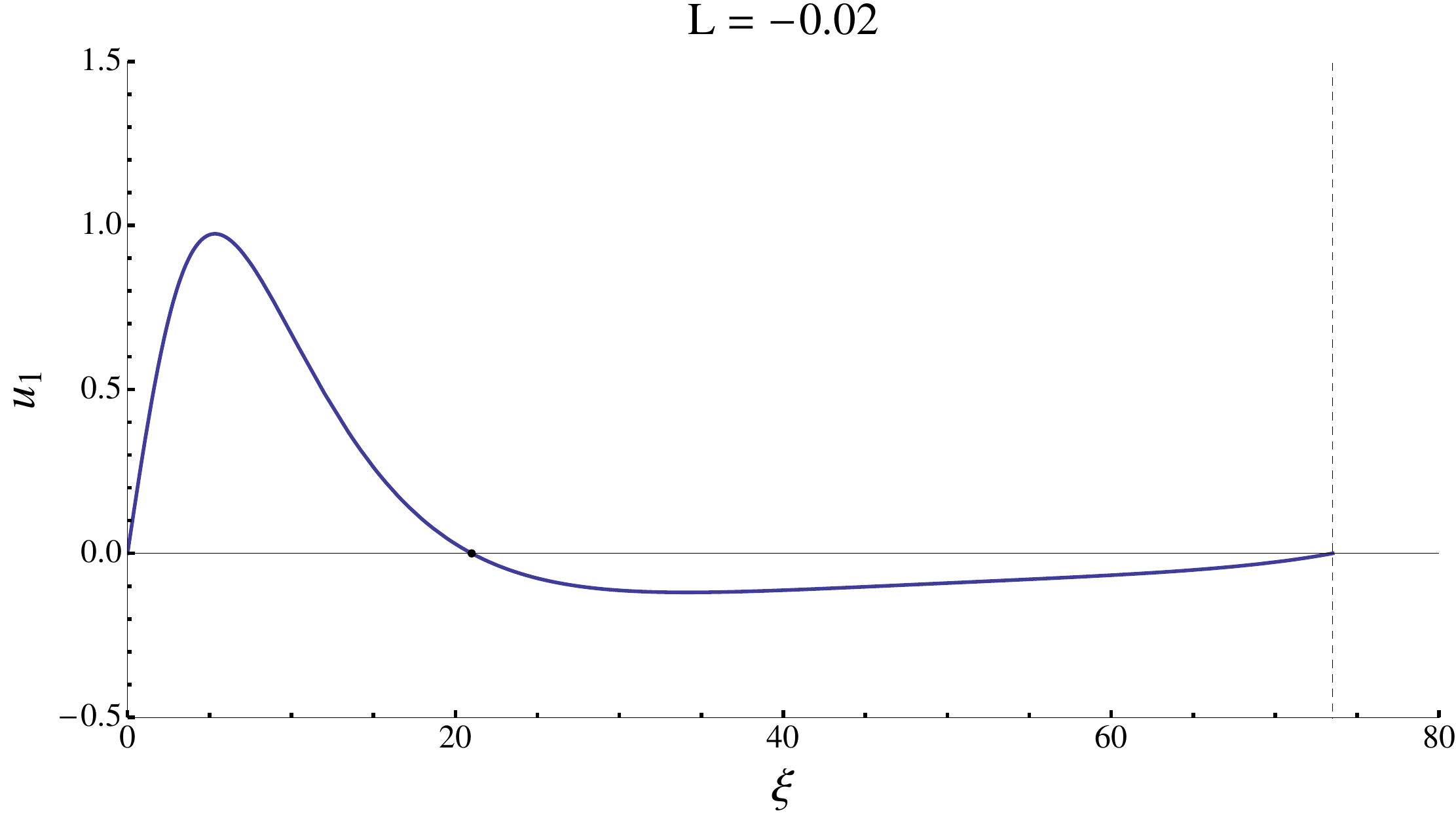}
 \includegraphics[width=9cm]{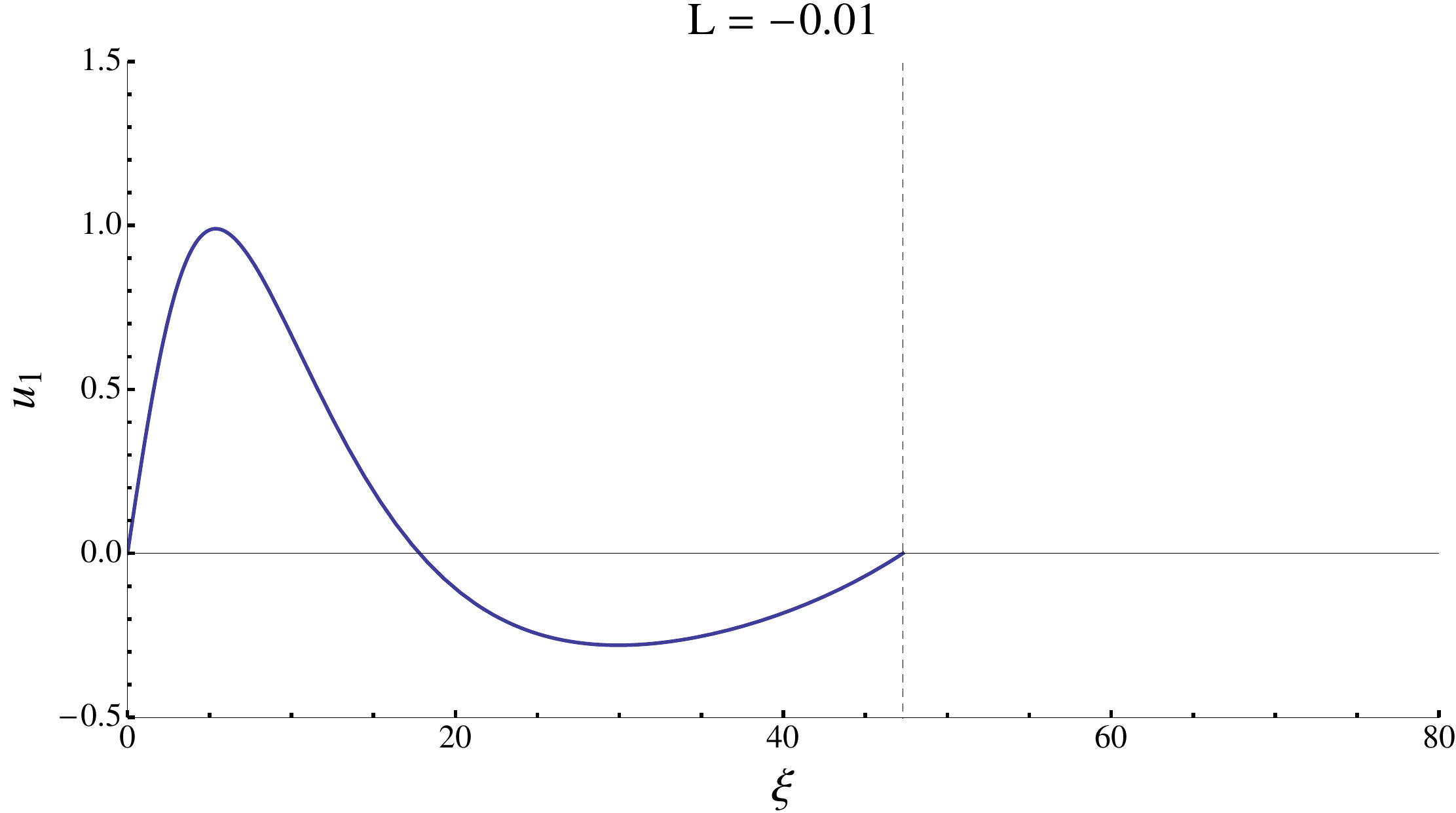}
 \includegraphics[width=9cm]{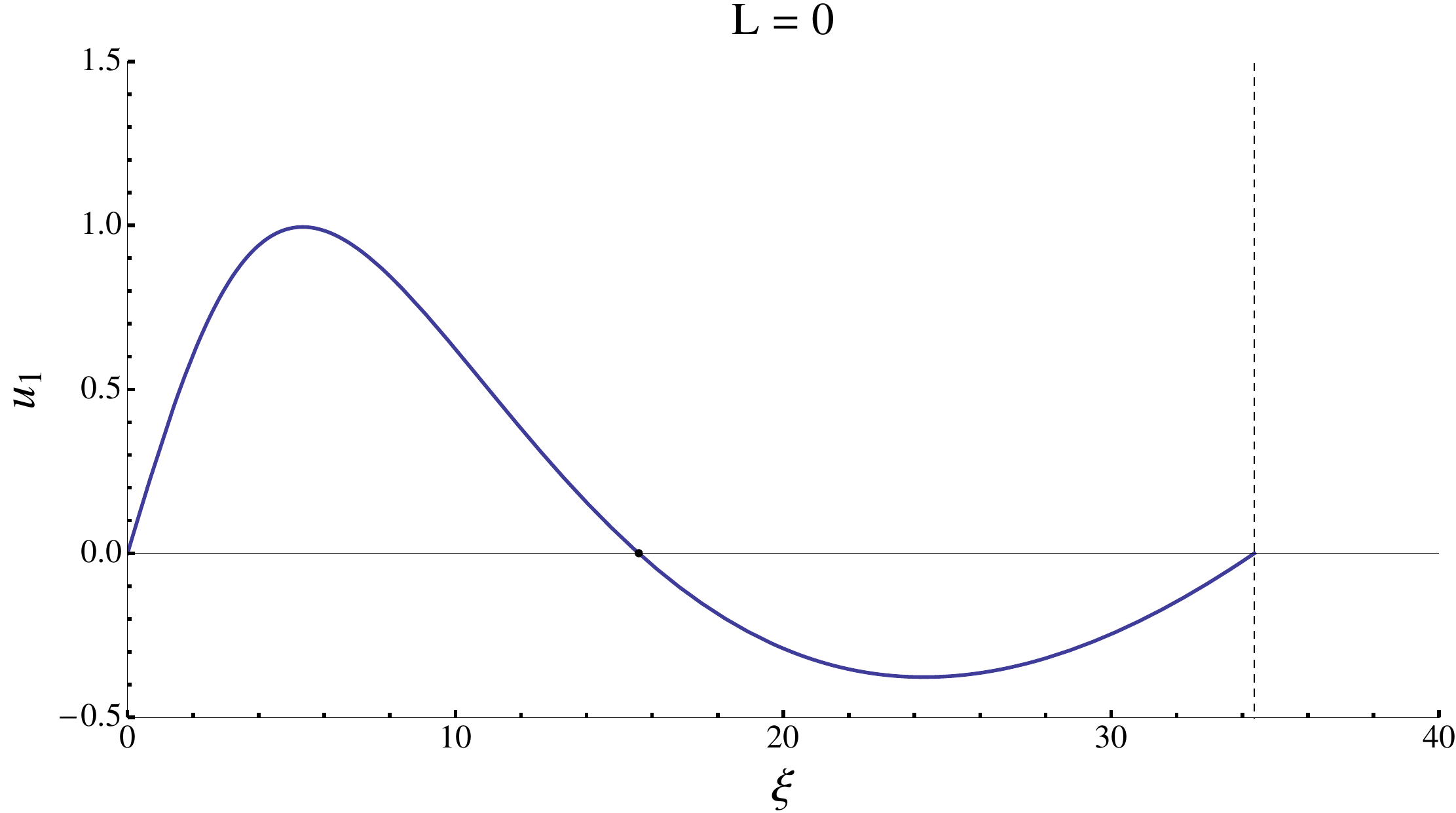}
 \includegraphics[width=9cm]{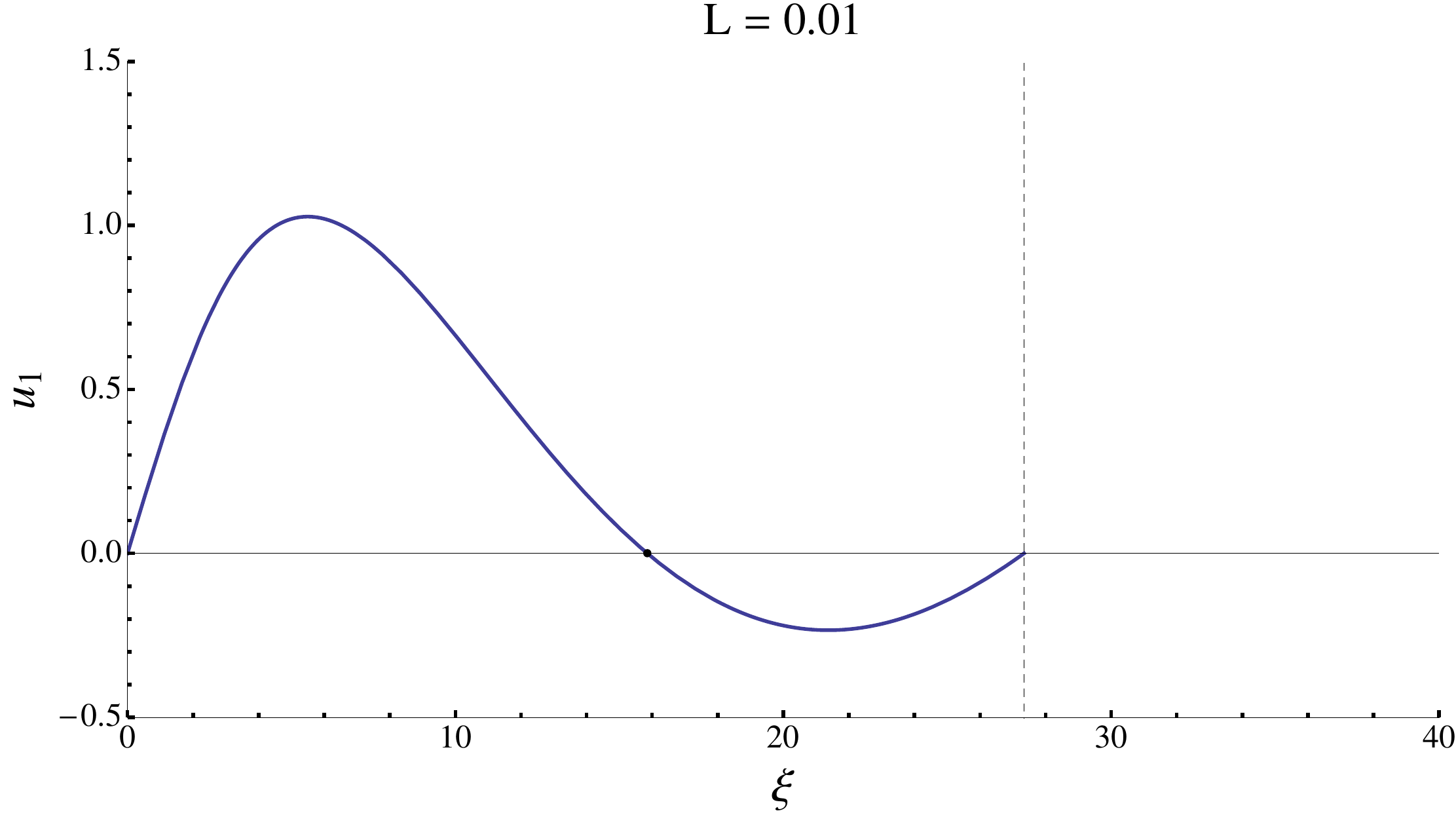}
 \includegraphics[width=9cm]{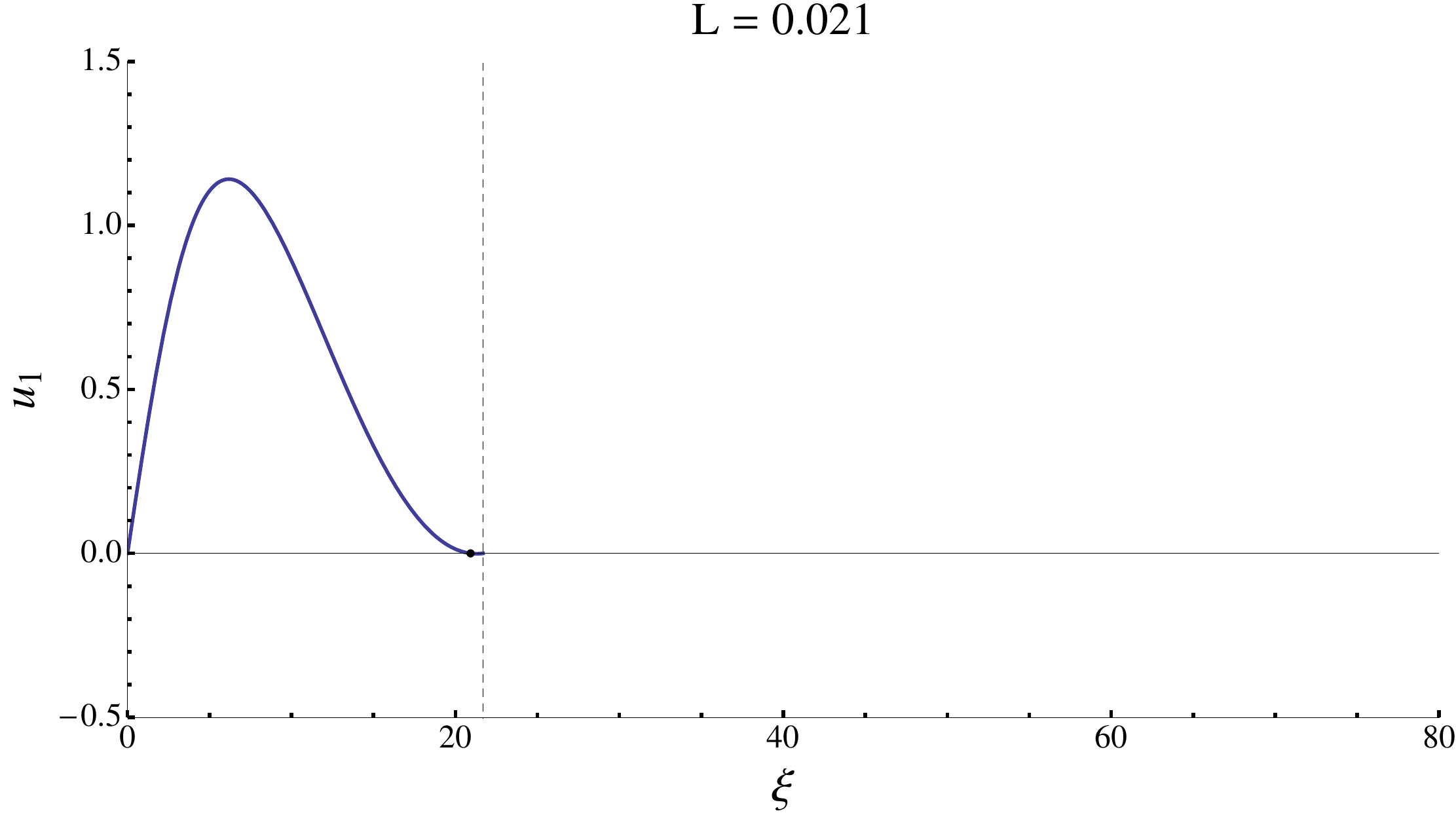}
 \includegraphics[width=9cm]{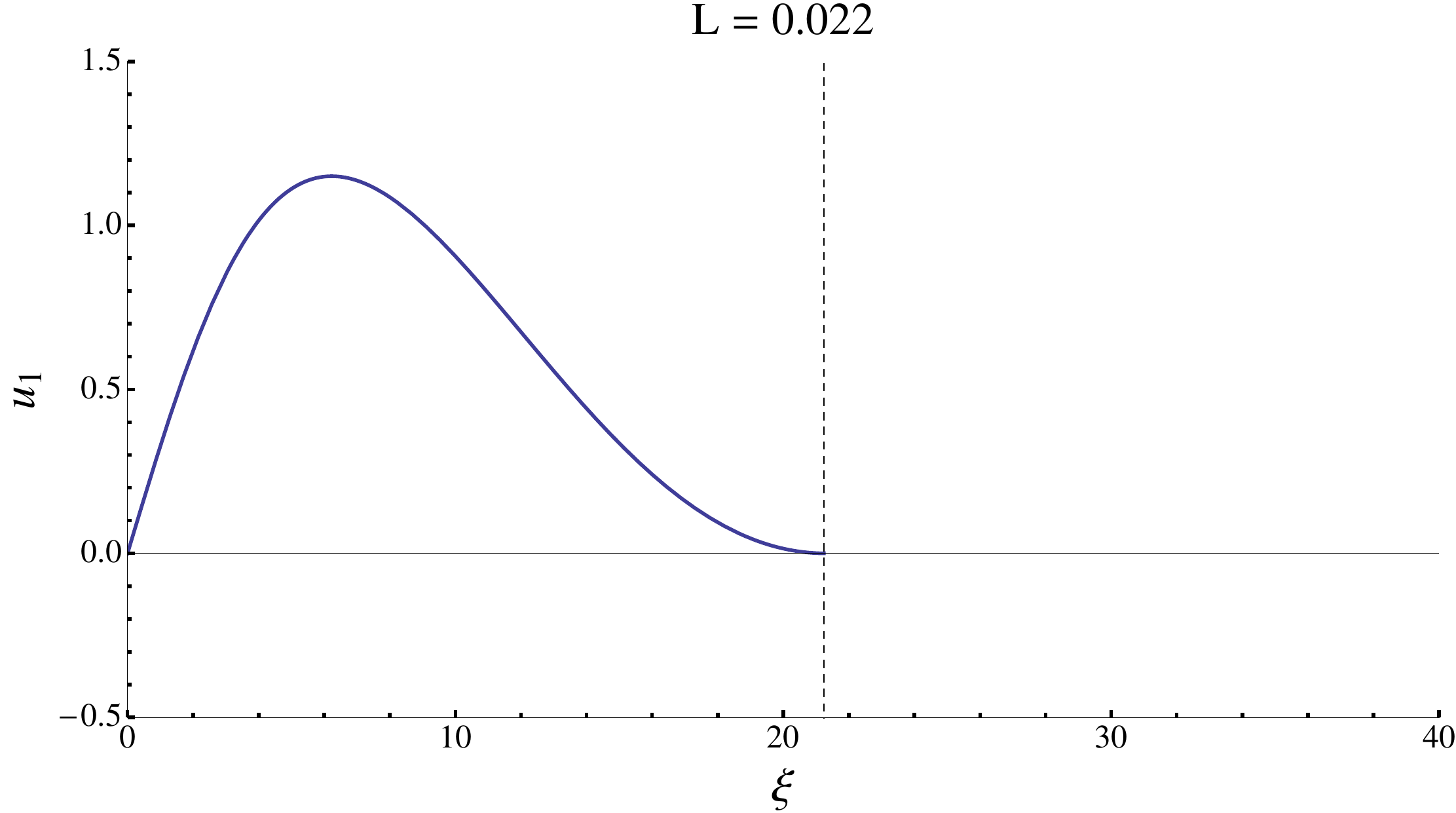}
    \includegraphics[width=9cm]{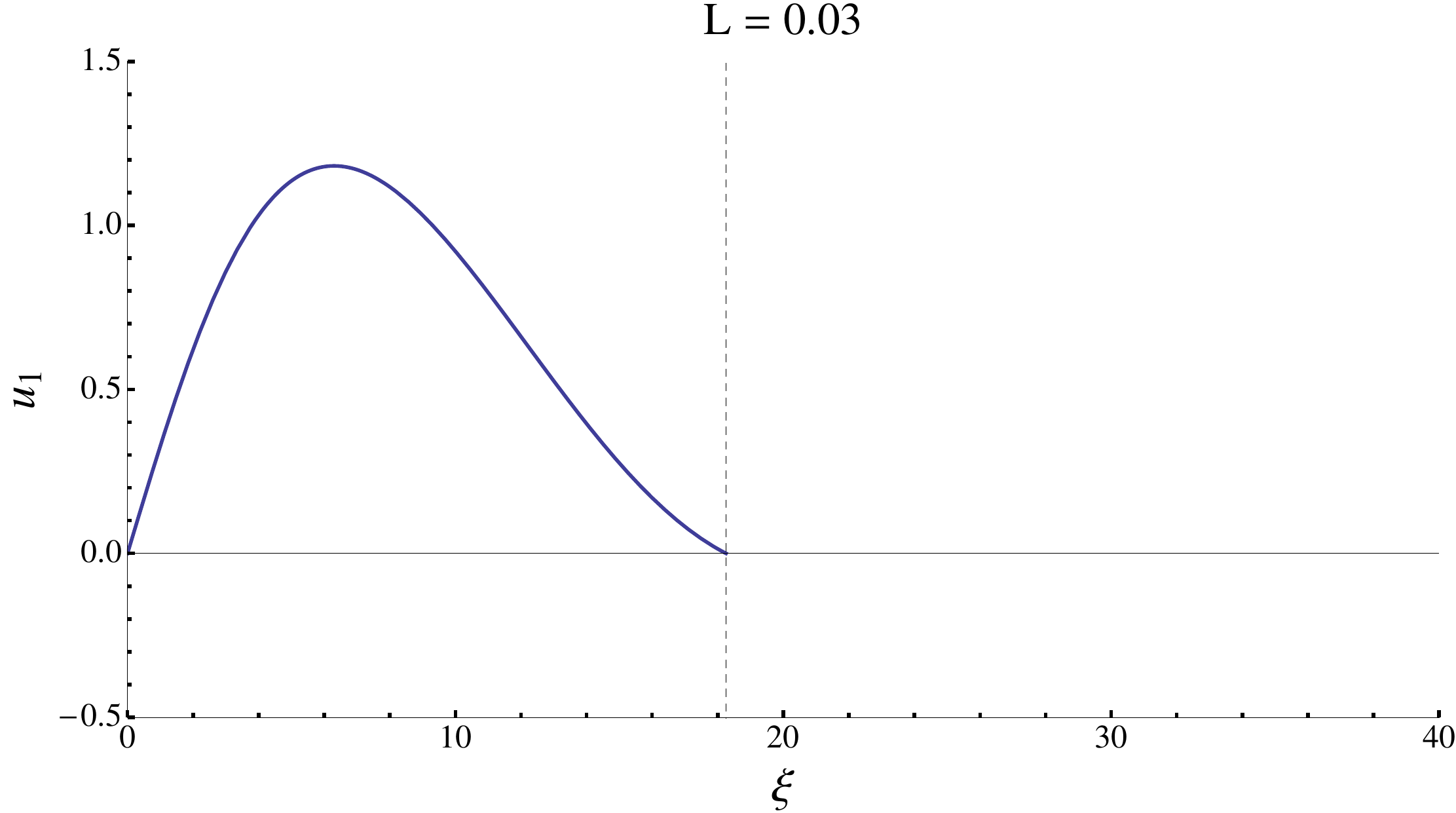}
    \includegraphics[width=9cm]{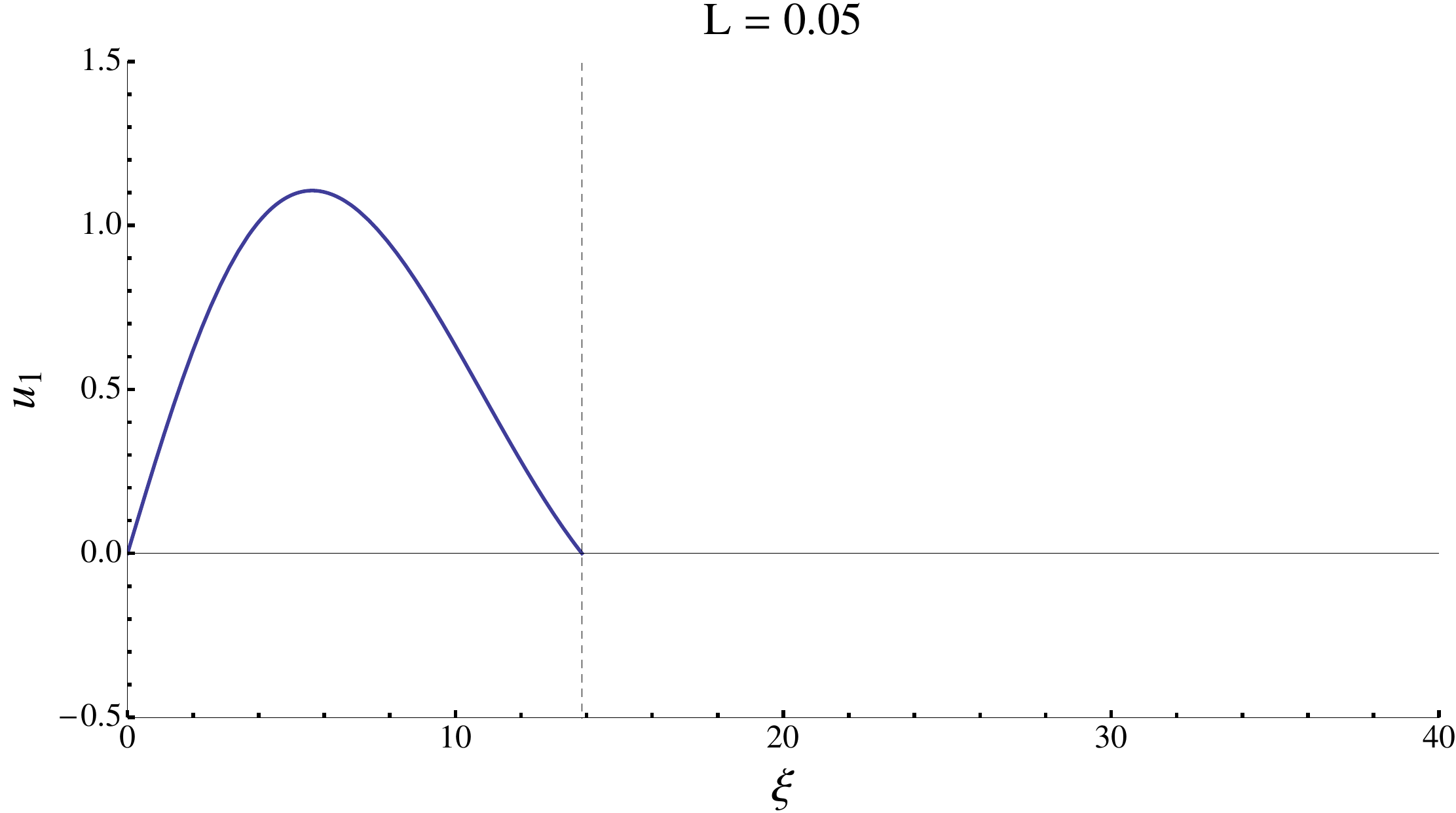}
    \caption{Velocity profiles of the normal modes for the constant $\{E,V\}$ case, obtained by solving Eq.~\eqref{idroEV1}. The modes should be truncated at a value $\xi=\Xi$ corresponding to the vertical dotted lines, in order to satisfy the boundary conditions \eqref{boundary}. Other zeros are displayed. Only modes of lowest $L$ for fixed $\Xi$ are shown. Note that the core-halo structure described in Subsection \ref{constantenergy} disappears when the velocity has no internal zeros, between $L=0.021$ and $L=0.022$.
            }
\label{fig:profvEV}
 \end{figure*}

            \begin{figure*}[h!]
            \subsection{constant $\{T, P\}$ profiles} \label{constantTPprofiles}
 \centering
 \includegraphics[width=9cm]{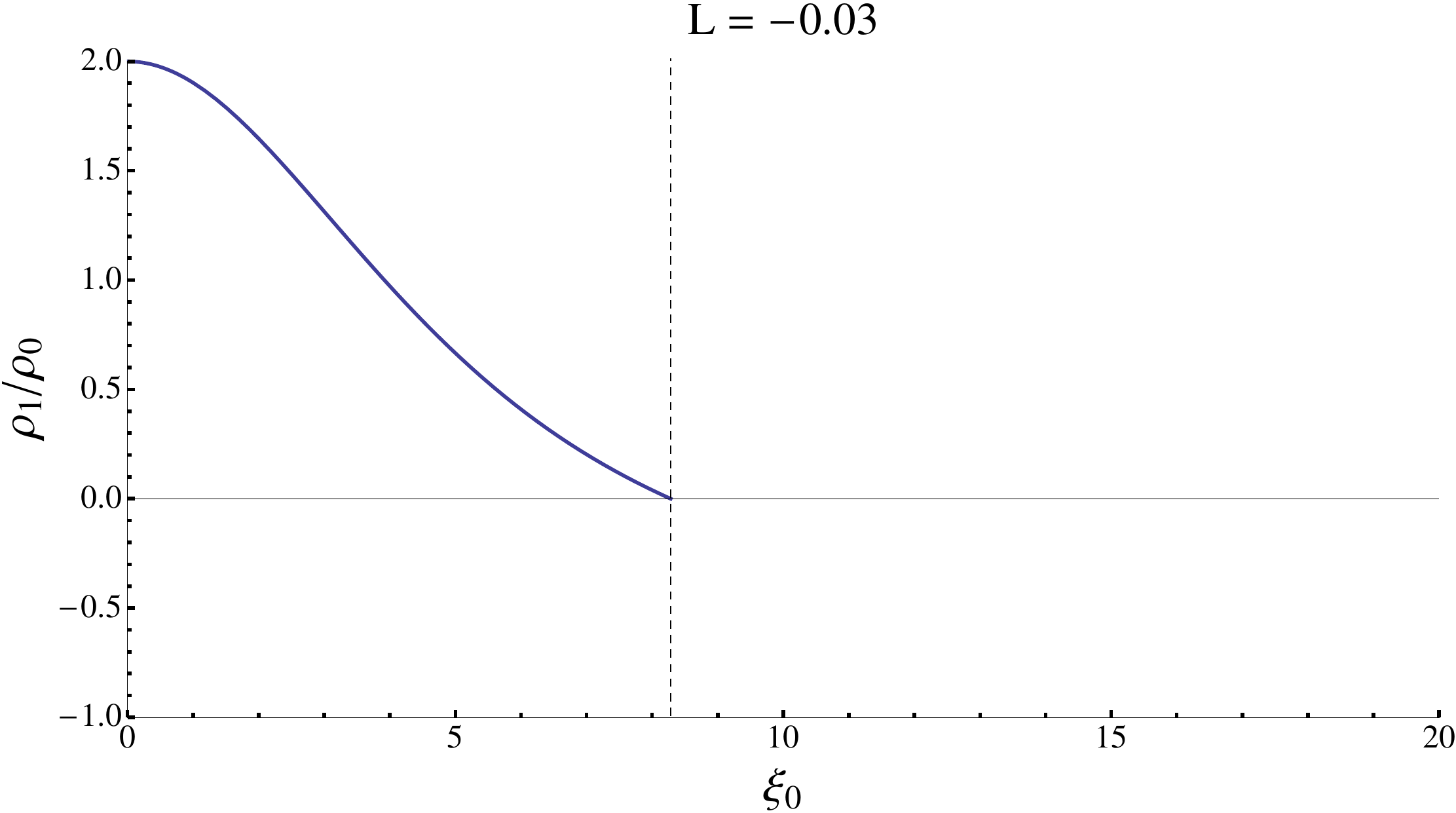}
 \includegraphics[width=9cm]{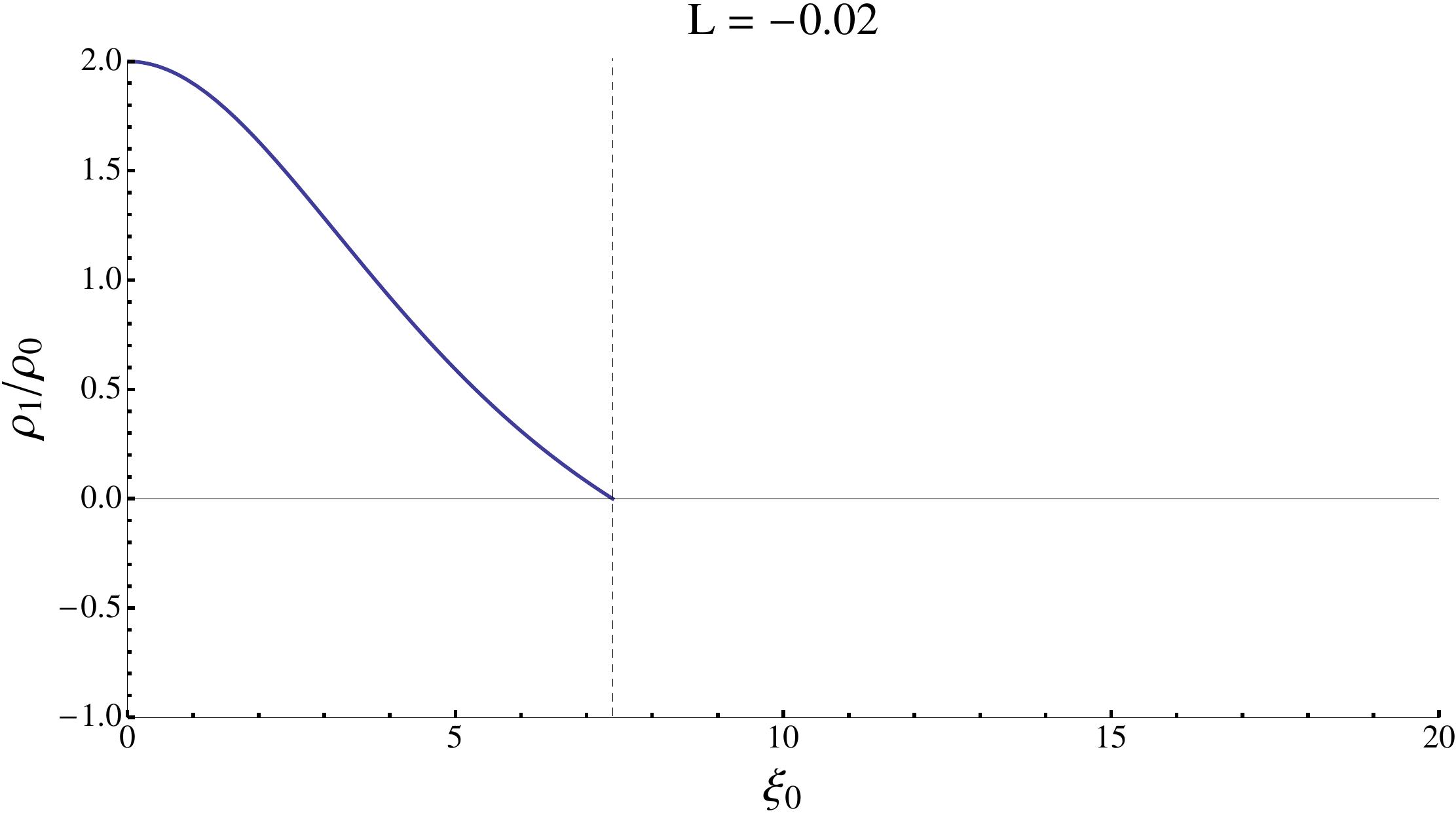}
 \includegraphics[width=9cm]{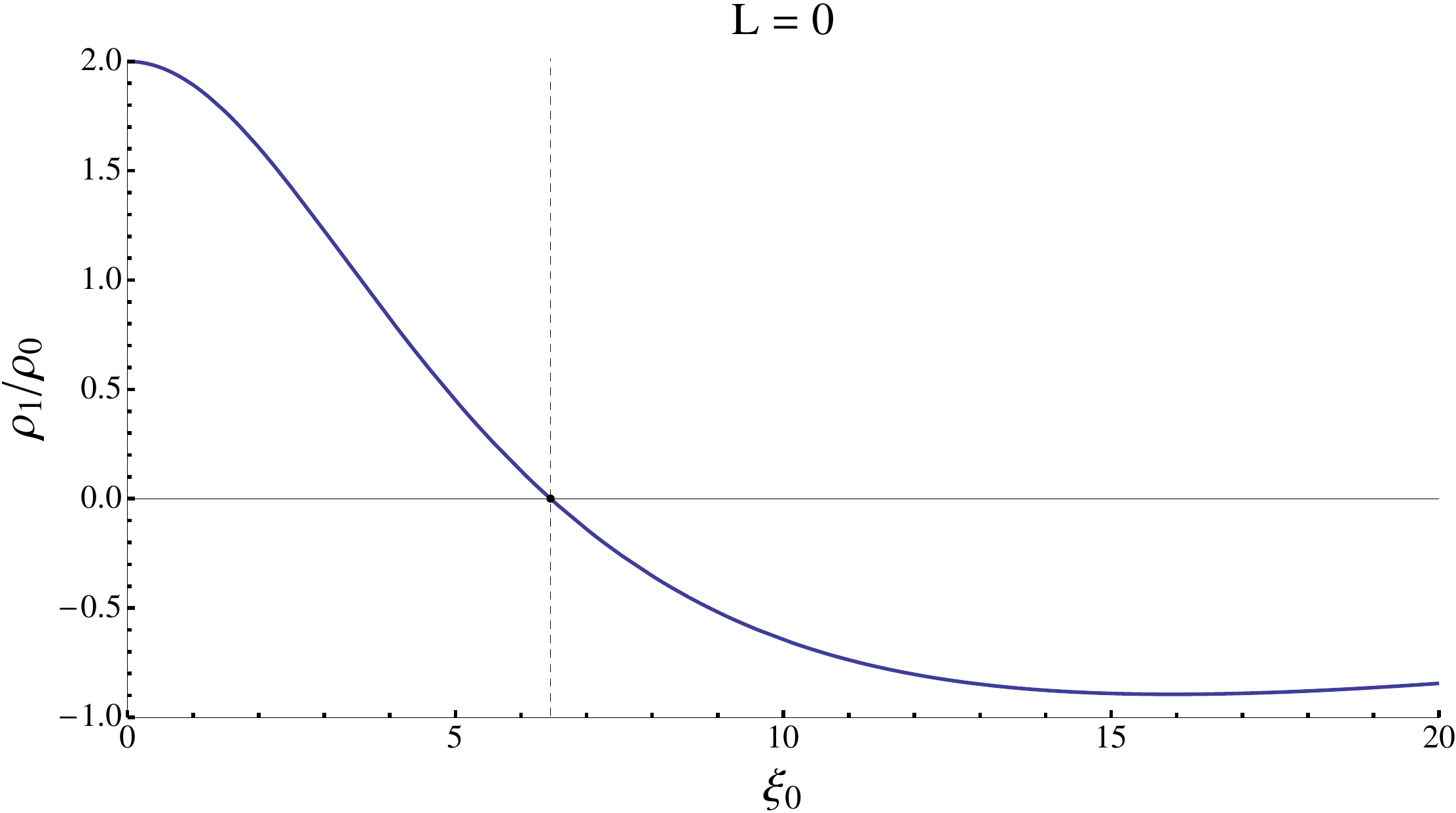}
 \includegraphics[width=9cm]{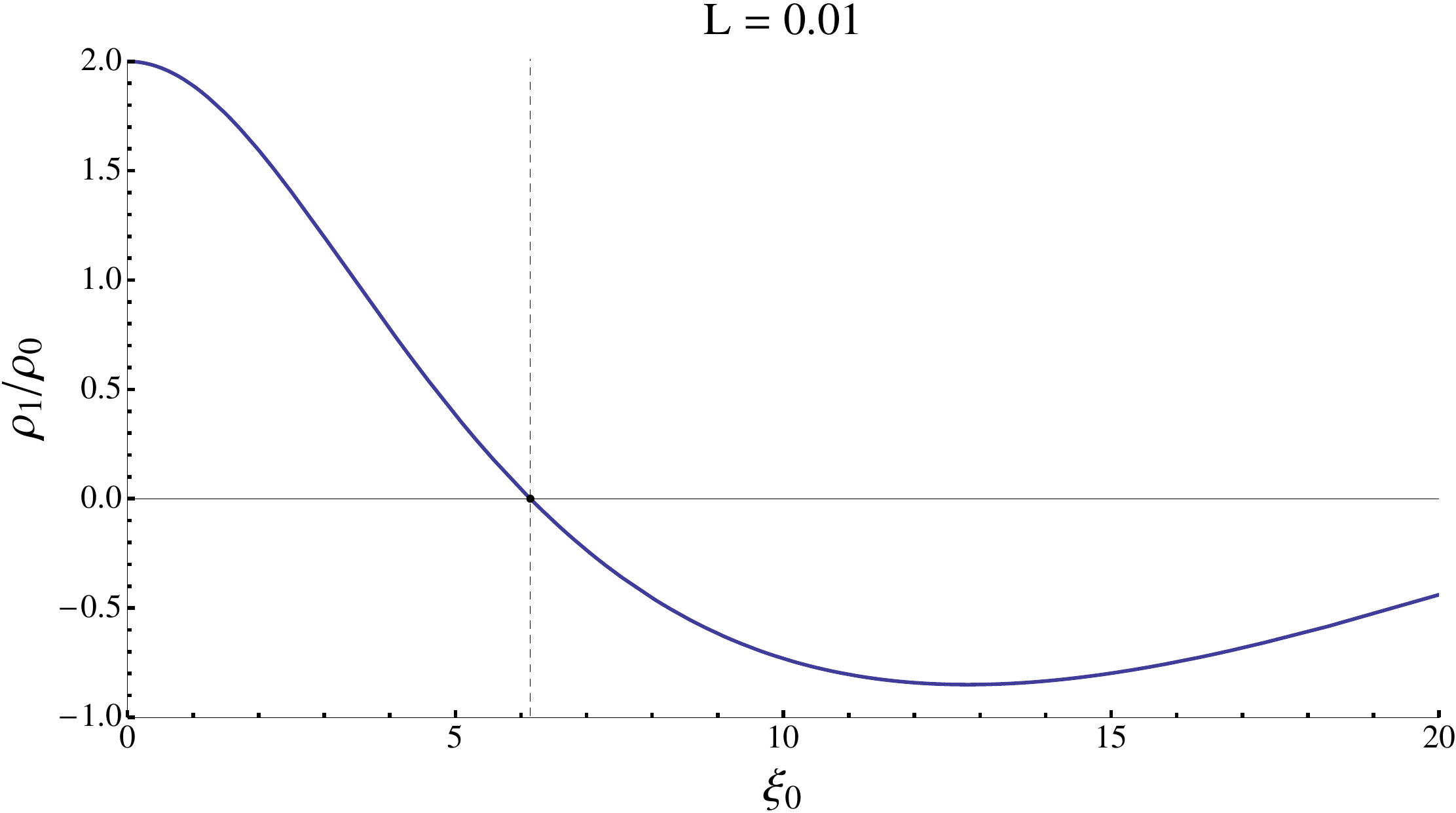}
  \includegraphics[width=9cm]{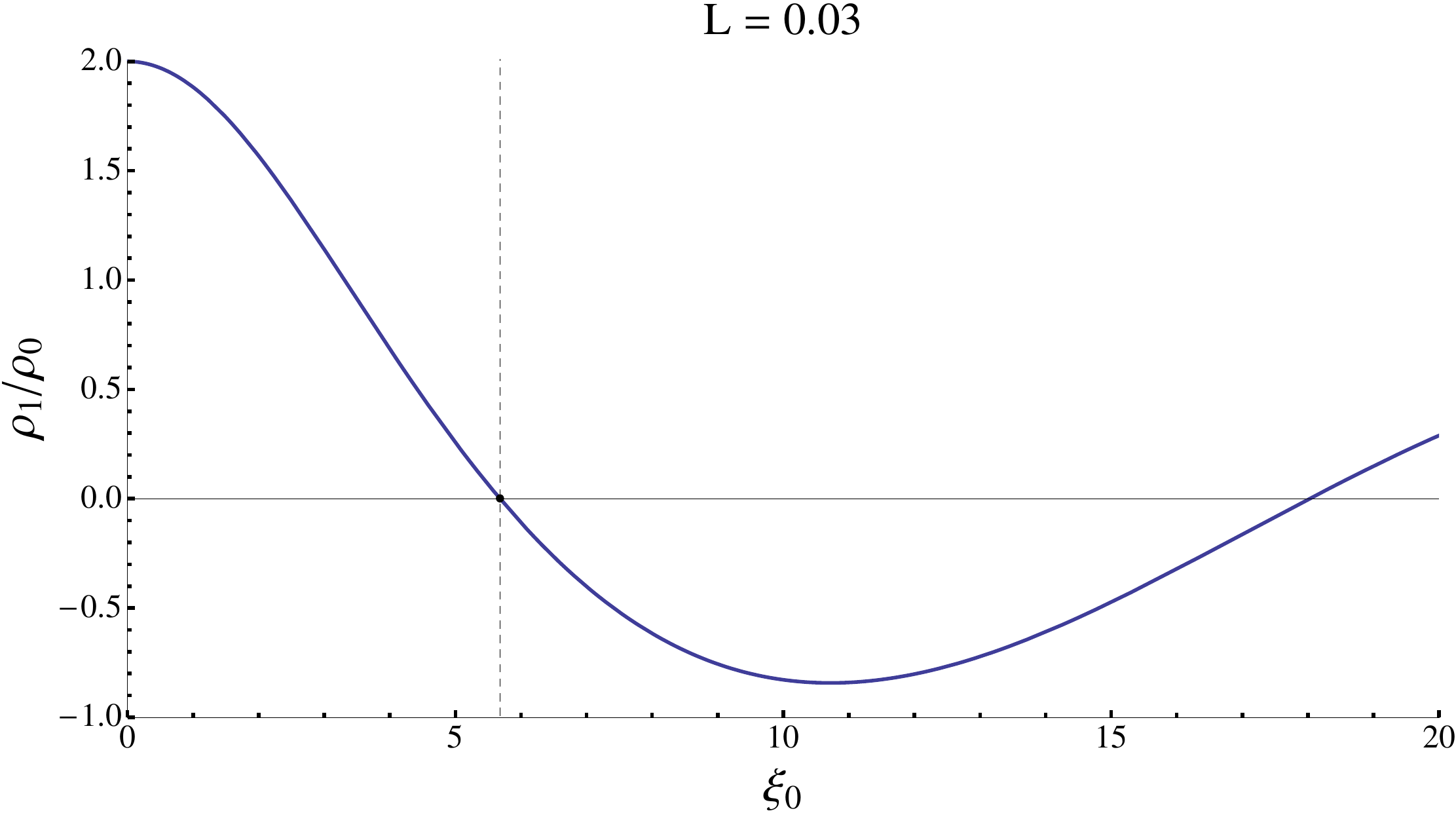}
 \includegraphics[width=9cm]{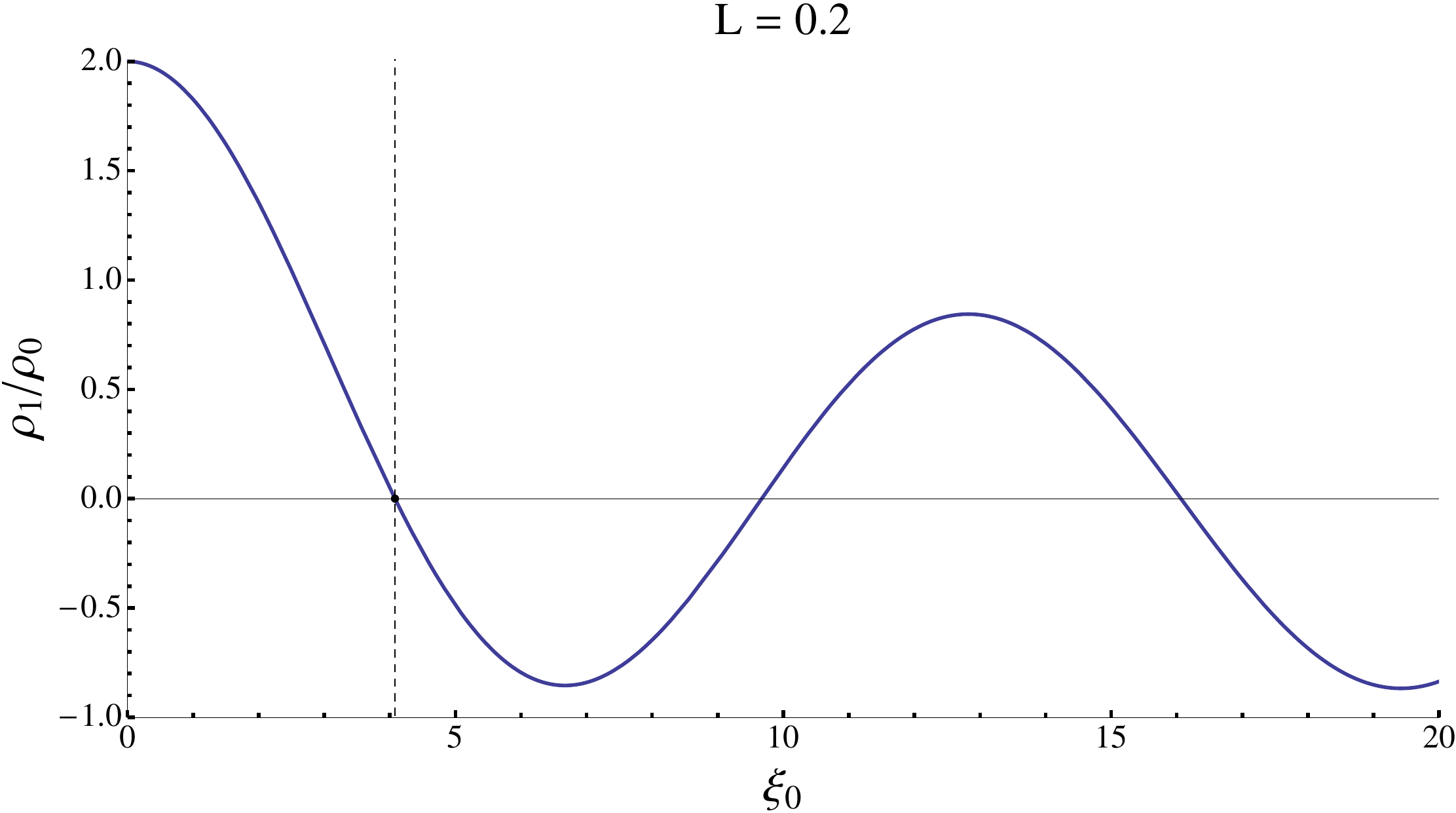}
    \caption{Relative density perturbation profiles $\rho_1(\xi_0)/\rho_0(\xi_0)$ of the normal modes for the constant $\{T,P\}$ case, obtained by solving Eq.~\eqref{perturbedTP}, calculated and displayed here in the Lagrangian representation. The modes should be truncated at a value $\xi_0=\Xi$ where the density profile vanishes, in order to satisfy the boundary conditions \eqref{condcontlagr}. The first zero, which represents the mode of minimum $L$ at given $\Xi$, is indicated by the vertical dotted line. A mode of higher $L$ for fixed $\Xi$ is shown in the $L=0.03$ case; for other cases higher modes can be identified in a similar way. 
            }
 \end{figure*}
 
 \clearpage

\section{Equations for the two-component case} \label{appendix2comp}

In this appendix we summarize the equations of the linear analysis for the two-component case and 
show some examples of the density profiles associated with the modes that characterize the onset of the instability. 
We denote by subscripts $A$ and $B$ the lighter and the heavier component, respectively.

The unperturbed states are the two-component self-gravitating truncated isothermal spheres 
considered by \cite{taff,lightman,yoshizawa,vega2comp,chava2comp}. The density profiles can be written as:
\begin{equation} \label{rho2comp}\begin{cases} 	\begin{alignedat}{1}
		 \rho_{A0}(\xi)=\rho_{A0}(0) e^{-(1+\frac{1}{\beta})\psi} \\ \rho_{B0}(\xi)=\rho_{B0}(0) e^{-(1+\beta)\psi}
	\end{alignedat}   &  \mbox{if } \xi\leq\Xi \\  & \\
	\begin{alignedat}{1}\rho_{A0}(\xi)=0 \\ \rho_{B0}(\xi)=0 \end{alignedat} & \mbox{if } \xi>\Xi~,
	\end{cases} 
\end{equation}
where  $\rho_{A0}$ and $ \rho_{B0}$ are respectively the density profiles of the lighter and heavier component, $\xi \equiv r/\lambda_2 $ is the dimensionless radial coordinate, where $ \lambda_2 \equiv \left[ {kT(1/m_A+1/m_B)}/{4\pi G \rho_0(0)} \right]^{1/2}$ and we denote by $\rho_0(\xi)\equiv \rho_{A0}(\xi)+\rho_{B0}(\xi)$ the total unperturbed density; 
$\Xi$ is the value of $\xi$ at the truncation radius, $\beta\equiv m_B/m_A$ is the ratio of the single-particle masses, $\psi$ is the solution of the following generalization of the Emden equation \eqref{emden}:
\begin{equation} \label{emden2comp} \frac{\mathrm{d} }{\mathrm{d}\xi}\left( \xi^2 \psi'\right) =\xi^2\left[\frac{1}{1+\alpha}e^{-(1+\frac{1}{\beta})\psi}
+ \frac{1}{1+\frac{1}{\alpha}}e^{-(1+\beta)\psi }\right], \end{equation}
\begin{equation}
\psi(0)=\psi'(0)=0,
\end{equation}
where $\alpha=\rho_{A0}(0)/\rho_{B0}(0)$ is the ratio of the unperturbed central densities. The symbol $'$ denotes derivative with respect to the argument $\xi$.

The linearized hydrodynamical equations, governing the evolution of the two-component fluid system for small deviations from the unperturbed states described above, are obtained by generalizing in a straightforward manner the steps leading from Eqs.~\eqref{idro1} to Eq.~\eqref{constTV}. 
The result, which generalizes Eq.~\eqref{constTV}, is the following system of equations that governs the evolution of radial perturbations:

\begin{equation} \label{2compTV} 
\begin{alignedat}{1}
L f_A  = &\left[ -\left(1+\frac{1}{\beta}\right)\psi' \left(\frac{2}{\xi}f_A+ f_A'\right) -  f_A'' -\frac{2}{\xi} f_A' 
+ \frac{2}{\xi^2}f_A\right]\frac{1}{1+\frac{1}{\beta}}  \\  & - \frac{1}{1+\frac{1}{\alpha}} e^{-(1+\frac{1}{\beta})\psi}(f_A+f_B) \\ 
L f_B  = &\left[ -\left(1+\beta\right)\psi' \left(\frac{2}{\xi}f_B+ f_B'\right) -  f_B'' -\frac{2}{\xi} f_B' + \frac{2}{\xi^2}f_B\right]\frac{1}{1+\beta} \\ & - \frac{1}{1+\alpha}e^{-(1+\beta)\psi} (f_A+f_B)~. \end{alignedat} \end{equation}
Here $L={\omega^2}/{4\pi G \rho_{0}(0)}$
represents the dimensionless (squared) eigenfrequency, 
$f_A(\xi)\equiv \rho_{A0}(\xi) u_{A1}(\xi)$ and $f_B(\xi)\equiv  \rho_{B0}(\xi) u_{B1}(\xi)$, where $u_{A1}$ and $u_{B1}$ are the radial velocity perturbations of the two components. 
The boundary conditions are:
\begin{equation}\label{condcont2compdin}
\begin{array}{l}
f_A(0)=f_B(0)=0\\
f_A(\Xi)=f_B(\Xi)=0~.
\end{array}
\end{equation}
Similarly to the one-component case, the two conditions at the center follow from requiring regularity and spherical symmetry, while the two conditions at the truncation radius satisfy the requirement that the radial velocities must vanish at the edge.

The system \eqref{2compTV} for $L=0$ is equivalent to the system that can be obtained by generalizing in a straightforward manner the thermodynamical analysis of \cite{chavanis1}. The latter analysis can be used to find the points for the onset of instability. This proves that the onset of instability occurs at the same values of $\Xi$ in the dynamical and in the thermodynamical approach.

\begin{figure*}[h!]
\centering
\includegraphics[width=9cm]{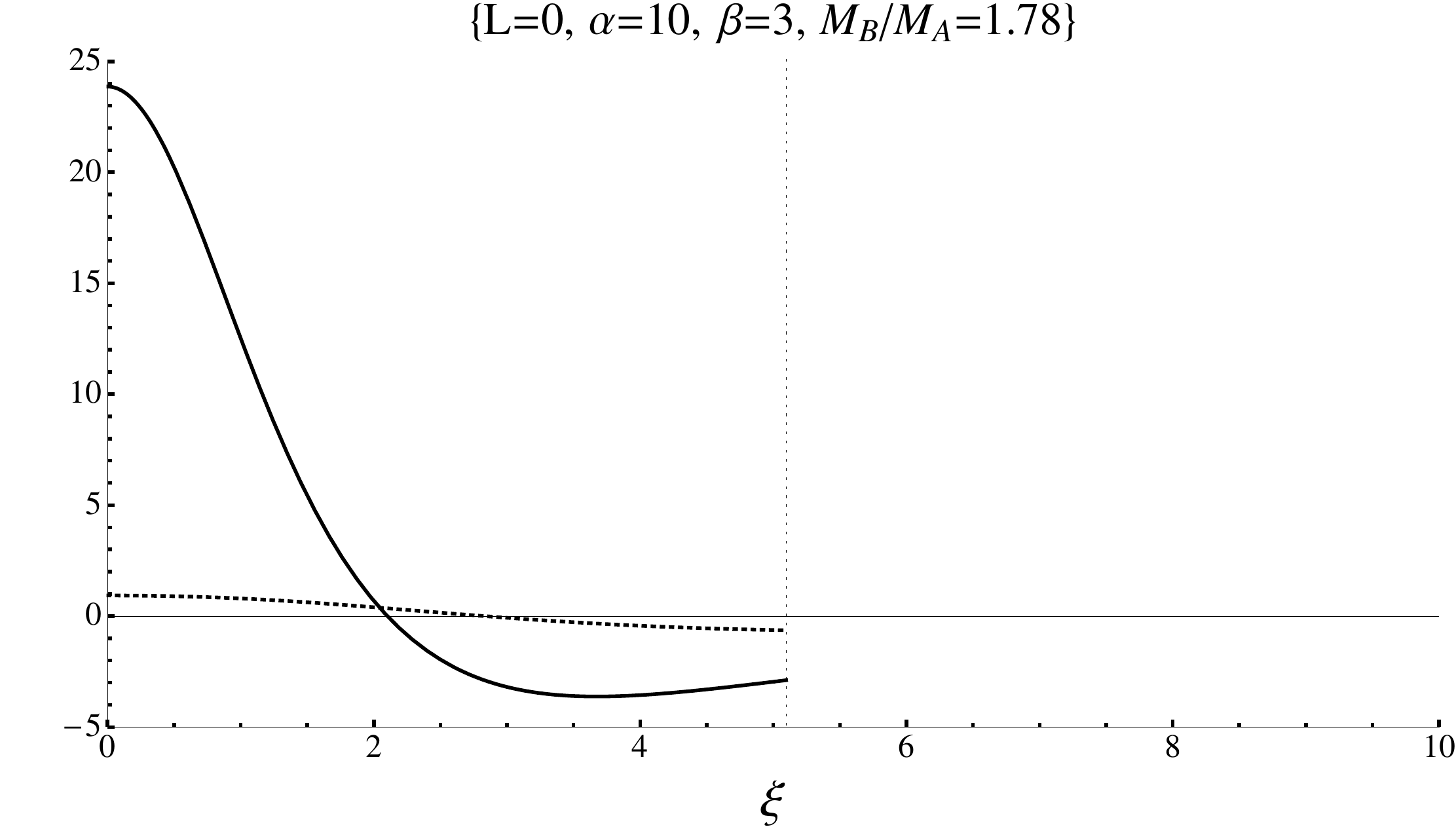}
\includegraphics[width=9cm]{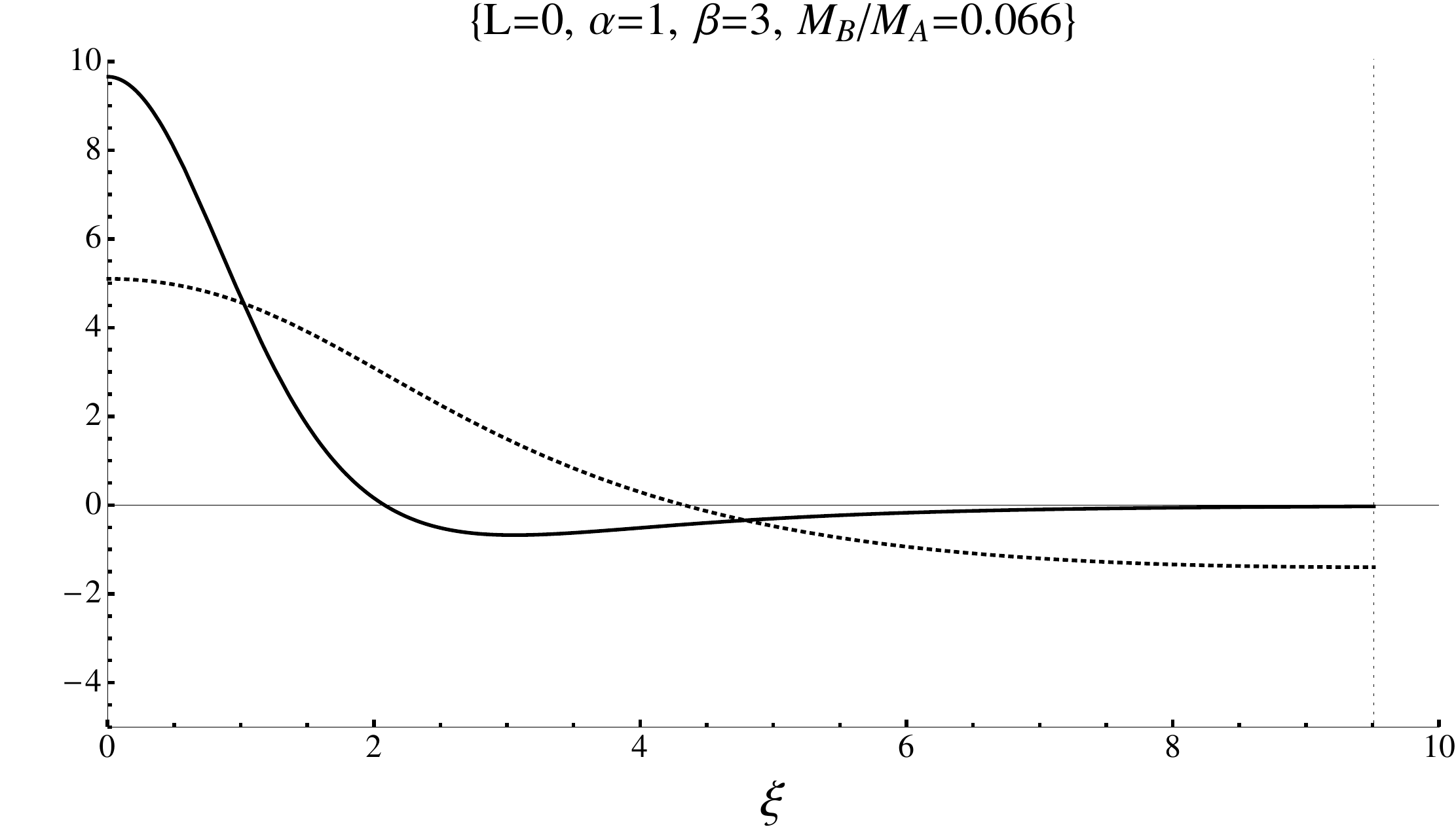}
\includegraphics[width=9cm]{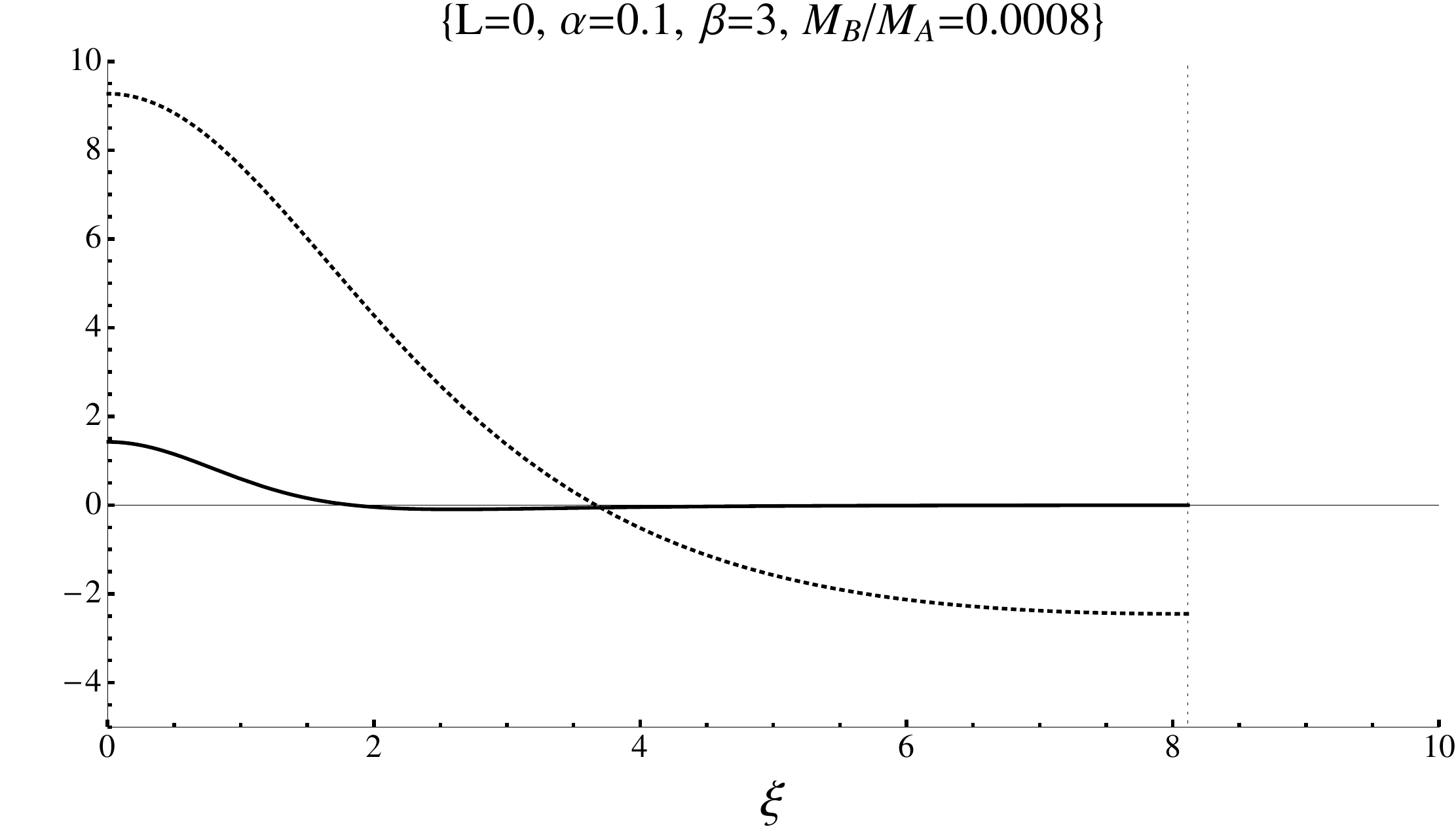}
   \caption{Relative density perturbation profiles $\rho_{1A}(\xi)/\rho_0(\xi)$ (lighter component, dotted line) and $\rho_{1B}(\xi)/\rho_0(\xi)$ (heavier component, solid line) of the normal modes for the two-component constant $\{T,V\}$ case, obtained by solving Eq.~\eqref{2compTV}. The plots show marginally stable modes ($L=0$) of minimum $L$ at given $\Xi$. They represent the density profiles that characterize the onset of the instability for fixed value of $\beta=m_B/m_A=3$ at different values of the total mass ratio $M_B/M_A$. Even for small values of $M_B/M_A$, the density perturbation of the heavier component dominates, thus suggesting that the heavier component is the more important driver of the instability. \label{fig:2comp}
               }
\end{figure*}

In Fig.~\ref{fig:2comp} we show the density profiles for the marginally stable modes ($L=0$) in three different situations, that is, with $\beta=3$ and three different values of $M_B/M_A$. The density perturbation of the heavier component is greater than the density perturbation of the lighter component even for small values of $M_B/M_A$, indicating that the heavier component is the more important driver of the instability.

\end{document}